%% file: sn-article.tex
\documentclass[pdflatex,sn-mathphys-ay, twocolumn]{sn-jnl}

\usepackage{geometry}
\usepackage{graphicx}%
\usepackage{multirow}%
\usepackage{amsmath,amssymb,amsfonts}%
\usepackage{amsthm}%
\usepackage{mathrsfs}%
\usepackage[title]{appendix}%
\usepackage{xcolor}%
\usepackage{textcomp}%
\usepackage{manyfoot}%
\usepackage{booktabs}%
\usepackage{algorithm}%
\usepackage{algorithmicx}%
\usepackage{algpseudocode}%
\usepackage{listings}%
\usepackage{tikz}
\usepackage{pgfplots}
\pgfplotsset{compat=1.18}
\usepgfplotslibrary{patchplots}
\usepackage{mathtools, cuted}
\usepackage{amsmath}
\usepackage[authoryear]{natbib}
\usepackage{newtxtext,newtxmath}

\usepackage{ulem}
\usepackage{structuralanalysis}
\usepackage{prettyref}
\newrefformat{fig}{Figure~\ref{#1}}
\newrefformat{eq}{Eq.~(\ref{#1})}
\newrefformat{tab}{Table~\ref{#1}}
\usetikzlibrary{automata,positioning,arrows}
\usepackage{caption}
\usepackage{subcaption}
\usepackage{float}
\newcommand{\RNum}[1]{\uppercase\expandafter{\romannumeral #1\relax}}
\usepackage{amsmath}
\usepackage{array}

\newtheorem{theorem}{Theorem}
\newtheorem{lemma}[theorem]{Lemma}
\newtheorem{corollary}{Corollary}
\newtheorem{identity}[corollary]{Identity}

\begin{document}

\title[Shape derivatives in bi-material level-set optimization]{Shape derivatives in bi-material level-set optimization with a precise interface: a comparative study}


\author*[1]{\fnm{Majd} \sur{Kosta}}\email{majd-costa@campus.technion.ac.il}

\author[1]{\fnm{Oded} \sur{Amir}}\email{odedamir@technion.ac.il}
\equalcont{These authors contributed equally to this work.}

\affil*[1]{\orgdiv{Faculty of Civil and Environmental Engineering}, \orgname{Technion -- Israel Institute of Technology}, \orgaddress{\street{} \city{Haifa}, \postcode{} \state{} \country{Israel}}}


\abstract{

In this study, we investigate and compare formulations for computing shape derivatives in bi-material level-set optimization with precise modeling of the interface. 
The level-set function is parameterized using B-splines, whose coordinates serve as design variables.
A precise mechanical model is obtained every design cycle, replicating the exact geometry of the bi-material design, using untrimming techniques and IGA on unstructured meshes.
Design sensitivities are formulated by either a ``discretize-then-differentiate'' or a ``differentiate-then-discretize'' approach. 
A detailed comparative study shows the limitations of the latter in terms of accuracy, specifically when stresses near the material interface dominate the stress field. 
The precise representation of the interface facilitates an accurate evaluation of interfacial stresses, and the consistent discretized sensitivities enable to minimize them directly -- highlighting the main advantage of our framework.
Furthermore, reducing the discretized approach by considering only interface control points and a selective set of adjacent control points provides an ideal trade-off between accuracy and numerical efficiency.
This lays the foundations for multi-material shape and topology optimization procedures, considering accurate responses on the interface. 
}




\maketitle

\section{Introduction}\label{sec1}

Topology optimization (TO) is a computational method for optimizing the material layout within a given design space, aiming to maximize the performance of a structure or mechanical part.
TO has developed rapidly with the improvement in computational capabilities over the last decades. 
Advancements in manufacturing technologies enable to fabricate multi-material components, leading to a growing interest in analyzing and optimizing multi-material structures.
Challenges arise when optimizing multi-material components while considering physical responses on the interface between material phases -- such as stress, pressure, contact etc. 
Such problems are sensitive to the representation of the boundary, so well-defined boundaries are crucial for the credibility of the optimization procedure.
Despite some recent advancements in the context of single-material topology optimization, optimizing the stress field in multi-material structures with precise evolving boundaries is still a challenge.

Density-based representations are the most common in stress-constrained topology optimization \citep[e.g.,][]{duysinx1998topology,duysinx1998new, bruggi2008mixed, le2010stress}. 
However, they suffer from a major drawback: the evolving layouts consist of smeared and jagged boundaries, with intermediate density elements near the interface, where it is most critical to capture the stress field accurately. 
\citet{duysinx1998topology} showed that the stresses of the post-processed body-fitted model exceed the stress constraint by up to 13\%. 
Slightly better accuracy has been demonstrated in more recent publications \citep[e.g.,][]{salazar2018adaptive, da2019stress,amir2021efficient}. 
Nevertheless, violations are expected to inflate in multi-material structures, due to the discontinuity of the stress field -- posing a challenge for stress-based optimization with multiple materials.

Another common representation in TO is based on the Level Set Method (LSM). 
The central idea of LSM for TO is to implicitly represent the interface between the material phases by level sets of a function of a higher dimension 
\citep{wang2003level,allaire2004structural,van2013level}.
Even though the level set provides an accurate definition of the interface, this accuracy typically does not propagate to the computation of stresses by Finite Element Analysis (FEA). 
Three distinct approaches can be found in the literature for mapping the level-set-based geometry to a mechanical model \citep{van2013level}: 
Ersatz material approach; immersed boundary techniques; and evolving body-fitted meshes. 
The Ersatz material approach leads to intermediate density elements on the boundary, hence it reproduces the difficulties associated with density-based procedures.
Immersed boundary methods -- mainly the Extended Finite Element Method (XFEM) \citep{van2007stress,noel2017shape,sharma2018stress}, Generalized Finite Element Method (GFEM) \citep{van2021interface} and cutFEM \citep{villanueva2017cutfem,andreasen2020level} -- account for the discontinuity of the design across the interface, but special treatment is needed for accurate stress evaluation.  
In body-fitted approaches, the boundaries are typically discretized linearly, using a triangular finite element mesh \citep{eschenauer1994bubble, schleupen2000adaptive, allaire2013mesh,  christiansen2014topology, christiansen2015combined, feppon2020topology, feppon2021body}. 
The linear discretization of the interface integrates the geometry and the analysis model when both are defined using triangulations (or tetrahedra in 3-D). 
When the geometry is represented by smooth, high-order functions, as usual in Computer Aided Design (CAD), the accuracy of the interface is compromised. 
Furthermore, low-order finite element meshes demonstrate low accuracy in capturing interfacial stresses, particularly stress concentrations. 
Hence, there is room for exploring more accurate representations, where the geometry is smooth and compatible with the analysis model.  

Recently, \citet{shakour2022stress} presented stress-constrained TO with precise and explicit geometry using untrimming and Iso Geometric Analysis (IGA) for single-material structures.
The key point of IGA is that it adopts the bases of CAD -- such as B-splines and T-splines -- to represent the geometry as well as the solution field for numerical analysis, leading to a seamless integration between the geometrical and mechanical models \citep{cottrell2009isogeometric}. 
While aiming at multi-material problems, our work follows an approach similar to  \citet{shakour2022stress}: the topology is defined following the LSM where a bi-cubic B-spline surface is used to parameterize the level-set function, leading to a crisp and explicit representation of the boundaries.
We adopt untrimming techniques and IGA on unstructured meshes to simulate the response according to the exact boundaries as they evolve during optimization.
The explicit and smooth boundary representation not only enhances the accuracy of the analysis, but also opens up space to apply a variety of approaches for sensitivity analysis (SA) on a well-defined material interface.
Therefore, the focus of the current work is on investigating the accuracy and efficiency of various techniques for computing the design sensitivities in this particular spline-based setup, where design variables are the parameters of the level set function and the analysis mesh replicates the zero level-set contour precisely. 

Various approaches for sensitivity analysis in similar, though not identical setups, can be found in the literature.
\citet{allaire2008minimum, allaire2011damage} suggested C\'ea's method for minimizing the compliance and the domain stresses in single and multi-material domains, using the Ersatz material approach on the boundaries. 
Later on, \citet{feppon2019shape, feppon2020topology, feppon2021body} introduced shape differentiation for applications of topology optimization in weakly coupled multi-physics problems with body-fitted meshes. 
In their work, the lift functional -- the aerodynamic force generated by the integration of pressure and shear stress distributions -- was reformulated as a volume integral. 
The issue of stress evaluation at the material interface in topology optimization for multi-material structures was also addressed by \citet{liu2020multi}. 
They utilized an interface-conforming finite element mesh and employed C\'ea's method for sensitivity analysis.
Nonetheless, the interfacial stress was approximated in the form of a narrow-band domain integral, where a smooth approximation of the Dirac delta function was utilized to describe the discontinuity along the material interface. 

To the authors' knowledge, none of the above-mentioned studies fully recovers the evaluation and optimization of stresses at the material interface.
This calls for a careful formulation of the design sensitivities that will allow to optimize the stress field on the interface between materials.
In this paper, we show how the combination of a level-set representation and IGA creates a variety of possibilities for computing shape derivatives.
We investigate three distinct approaches and demonstrate how the choice of formulation affects their accuracy and computational efficiency, laying the foundations for consistent shape and topology optimization on smooth, precise evolving boundaries. 

The remainder of the paper is organized as follows.
A concise introduction to the LSM and the meshing procedure is presented in Section \ref{sec:preliminaries}.
The problem formulation and detailed sensitivity analyses are provided in Section \ref{sec:formulation}.
The core of the article is Section \ref{sec:insights}, where we investigate, compare and discuss the various formulations. 
Numerical examples in Section \ref{sec:examples} show the performance of the various formulations within a complete shape optimization process.
Finally, concluding remarks and a discussion are presented in Section \ref{sec:conclusion}. 

\section{Preliminaries}\label{sec:preliminaries}
In this study, we introduce bi-material LSM-based shape optimization with a precise boundary representation using IGA. 
In this section, we provide a brief introduction to the underlying methodologies that form the basis of this work. 
For detailed descriptions of the LSM and its application to shape and topology optimization, the readers are referred to \citet{osher1988fronts, sethian1999advancing, osher2001level,wang2003level,allaire2004structural,van2013level, wei201888}.
Further information on IGA can be found in \citet{hughes2005isogeometric,cottrell2009isogeometric} and a detailed account of the method we use for precise evolving boundaries can be found in \citet{shakour2021topology,shakour2022stress,shakur2024isogeometric}.
Note that the term ``shape optimization'' is used throughout this article, but the design sensitivities can be used  without further modifications for topology optimization as well, following the concepts of LSM-based topology optimization \citep[e.g.,][]{allaire2021shape}.   
To stay in line with the terminology defined by \citet{shakour2021topology}, herein the term ``geometrical model'' refers to the geometry of the optimized structure as it evolves during optimization, while the term ``mechanical model'' refers to the simulation model for IGA.  

\subsection{The level set method}

Aiming to analyze the motion of shapes and surfaces, the LSM has been investigated in the area of topology optimization for the last few decades.
Commonly, the LSM implicitly defines the interface between the material phases by a Level Set Function (LSF) $\phi$ in a higher dimension, as

\begin{equation}
    \begin{aligned}
    \phi(\mathbf{x}) > 0 & \Leftrightarrow \mathbf{x} \in \Omega_1 && \text{(phase 1)} \\
        \phi(\mathbf{x}) = 0 & \Leftrightarrow \mathbf{x} \in \Gamma && \text{(interface)} \\
        \phi(\mathbf{x}) < 0 & \Leftrightarrow \mathbf{x} \in (D\setminus \Omega_1) = \Omega_2 && \text{(phase 2)} \\
    \end{aligned}
\end{equation}
where $\Omega_i$ and $\Gamma$ are the material phase $i$ and the material interface, respectively; $D$ is the design domain where $\phi$ is defined and $\mathbf{x}$ is a point in $D$.
In principle, the material interface in both geometrical and mechanical models is defined by $\phi(\mathbf{x}) = 0$. 
A slightly different choice is taken in this study, as will be described in the next section.
For design domains $D\in\mathbb{R}^2$, the LSF is parameterized by a bi-cubic B-spline surface in $\mathbb{R}^3$, yielding structural boundaries and material interface defined as cubic spline curves.

\subsection{Generating the geometrical and mechanical models}
\label{sec:gener}

The process for creating both the geometrical and mechanical models is based on the methodologies presented by \citet{shakour2021topology,shakour2022stress} for single-phase structures and \citet{shakur2024isogeometric} for multi-material structures. 
The detailed procedure is thoroughly described in the latter work, so we offer only a brief summary of the key points herein.
The procedure is divided into two main phases -- first producing the geometrical mesh and subsequently the mechanical mesh.

\subsubsection{The geometrical model}
\begin{figure*}[t!] 
\centering
\subfloat[]{\label{fig:FirstDomain} 
\includegraphics[width=0.33\textwidth]{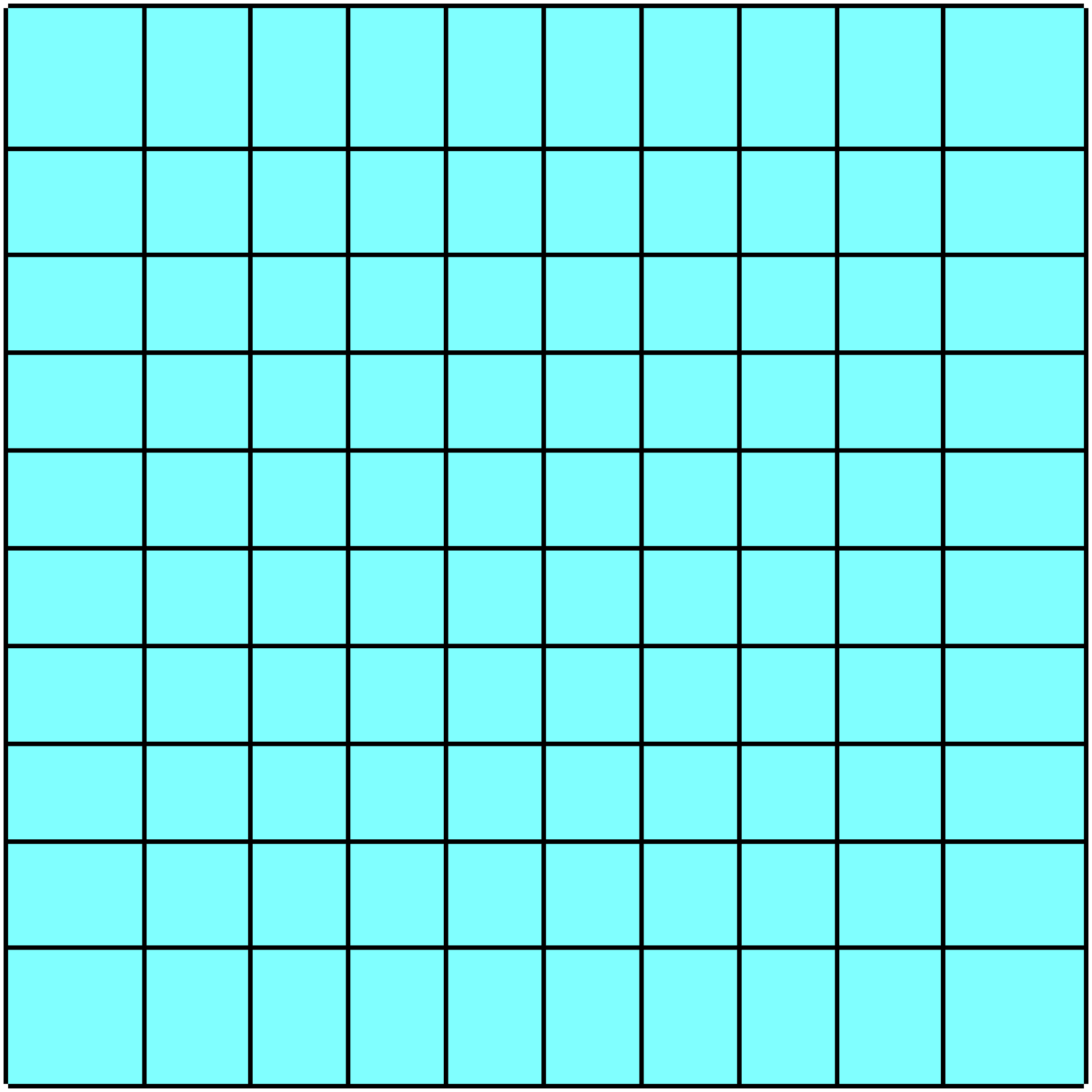}}
\subfloat[]{\label{fig:FirstMesh}
\includegraphics[width=0.33\textwidth]{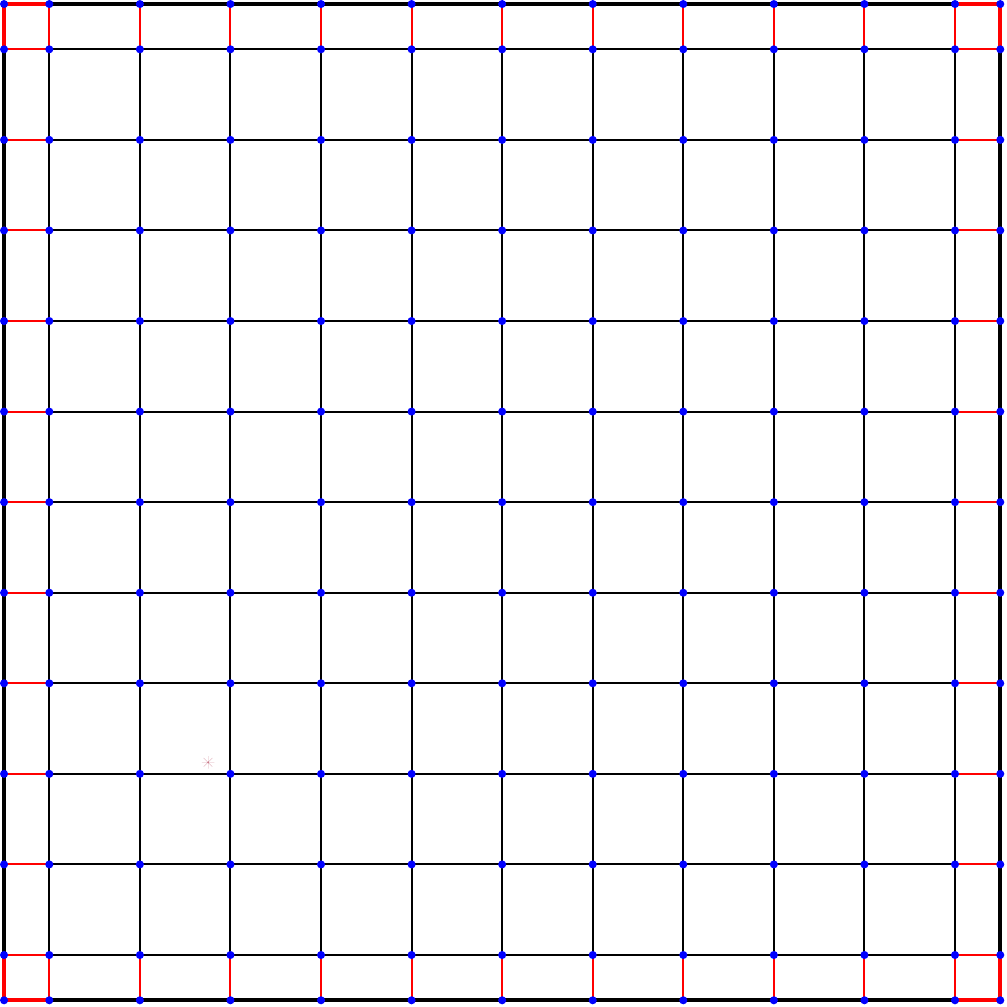}}
\subfloat[]{\label{fig:WithBound}
\includegraphics[width=0.33\textwidth]{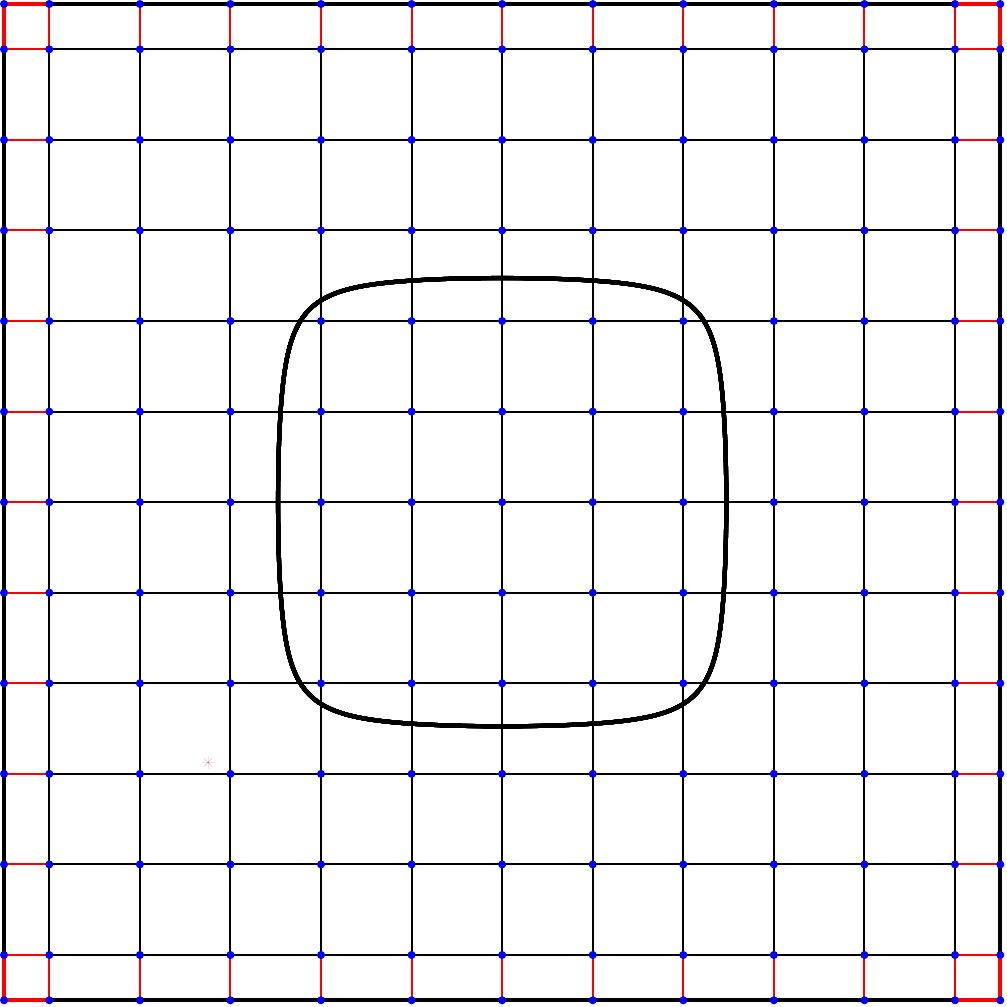}}\\
\subfloat[]{\label{fig:TrimmedB}
\includegraphics[width=0.33\textwidth]{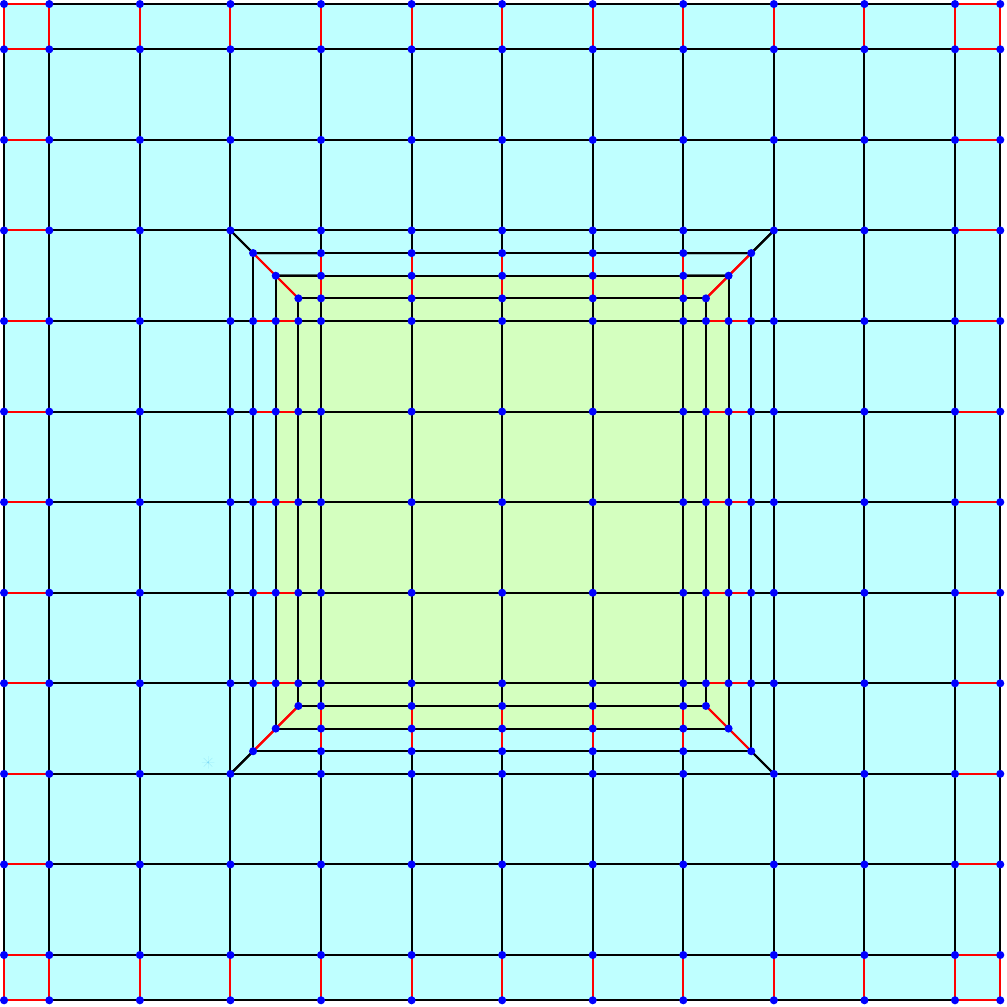}} 
\subfloat[]{\label{fig:Trimmed}
\includegraphics[width=0.33\textwidth]{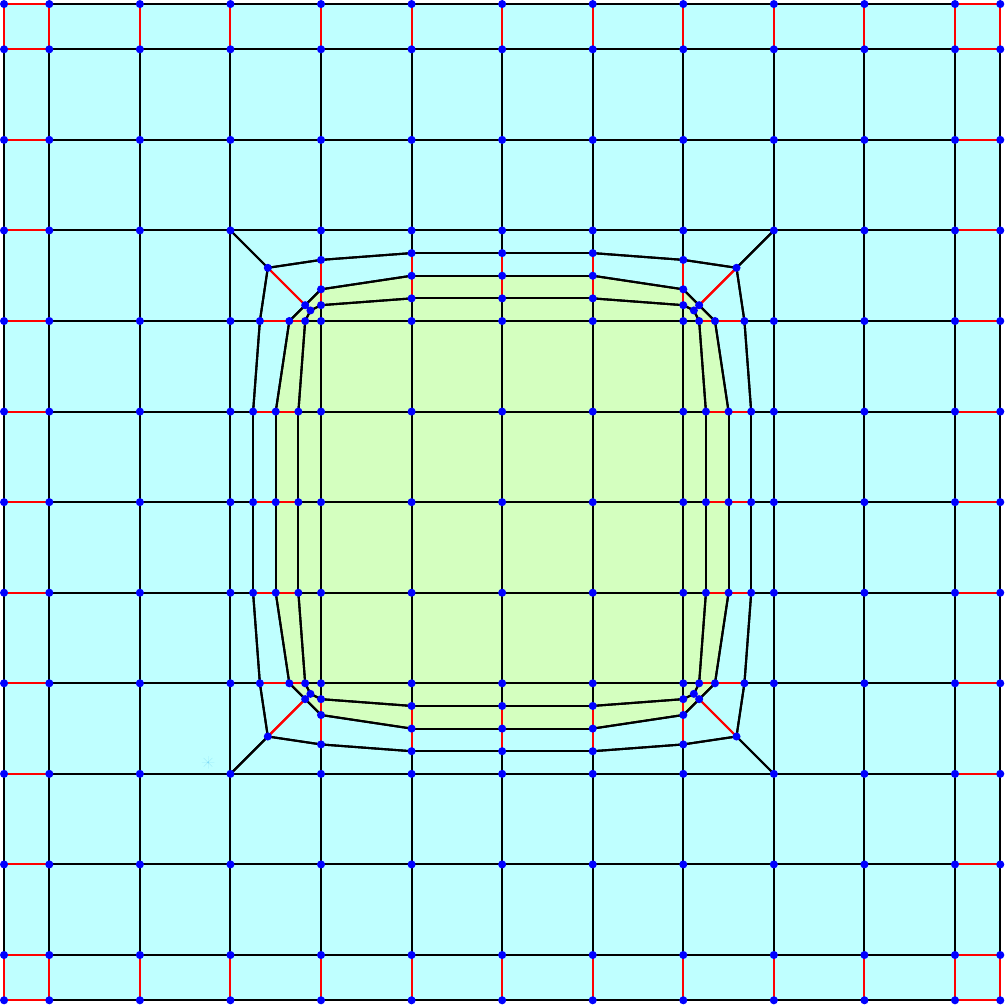}} 
\subfloat[]{\label{fig:Trimmed_AfterLap}
\includegraphics[width=0.33\textwidth]{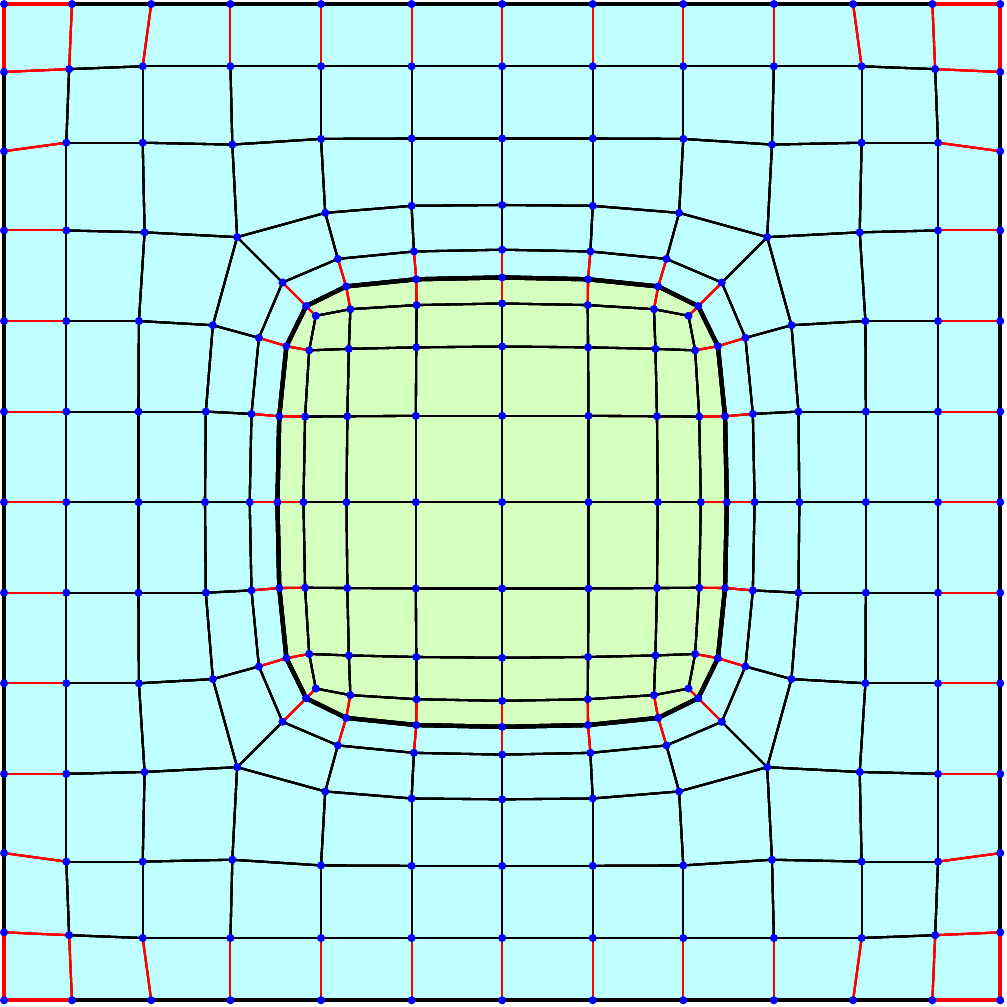}} 
\caption{Generating the geometrical model: Panels (a) and (b) present the initial physical domain and control mesh of the bi-cubic B-spline surface, respectively. In panel (c), the zero-level contour of the LSF is sketched on the same control mesh as in (b). By adding new control points, the cut edges are divided into four equally spaced edges as shown in (d), with a clear definition of the two-phase structure. Panel (e) presents some mesh manipulations by changing the location of the newly added control points, and panel (f) shows the final geometrical design after Laplacian smoothing.}
\label{fig:Geometrical}
\end{figure*}

Using untrimming techniques \citep{sederberg2008watertight}, and the zero-level contour of the LSF, the spline-based topology of the geometrical model is produced. 
This process is executed in the following steps:
\begin{enumerate}
     \item Define the initial design domain using a bi-cubic B-spline surface, \prettyref{fig:FirstDomain}. 
     The control mesh of the B-spline LSF is presented in \prettyref{fig:FirstMesh}.
    \item On the control mesh, sketch the zero-level contour of the LSF which defines the trimmed geometry, as in \prettyref{fig:WithBound}. 
    \item Divide each trimmed edge into four equally spaced edges by adding new control points along the same edge, as in \prettyref{fig:TrimmedB}.
    This yields a clear definition of the two-phase structure.
    \item Fit the control polygon by updating the locations of the newly added control points (step 3) based on a linear interpolation of the LSF and the original control points, \prettyref{fig:Trimmed}.
    \item Apply Laplacian smoothing to all control points, including the interfacial ones. This increases the quality of the mesh and avoids high curvatures that might lead to high stress concentrations, \prettyref{fig:Trimmed_AfterLap}.
\end{enumerate}

Once Step 5 is completed, the geometrical model is obtained. 
The final set of control points in this model is denoted by $\mathbf{P}^{Gl}$. 
These control points are a function of the control points of the level-set function, namely $\mathbf{P}^{Gl} \equiv \mathbf{P}^{Gl}(\boldsymbol{\alpha})$, where a subset of $\boldsymbol{\alpha}$ will serve as the design variables for optimization.
As described by \citet{shakur2024isogeometric}, this correlation is differentiable and can be expressed via a linear operator $\mathbf{P}^{Gl} = \left(\mathsf{L}^G\mathsf{A}\right)\boldsymbol{\alpha}$, where $\mathsf{L}^G$ represents the Laplacian smoothing matrix and $\mathsf{A}$ includes the first four steps. 
The mesh of the geometrical model is not necessarily suitable for analysis. 
Some adjustments are needed to guarantee analysis-suitability, as described next.

\subsubsection{The mechanical model}
\label{Sec:mechanical}
The mechanical model is obtained from the geometrical model by carefully performing \textit{h}-refinement. 
The refinement is applied on different levels based on the location of the element: 
for boundary and interface elements, two \textit{h}-refinement levels are applied, to ensure the minimal distance between extraordinary control points -- that might exist near the boundaries -- and thus fulfill an analysis-suitability requirement. 
One level of \textit{h}-refinement is applied to the neighbors of the boundary and interface elements. 
As for the rest of the elements, they remain unmodified.
Finally, another Laplacian smoothing is applied to the mechanical mesh, for all control points excluding those on the boundary or interface, to eliminate any skewed or thin elements. 

\begin{figure}[tbp] 
\centering
\subfloat[]{\label{fig:FinalMech} 
\includegraphics[width=0.45\textwidth]{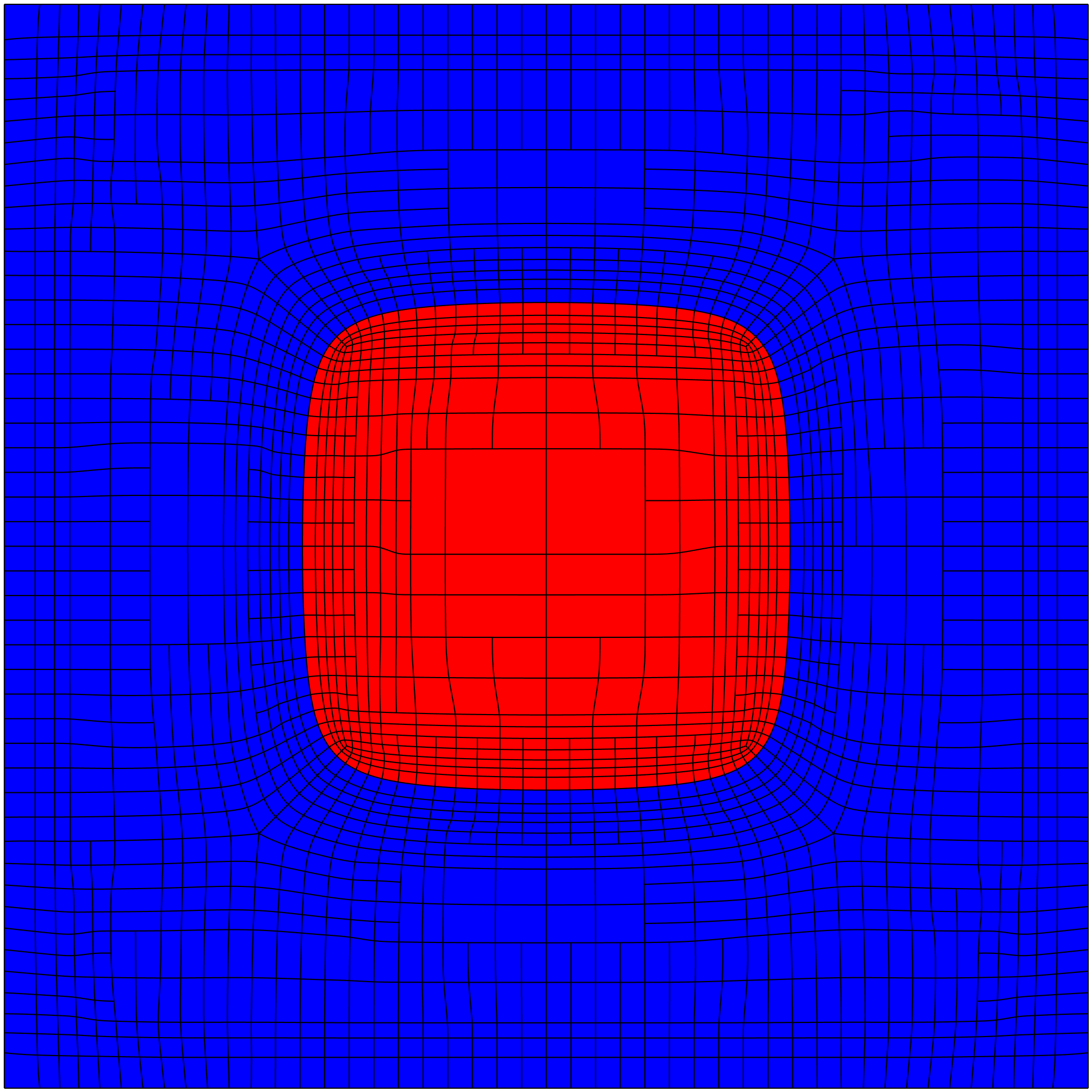}} \\
\subfloat[]{\label{fig:FinalMesh}
\includegraphics[width=0.45\textwidth]{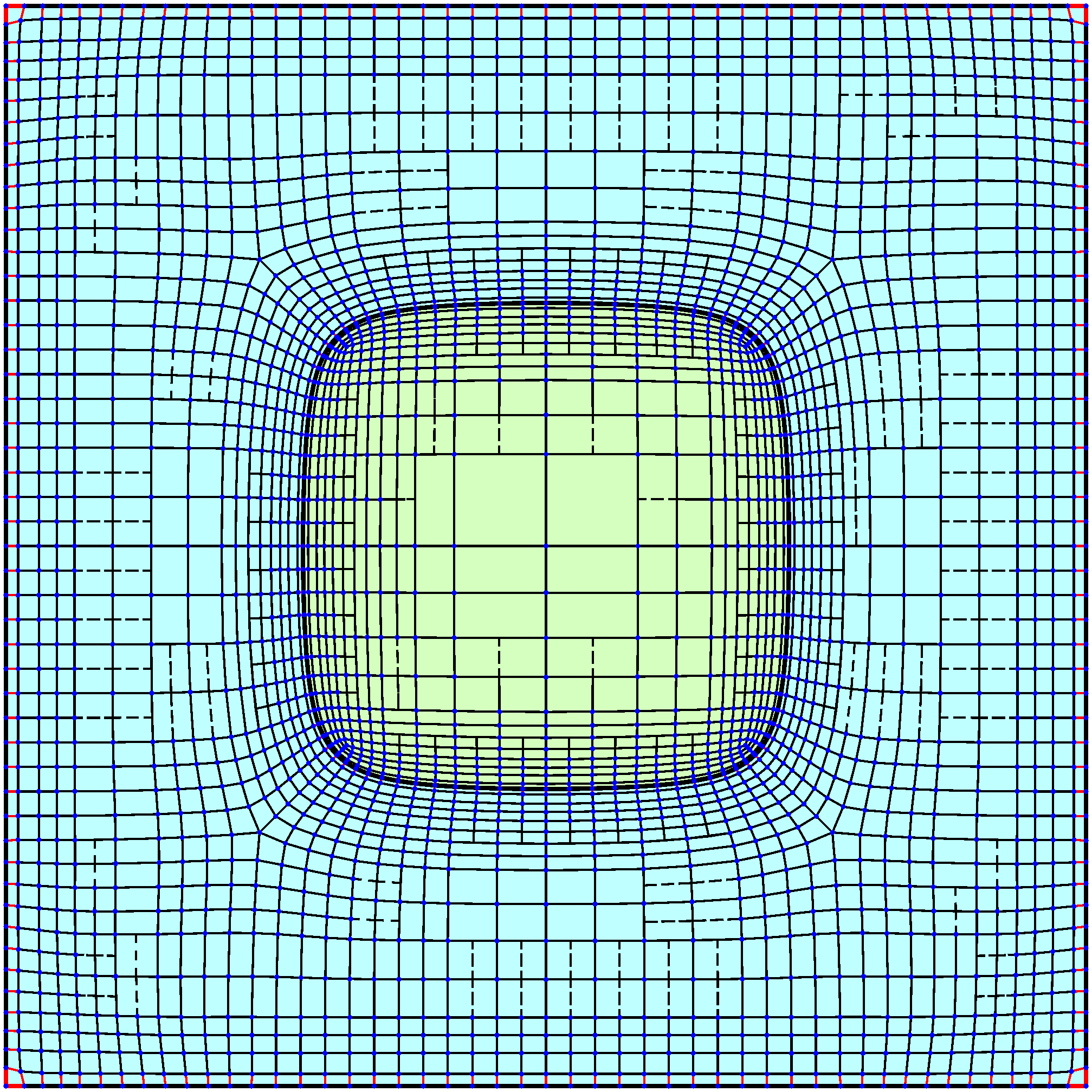}}
\caption{The mechanical model corresponding to the geometry of \prettyref{fig:Geometrical} . Panels (a) and (b) present the bi-material mechanical design and the mechanical mesh, respectively, after double \textit{h}-refinement as discussed in Section \ref{Sec:mechanical}. Face extensions are represented with dashed black lines.}
\label{fig:Mechanical}
\end{figure}

The final mechanical design -- consisting of a matrix (blue) material and inclusion (red) material -- and the final mechanical mesh are presented in \prettyref{fig:Mechanical}, corresponding to the geometry from \prettyref{fig:Geometrical}.
The final set of control points in the mechanical mesh is denoted by $\mathbf{P}^M$.
Mechanical control points are linearly related to the geometrical control points, and this relationship can be expressed as $\mathbf{P}^M = \mathsf{T}\mathbf{P}^{Gl}$.
Because the relation between the geometrical and mechanical meshes is only through \textit{h}-refinement, and because boundary and interface control points are not modified in the process, both models represent the same physical domain.
It is important to note that both models are physically identical but they do not replicate exactly the zero-level contour of the LSF.
This has no implication on the accuracy of the suggested method, since the main point is to have a seamless integration between the geometrical and the mechanical models, while the LSF is merely an auxiliary tool to generate the underlying shape of the interface.

\subsection{Linear elasticity}
In this work, we apply IGA to linear elasticity problems, where the aim is to find the structural response -- i.e., the displacements due to external forces. 
As IGA can be seen as a particular form of FEA, the discretization leads to a linear algebraic system of equations 
\begin{equation}
    \mathbf{Ku = f}\quad,
\end{equation}
where $\mathbf{K}$ is the global stiffness matrix, $\mathbf{u}$ denotes the displacement vector and $\mathbf{f}$ is the external load vector.
Both the stiffness matrix and the load vector are generated by assembling the contributions of the local stiffness matrices $\mathbf{k}_e$ and the local external force vectors $\mathbf{f}_e$ that are computed on patches in a manner similar to standard FEA,
\begin{equation}
\label{eq:elemental}
     \begin{aligned}
         & \mathbf{k}_e = \int_{\Omega_e} \mathbf{B}^T\mathbf{CB} d\Omega_e \quad ,\\
         & \mathbf{f}_e = \int_{\Omega_e} \mathbf{B^TF} d\Omega_e \quad ,
     \end{aligned}
\end{equation}
where $\mathbf{B}$ is the strain-displacement matrix; $\mathbf{C}$ is the elasticity matrix; $\mathbf{F}$ is the domain load vector; and $\Omega_e$ is the physical domain of a single patch.

To compute the terms in \eqref{eq:elemental}, B\'ezier extraction is carried out as described by \citet{scott2013isogeometric}. 
Afterwards, each patch is transformed into a parametric domain $\overline{\Omega}_e$, which is defined by the parameters $\overline{\xi}$ and $\overline{\eta}$, $[0, 1] \times [0, 1]$. 
To maintain consistency with the work of \cite{shakour2022stress}, the integration in \prettyref{eq:elemental} is performed over the integration domain $\hat{\Omega}_e$ that is defined by the parameters $\hat{\xi}$ and $\hat{\eta}$, $[-1, 1] \times [-1, 1]$. 
Therefore, two transformations are applied to compute the local vectors and matrices for each single patch. 
The Jacobian matrices for those transformations are denoted by $\mathbf{J}_1$ and $\mathbf{J}_2$, respectively. 
Hence, the integrations in \prettyref{eq:elemental} are rewritten as
\begin{equation}
\label{eq:elemental_J}
     \begin{aligned}
         & \mathbf{k}_e = \int_{\Omega_e} \mathbf{B}^T\mathbf{CB} |\mathbf{J}_1||\mathbf{J}_2| d\hat{\Omega}_e \\
         & \mathbf{f}_e = \int_{\Omega_e} \mathbf{B^TF}|\mathbf{J}_1||\mathbf{J}_2| d\hat{\Omega}_e
     \end{aligned}
     \quad .
\end{equation}

\section{Problem formulation and sensitivity analysis}\label{sec:formulation}
In this section, we present the problem formulation and the framework for analytical sensitivity analysis. 
We use a simple weighted objective of compliance and an aggregated stress measure to demonstrate the various formulations for sensitivity analysis.

\subsection{Problem formulation}
The optimization problem formulation is given by
\begin{equation}
\label{eq:formulation2}
\begin{aligned}
& \underset{\mathbf{z}_{cp}}{\text{minimize}}
& & f = \left(1-\omega\right) \frac{f_c}{f_{c_0}} + \omega \frac{f_{\sigma_i}}{f_{\sigma_i}^{*}}\\
& \text{subject to}
& & V_i \leq V_i^*, \\ 
& & & -1 \leq \mathbf{z}_{cp} \leq  1, 
\end{aligned}
\end{equation}
where the design variables $\mathbf{z}_{cp}$ are the $z$ coordinates of the control points that parameterize the LSF; $\omega$ is a weighting factor, $0 \leq \omega \leq 1$; the compliance functional $f_c = \mathbf{f}^T\mathbf{u}$ is normalized with respect to the compliance of the initial design $f_{c_0}$; $f_{\sigma_i}$ is an aggregated measure of the stress in material phase $i$, and can represent various stress-based design goals, as elaborated later; $f_{\sigma_i}^{*}$ is a normalization parameter for $f_{\sigma_i}$; $V_i$ and $V_i^*$ are the volumes of materials $i = 1,2$ and the maximum allowed volume of the same material, respectively; and the box constraints ensure a reasonable curvature of the LSF.

The purpose of incorporating $f_{\sigma_i}$ in the objective is to reduce the maximal stress at a certain region -- in the domain, on the interface, or both. 
Since the maximal stress, denoted hereafter $\hat{\sigma}_i$, is non-differentiable, we consider $p$-norm smooth approximations encoded as $f_{\sigma_i}$.  
In this work, three distinct definitions of $f_{\sigma_i}$ are used, to maintain compatibility with the various formulations of sensitivity analysis.
The first approximation is computed using a sum of stress evaluations at discretization points throughout a material domain, 
\begin{equation}
\label{eq:Pnorm1}
    f^{\RNum{1}}_{\sigma_i}= \left(\sum^{N_{\Omega_i}}_{j=1} \sigma_{vm,ij}^p\right)^{1/p},
\end{equation}
where $N_{\Omega_i}$ is the number of computational points; $\sigma_{vm,ij}$ is the von Mises stress of phase $i$ at the computational point $j$; and $p$ is a sufficiently large number. 
Based on the mesh obtained according to Section \ref{Sec:mechanical}, the von Mises stresses are computed at the center of each patch 
so the total number of computational points is equal to the number of patches inside the domain of phase $i$. 
More refined evaluations of the stress can be obtained by adding points inside patches, without affecting the generality of our formulation. 
Alternatively, the $p$-norm approximation can be formulated as a domain integration,
\begin{equation}
\label{eq:Pnorm2}
    f^{\RNum{2}}_{\sigma_i} = \left(\int_{\Omega_i} \sigma_{vm,ij}^p\right)^{1/p},
\end{equation}
where the integral in \prettyref{eq:Pnorm2} is computed numerically using Gauss quadrature. 
This stress measure is suitable for ``differentiate-then-discretize'' approaches to level-set shape optimization \citep[e.g,][]{allaire2008minimum}. 
The third approximation is a discrete summation of interfacial stresses, computed as follows:
\begin{equation}
\label{eq:Pnorm3}
    f^{\RNum{3}}_{\sigma_i}= \left(\sum^{N_\Gamma}_{j=1} \sigma_{vm,ij}^p\right)^{1/p}
\end{equation}
where $N_\Gamma$ is the number of computational points on the interface. 
Defining a stress functional based on the continuous integration of interfacial stresses -- analogous to $f^{\RNum{2}}_{\sigma_i}$ -- falls beyond the scope of this paper:  
while ``differentiate-then-discretize'' methods have been used for interfacial functionals directly dependent on the state variable $u$ \citep[e.g.][]{cea1986conception, allaire2004structural}, the authors are not aware of previous studies that addressed functionals dependent on the spatial derivative of $u$, such as the boundary integration of interfacial stresses. 
In subsequent sections, we will demonstrate numerical challenges and limitations, that may affect the applicability of the ``differentiate-then-discretize'' approach on such functionals.

Without loss of generality, we limit the discussion to plane stress situations. 
For readability of our derivations, we recall that the von Mises stress at a specific computation point $j$ is given by

\begin{equation}
    \sigma_{vm,j} = \sqrt{\boldsymbol{\sigma}^T_j\mathbf{V}\boldsymbol{\sigma}_j}, 
\end{equation}
where
\begin{equation}
   \mathbf{V} = \begin{bmatrix}
        1 & -0.5 & 0\\
        -0.5 & 1 & 0\\
        0 & 0 & 3
    \end{bmatrix},
\end{equation}
and
\begin{equation}
    \boldsymbol{\sigma}_j = \mathbf{C}\mathbf{B}_j\mathbf{u} \quad .
\end{equation}

\subsection{Sensitivity analysis}
In this section, we introduce three different types of sensitivity analyses, with the intention to compare their suitability for deriving the functionals $f_c$ and $f_{\sigma_i}$.
Initially, we establish a universal function influenced by the shape and topology of the structure, together with the state variables (displacements in the current context).
The universal function is expressed as $R = R(\mathbf{z}_{cp}, \mathbf{u}(\mathbf{z}_{cp}))$, where $R$ could denote compliance, stress, or volume.
The sensitivity of the function with respect to the design variables $\mathbf{z}_{cp}$ is given by 
\begin{equation}
\label{eq:SA}
    \frac{dR}{d\mathbf{z}_{cp}} = \frac{\partial R}{\partial \mathbf{u}}\frac{d\mathbf{u}}{d\mathbf{z}_{cp}} +  \frac{\partial R}{\partial \mathbf{z}_{cp}}
\end{equation}
where $\frac{d}{d\mathbf{z}_{cp}}$ and $\frac{\partial}{\partial \mathbf{z}_{cp}}$ represent the total and partial derivatives with respect to $\mathbf{z}_{cp}$, respectively. 
These sensitivities are determined using the chain rule 
\begin{equation}
    \label{eq:dcdot_dzcp}
    \frac{d(~\cdot~)}{d\mathbf{z}_{cp}} = \frac{d~\cdot~}{d\mathbf{P}^{M}}\frac{d\mathbf{P}^{M}}{d\mathbf{z}_{cp}},
\end{equation}
where the control points of the mechanical model $\mathbf{P}^{M}$ are treated as intermediate variables. 

The availability of an explicit parameterization of the level set function and of the interface in the mechanical model, raises the question: 
should we discretize first and then differentiate, or vice versa?
We note that classical level set optimization procedures use an implicit representation of the level set function and follow a ``differentiate-then-discretize'' approach, also when a body-fitted mesh is used for mechanical simulation  \citep{allaire2021shape}.
In the subsequent three sections, we will delve into three distinct formulations of the sensitivity analysis, distinguished by the sequence of steps in the process and their definitions of intermediate design variables. 
In the first type, we first discretize and subsequently differentiate the discretized model, considering all control points of the mechanical model as intermediate design variables. 
In the second type, we follow classical shape optimization where differentiation is conducted on a continuum domain before discretization. 
Consequently, only control points of the moving boundary -- the interface between material phases -- are considered as intermediate design variables as they parameterize the shape of the moving boundary.
The third type can be classified as ``parameterized shape SA'', concentrating solely on control points of the interface or a carefully chosen subset of control points proximal to the interface. 
This type belongs to the ``discretize-then-differentiate'' class in which design sensitivities are derived on the discretized model.

Finally, we note that design sensitivities of the volume constraint are formulated with the second type only, because the volume is independent of the discretization.
Hence, the scope of our discussion is limited to the more intriguing functionals: compliance and stress.

\subsubsection{Sensitivity analysis on the discretized domain }
\label{sec:FirstSA}
As mentioned above, herein we ``discretize-then-differentiate'', meaning that we derive the discretized solution that is obtained by IGA. 
The IGA model is defined by the control points of the mechanical model $\mathbf{P}^{M}$ which are related to the design variables $\mathbf{z}_{cp}$ by a series of differentiable operations. 
The adjoint method is adopted to eliminate the implicit sensitivity of the state variables with respect to the intermediate design variables, namely $\frac{d\mathbf{u}}{d\mathbf{P}^{M}}$ in Eqs.~\eqref{eq:SA}, \eqref{eq:dcdot_dzcp}.
The response functional is evaluated on the discretized model represented by $\mathbf{P}^{M}$, so an augmented functional is defined as
\begin{equation}
    \hat{R} = R\left(\mathbf{P}^{M},\mathbf{\mathbf{u}}(\mathbf{P}^{M})\right) - \boldsymbol{\lambda}^{T}\left(\mathbf{K}(\mathbf{P}^{M})\mathbf{u}(\mathbf{P}^{M}) - \mathbf{f}\right)
    \label{eq:Augmented}
\end{equation}
where we introduce the adjoint vector $\boldsymbol{\lambda}$. 
By differentiating \prettyref{eq:Augmented} w.r.t.~$\mathbf{P}^{M}$ and considering an external load independent of the design variables, we obtain

\begin{equation}
    \frac{d\hat{R}}{d\mathbf{P}^{M}} = \frac{\partial R}{ \partial\mathbf{u}}\frac{d\mathbf{u}}{d\mathbf{P}^{M}} +  \frac{\partial R}{\partial \mathbf{P}^{M}} - \boldsymbol{\lambda}^{T}\frac{d \mathbf{K}}{d \mathbf{P}^{M}}\mathbf{u} - \boldsymbol{\lambda}^{T}\mathbf{K}\frac{d\mathbf{u}}{d\mathbf{P}^{M}} \quad .
\end{equation}
To eliminate the implicit derivative ($\frac{d\mathbf{u}}{d\mathbf{P}^{M}}$), the adjoint vector must satisfy
\begin{equation}
\label{eq:AdEq}
    \mathbf{K}^T\boldsymbol{\lambda} = \left(\frac{\partial R}{\partial \mathbf{u}}\right)^T \quad.
\end{equation}
Hence, the sensitivity of the augmented functional is narrowed down to

\begin{equation}
    \label{eq:Adjoint}
    \frac{d\hat{R}}{d\mathbf{P}^{M}} = \frac{\partial R}{\partial \mathbf{P}^{M}}  - \boldsymbol{\lambda}^{T}\frac{d \mathbf{K}}{d \mathbf{P}^{M}}\mathbf{u}
\end{equation}
and the sensitivity with respect to the design variables is evaluated via the chain rule,

\begin{equation}
    \label{eq:chainrule}
    \begin{aligned}
    \frac{d\hat{R}}{d\mathbf{z}_{cp}} = \frac{d\hat{R}}{d\mathbf{P}^{M}}\frac{d\mathbf{P}^{M}}{d\mathbf{z}_{cp}}
    \end{aligned}
\end{equation}

The second part of \prettyref{eq:chainrule}, $\frac{d\mathbf{P}^{M}}{d\mathbf{z}_{cp}}$ is computed based on the analytical relations given in Section \ref{sec:gener}. 
For a detailed description, the reader is referred to \cite{shakour2022stress}. 
This leaves us with the the need to compute $\frac{\partial R}{\partial \mathbf{P}^{M}}$, $\frac{\partial R}{\partial \mathbf{u}}$ and $\frac{d \mathbf{K}}{d \mathbf{P}^{M}}$. Based on \prettyref{eq:elemental_J}, the derivative of the stiffness matrix is given by:

\begin{align}
              \frac{d \mathbf{K}}{d \mathbf{P}^{M}}  & =  \int_{\hat{\Omega}}  \mathbf{B}^T \mathbf{C} \frac{d \mathbf{B}}{d \mathbf{P}^M}|\mathbf{J}_1||\mathbf{J}_2|
           + \\ & \frac{d \mathbf{B}^T}{d \mathbf{P}^M} \mathbf{C} \mathbf{B}|\mathbf{J}_1||\mathbf{J}_2| + \mathbf{B}^T \mathbf{CB}\frac{d |\mathbf{J}_1|}{d \mathbf{P}^M}|\mathbf{J}_2| d \hat{\Omega}. 
\end{align}
Since the compliance is self-adjoint, its final sensitivity analysis is:
\begin{equation}
\label{eq:SA_comp}
    \frac{df_c}{d\mathbf{P}^M} = -\mathbf{u}^T\frac{d\mathbf{K}}{d\mathbf{P}^M}\mathbf{u}, 
\end{equation}

Because we derive the discretized model, the derivation suits a stress measure that is computed using a summation of discrete computational points. 
Hence, this type of SA is applicable for both stress functionals $f^{\RNum{1}}_{\sigma_i}$ and $f^{\RNum{3}}_{\sigma_i}$ (see Eqs.~(\ref{eq:Pnorm1}),(\ref{eq:Pnorm3})). 
Analytically, the formulation is the same for both measures, so we present the expressions for $f^{\RNum{1}}_{\sigma_i}$, recognizing that they apply equally to $f^{\RNum{3}}_{\sigma_i}$ without loss of generality. 
The partial derivative of $f^{\RNum{1}}_{\sigma_i}$ is given by 

\begin{equation}
    \frac{\partial f^{\RNum{1}}_{\sigma_i}}{\partial \mathbf{P}_{ij}^M} = \frac{\partial f^{\RNum{1}}_{\sigma_i}}{\partial \sigma_{vm,ij} } \frac{\partial \sigma_{vm,ij}}{\partial\boldsymbol{\sigma}_{ij}}\frac{\partial\boldsymbol{\sigma}_{ij}}{\partial \mathbf{P}_{ij}^M},
\end{equation}
where $\boldsymbol{\sigma}_{ij}$ is the stress vector at the hosting patch $j$ of phase $i$; 
and the subscript $\Box_{ij}$ represents the quantity that is associated with patch $j$ of phase $i$, e.g., $\mathbf{P}_{ij}^M$ is the set of mechanical control points that are associated with patch $j$ of phase $i$. 
Furthermore,

\begin{equation}
\label{eq:dSig}
\begin{aligned}
    & \frac{\partial f^{\RNum{1}}_{\sigma_i}}{\partial \sigma_{vm,ij} } = 
    \left(f^{\RNum{1}}_{\sigma_i}\right)^{1-p}\sigma_{vm,ij}^{p-1}, \\
    & \frac{\partial \sigma_{vm,ij}}{\partial\boldsymbol{\sigma}_{ij}}  = \frac{1}{2}\left(\boldsymbol{\sigma}_{ij}^T \mathbf{V} \boldsymbol{\sigma}_{ij}\right)^{-\frac{1}{2}}2\boldsymbol{\sigma}_{ij}^T\mathbf{V} & \\ &\quad\quad\quad\,\,=\frac{\boldsymbol{\sigma}_{ij}^T\mathbf{V}}{\sigma_{vm,ij}}, \\
    & \frac{\partial\boldsymbol{\sigma}_{ij}}{\partial \mathbf{P}_{ij}^M} = \mathbf{C} \frac{d \mathbf{B}_{ij}}{d \mathbf{P}_{ij}^M}\mathbf{u}_{ij},
\end{aligned}
\end{equation}
which finally yields:
\begin{equation}
    \label{eq:dSigmadP}
    \frac{df^{\RNum{1}}_{\sigma_i}}{d\mathbf{P}^M} = \left(f^{\RNum{1}}_{\sigma_i}\right)^{1-p}\left(\sum_{j=1}^N \sigma_{vm,ij}^{p-2}{\boldsymbol{\sigma}_{ij}^{T}}{\mathbf{V}}\left(\mathbf{C}\frac{d \mathbf{B}_{ij}}{d \mathbf{P}_{ij}^M}\mathbf{u}_{ij}\right)\right) \quad .
\end{equation}
As for the derivative with respect to the displacements, it is given by:
\begin{equation}
\label{eq:dSigdUT}
    \frac{\partial f^{\RNum{1}}_{\sigma_i}}{\partial \mathbf{u}_{ij}} = \frac{\partial f^{\RNum{1}}_{\sigma_i}}{\partial \sigma_{vm,ij} } \frac{\partial \sigma_{vm,ij}}{\partial\boldsymbol{\sigma}_{ij}}\frac{\partial\boldsymbol{\sigma}_{ij}}{\partial \mathbf{u}_{ij}},
\end{equation}
where
\begin{equation}
\label{eq:dSigdU}
    \frac{\partial\boldsymbol{\sigma}_{ij}}{\partial \mathbf{u}_{ij}} = \mathbf{CB}_{ij}.
\end{equation}
Combining Eqs.~\eqref{eq:dSig}, \eqref{eq:dSigdUT} and \eqref{eq:dSigdU} yields:
\begin{equation}
    \label{eq:dSigmadU}
    \frac{df^{\RNum{1}}_{\sigma_i}}{d\mathbf{u}} = \left(f^{\RNum{1}}_{\sigma_i}\right)^{1-p}\left(\sum_{j=1}^{N}\sigma_{vm,{ij}}^{p-2}\left(\mathbf{CBu}_{ij}\right)^T\mathbf{VCB}_{ij}\right)
\end{equation}

\subsubsection{Shape derivatives by C\'ea's method}
\label{sec:deriv_Cea}
In the second type of SA, we follow C\'ea's Lagrangian method \citep{cea1986conception}. 
We choose to explore this approach because it was already applied in closely related studies for a multi-material setting with a body-fitted mesh, where the material interface is discretized using a piece-wise linear mesh \citep{allaire2014multi,liu2020multi}. 
Furthermore, it is computationally more efficient than the method of Section \ref{sec:FirstSA} because it is based on boundary integrals.
Another interesting aspect is that Green's theorem can be applied precisely because our mechanical model consists of a smooth and explicit interface representation. 

In this section, we outline the Lagrangian derivation in a detailed step-by-step manner. 
As noted previously, differentiation occurs prior to discretization in this method. 
A crucial aspect of this technique is its use of Green's theorem to convert domain integrals into boundary integrals, significantly reducing the computational effort. 
Consequently, the stress functional is calculated as a domain integration instead of a summation of discrete points, i.e., $f^{\RNum{2}}_{\sigma_i}$ in \prettyref{eq:Pnorm2}.
For clarity, we first recall the following lemmas and identities that will be utilized in our subsequent derivations. 

\begin{lemma}
\label{lem:lemma1}
As presented in \citet{allaire2008minimum}, for a smooth function $f(x)$, define
\begin{equation*}
\begin{aligned}
        J_{\text{vol}} = \int_\Omega f(x) \, d\Omega, &&&& J_{\text{surf}} = \int_\Gamma f(x) \, d\Gamma.
\end{aligned}
\end{equation*}
The shape derivative of these two functions is
\begin{equation*}
    \begin{aligned}
     &J'_{\text{vol}} = \int_\Gamma\ f(x)\cdot n \, d\Gamma, \\
     &J'_{\text{surf}} = \int_\Gamma \left(\frac{\partial}{\partial n} + H \right)f(x) \cdot n \, d\Gamma,
    \end{aligned}
\end{equation*}
where $\Gamma = \partial\Omega$; $n$ is the unit vector normal to $\Gamma$ and $H$ is the mean curvature of $\Gamma$.
\end{lemma}

\begin{lemma}
\label{lemma:Green}
Using Green's theorem, the following formula could be deduced:
\begin{equation*}
\begin{aligned}
    \int_\Omega div\left(\sigma\left(v\right) \right)\cdot w d\Omega &= \int_\Gamma \sigma\left(v\right)n\cdot w d\Gamma \\& - \int_\Omega \epsilon\left(w\right)\sigma\left(v\right) d\Omega,
\end{aligned}
\end{equation*}
where $v, w \in C^1$ are vectors defined over $\Omega$. An immediate outcome of this formula is 
\begin{equation*}
\frac{\partial}{\partial w} \int_\Omega \epsilon\left(w\right)\sigma\left(v\right) d\Omega = \int_\Gamma \sigma\left(v\right)n d\Gamma  - \int_\Omega div\left(\sigma\left(v\right) \right) d\Omega \quad .
\end{equation*}
\end{lemma}

\begin{identity}
\textit{As presented by \citet{allaire2011damage}, for given displacements $u$ and $v$, if $v=0$ on $\Sigma$, hence:}
\begin{equation*}
\begin{aligned}
        \sigma(u)n\cdot\frac{\partial v}{\partial n} = 2\left(\sigma(u)n\right)\cdot\left(\epsilon(v)n\right) - \sigma_{nn}(u)\epsilon_{nn}(v) \:\: \text{on} \quad \Sigma \quad,
\end{aligned}
\end{equation*}
\textit{which also implies that}
\begin{equation*}
\begin{aligned}
         \sigma(u)n\cdot & \frac{\partial (v_1 - v_2)}{\partial n} = 2\left(\sigma(u)n\right)\cdot\left(\epsilon(v_1)n -  \epsilon(v_2)n\right) \\& - \sigma_{nn}(u)\left(\epsilon_{nn}(v_1) - \epsilon_{nn}(v_2)\right) \:\: \text{on} \quad \Sigma \quad ,
\end{aligned}
\end{equation*}
{for $v_1 = v_2$ on $\Sigma$.}
\end{identity}

The derivation of the shape sensitivity analysis involves introducing the Lagrangian typically as an augmented function, composed of the sum of the objective function and constraints multiplied by appropriate Lagrange multipliers. 
In the context of shape optimization, the state equation is also treated as a constraint. 
Shape optimization in a multi-material setting is somewhat more complicated than its single-phase counterpart.
The reason for this is that the derivative of the solution field $u$ of the state equation is discontinuous at the material interface. 
To overcome this, we introduce an updated state equation that upholds the transmission conditions, following a similar approach as in \citet{allaire2011damage, liu2020multi},

\begin{equation}
\label{eq:UpdatedState_1}
\begin{aligned}
&-div\left(\boldsymbol{\sigma}_1\left(\mathbf{u}_1\right)\right) = \mathbf{f}_1 && \text{in} \quad\Omega^1, \\ 
&\mathbf{u}_1 = 0 && \text{on} \quad\Gamma_D^1, \\ 
&\boldsymbol{\sigma}_1\left(\mathbf{u}_1\right)\mathbf{n}_1 = \mathbf{g}_1 && \text{on} \quad\Gamma_N^1, \\ 
&\mathbf{u}_1 = \mathbf{u}_2 && \text{on} \quad\Sigma, \\ 
&\boldsymbol{\sigma}_1\left(\mathbf{u}_1\right)\mathbf{n}_1 + \boldsymbol{\sigma}_2\left(\mathbf{u}_2\right)\mathbf{n}_2 = 0 && \text{on}\quad\Sigma, \\ 
\end{aligned}
\end{equation}

and 

\begin{equation}
\label{eq:UpdatedState_2}
\begin{aligned}
&-div\left(\boldsymbol{\sigma}_2\left(\mathbf{u}_2\right)\right) = \mathbf{f}_2 && \text{in} \quad\Omega^2, \\ 
&\mathbf{u}_2 = 0 && \text{on} \quad\Gamma_D^2, \\ 
&\boldsymbol{\sigma}_2\left(\mathbf{u}_2\right)\mathbf{n}_2 = \mathbf{g}_2 && \text{on} \quad\Gamma_N^2, \\ 
&\mathbf{u}_1 = \mathbf{u}_2 && \text{on} \quad\Sigma, \\ 
&\boldsymbol{\sigma}_1\left(\mathbf{u}_1\right)\mathbf{n}_1 + \boldsymbol{\sigma}_2\left(\mathbf{u}_2\right)\mathbf{n}_2 = 0 && \text{on} \quad\Sigma, \\ 
\end{aligned}
\end{equation}
where $\mathbf{n} = \mathbf{n}_2 = -\mathbf{n}_1$ denotes the outward normal direction to the interface $\Sigma$; and $\Gamma^i_D$ and $\Gamma^i_N$ are the Dirichlet and Neumann boundary conditions for phase $i$. 
Subsequently, the Lagrangian is defined as

\begin{equation}\
    \begin{aligned}
        L &=  \hat{J} +  \sum_{i=1}^2 \int_{\Omega^i}\boldsymbol{\lambda}_{1i}\cdot\left(div\left(\boldsymbol{\sigma}_i\left(\mathbf{u}_i\right) \right) + \mathbf{f}_i\right) d\Omega \\
        &+ \sum_{i=1}^2 \int_{\Gamma^i_{N}}\boldsymbol{\lambda}_{2i}\cdot\left(\boldsymbol{\sigma}_i\left(\mathbf{u}_i\right)\cdot \mathbf{n}_i - \mathbf{g}_i\right) d\Gamma \\
        &+ \sum_{i=1}^2 \int_{\Gamma^i_{D}}\boldsymbol{\lambda}_{3i} \mathbf{u}_i d\Gamma \\
        &+ \int_{\Sigma} \boldsymbol{\lambda}_4\cdot\left(\mathbf{u}_1 - \mathbf{u}_2\right) d\Gamma \\
        &+ \int_{\Sigma} \boldsymbol{\lambda}_5\cdot\left(\boldsymbol{\sigma}_1\left(\mathbf{u}_1\right)\mathbf{n}_1 +  \boldsymbol{\sigma}_2\left(\mathbf{u}_2\right)\mathbf{n}_2 \right) d\Gamma
    \end{aligned}
\end{equation}
where $\hat{J}$ is the cost function, denoting the compliance or the stress functional; $\boldsymbol{\lambda}_{1i}, \boldsymbol{\lambda}_{2i}, \boldsymbol{\lambda}_{3i}, (i=1,2)$ and $\boldsymbol{\lambda}_{4}, \boldsymbol{\lambda}_{5}$ denote the Lagrangian multipliers. 
Exploiting Lemma \ref{lemma:Green}, the Lagrangian can be re-written as
\begin{equation}
\label{eq:Lagrangian}
    \begin{aligned}
        L & =  \hat{J} +  \sum_{i=1}^2 \int_{\Omega^i}\left(-\boldsymbol{\sigma}\left(\mathbf{u}_i\right)\boldsymbol{\epsilon}\left(\boldsymbol{\lambda}_{1i}\right) + \mathbf{f}_i \right) d\Omega \\
        & + \sum_{i=1}^2 \int_{\Gamma^i_{N}}\boldsymbol{\lambda}_{2i}\cdot\left(\boldsymbol{\sigma}_i\left(\mathbf{u}_i\right)\cdot \mathbf{n}_i - \mathbf{g}_i\right) + \boldsymbol{\lambda}_{1i}\boldsymbol{\sigma}\left(\mathbf{u}_i\right)\cdot \mathbf{n}_i \;d\Gamma \\  
        &+ \sum_{i=1}^2  \int_{\Gamma^i_{D}}\boldsymbol{\lambda}_{3i} \mathbf{u}_i + \boldsymbol{\lambda}_{1i}\boldsymbol{\sigma}\left(\mathbf{u}_i\right)\cdot \mathbf{n}_i \; d\Gamma \\
        & +\int_{\Sigma} \boldsymbol{\lambda}_4\cdot\left(\mathbf{u}_1 - \mathbf{u}_2\right) d\Gamma \\ 
        & +\int_{\Sigma} \boldsymbol{\lambda}_5\cdot\left(\boldsymbol{\sigma}_1\left(\mathbf{u}_1\right)\mathbf{n}_1 + \boldsymbol{\sigma}_2\left(\mathbf{u}_2\right)\mathbf{n}_2 \right) d\Gamma \\
        & + \int_{\Sigma} \boldsymbol{\lambda}_{1i}\boldsymbol{\sigma}\left(\mathbf{u}_i\right)\cdot \mathbf{n}_i \;d\Gamma.
    \end{aligned}
\end{equation}
Differentiation of the Lagrangian as shown in \prettyref{eq:Lagrangian} with respect to the Lagrange multipliers results in the revised state equations. 
Enforcing $\frac{\partial L}{\partial \mathbf{u}_i} = 0$ establishes the connections between the Lagrange multipliers and the adjoint equations, which yields the following Lagrangian,
\begin{equation}
\label{eq:UpdLag}
        \begin{aligned}
        L = & \hat{J} +  \sum_{i=1}^2 \int_{\Omega^i}\left(-\boldsymbol{\sigma}\left(\mathbf{u}_i\right)\boldsymbol{\epsilon}\left(\boldsymbol{\lambda}_{1i}\right) + \mathbf{f}_i \right) d\Omega \\
        & +\sum_{i=1}^2 \int_{\Gamma^i_{N}}\boldsymbol{\lambda}_{1i} \mathbf{g}_i   \;d\Gamma \\ & + \sum_{i=1}^2  \int_{\Gamma^i_{D}}\left(\sigma(\boldsymbol{\lambda}_{1i})\cdot \mathbf{n}_i - \partial \hat{J}_{i,\text{surf}} \right) \mathbf{u}_i + \boldsymbol{\lambda}_{1i}\boldsymbol{\sigma}\left(\mathbf{u}_i\right)\cdot \mathbf{n}_i \;d\Gamma \\
        & +\frac{1}{2} \int_{\Sigma} \left(\mathbf{u}_1 - \mathbf{u}_2\right)\left(\boldsymbol{\sigma}(\boldsymbol{\lambda}_{11})\cdot \mathbf{n}_1 - \boldsymbol{\sigma}(\boldsymbol{\lambda}_{12})\cdot \mathbf{n}_2\right) d\Gamma \\
         & + \frac{1}{2} \int_{\Sigma} \left(\boldsymbol{\lambda}_{11} - \boldsymbol{\lambda}_{12} \right)
        \cdot\left(\boldsymbol{\sigma}_1\left(\mathbf{u}_1\right)\cdot\mathbf{n}_1 - \boldsymbol{\sigma}_2\left(\mathbf{u}_2\right)\cdot\mathbf{n}_2 \right) d\Gamma \\
        & +\frac{1}{2} \int_{\Sigma} \left(\mathbf{u}_1 - \mathbf{u}_2\right)\left(-\partial \hat{J}_{1,\text{surf}} + \partial \hat{J}_{2,\text{surf}}  \right) d\Gamma,
    \end{aligned}
\end{equation}
where $\partial \hat{J}_{1,\text{surf}} = \boldsymbol{\sigma}\left(\mathbf{W}_i\right)\cdot \mathbf{n}$; $\mathbf{W}_i = \boldsymbol{\chi}_i\mathbf{C}$ and $\boldsymbol{\chi}_i =  \frac{\partial \hat{J}}{\partial\boldsymbol{\sigma}_{ij}}$.
The proof is postponed to Appendix \ref{Appendix_Lag}.
What is left for us is to compute the shape derivatives of the Lagrangian. 
This can be done using Lemma \ref{lem:lemma1}. 
For simplicity, we will break down the derivations of each integral alone as we did before, starting from the state equation. The shape derivative of the state equation is simply
\begin{equation}
\begin{aligned}
         L'_1 &= \int_{\Sigma}- \boldsymbol{\sigma}\left(\mathbf{u}_1\right)\boldsymbol{\epsilon}\left(\boldsymbol{\lambda}_{11}\right)\cdot \mathbf{n}_1 - \boldsymbol{\sigma}\left(\mathbf{u}_2\right)\boldsymbol{\epsilon}\left(\boldsymbol{\lambda}_{12}\right)\cdot \mathbf{n}_2 \,d\Gamma\\ & = \int_{\Sigma} \boldsymbol{\sigma}\left(\mathbf{u}_1\right)\boldsymbol{\epsilon}\left(\boldsymbol{\lambda}_{11}\right)\cdot \mathbf{n}  -\boldsymbol{\sigma}\left(\mathbf{u}_2\right)\boldsymbol{\epsilon}\left(\boldsymbol{\lambda}_{12}\right)\cdot \mathbf{n} \,d\Gamma,
\end{aligned}
\end{equation}
where 
\begin{equation}
\begin{aligned}
\boldsymbol{\sigma}\left(\mathbf{u}_i\right) & \boldsymbol{\epsilon}\left(\boldsymbol{\lambda}_{1i}\right) = \boldsymbol{\sigma}_{nn}\left(\mathbf{u}_i\right)\boldsymbol{\epsilon}_{nn}\left(\boldsymbol{\lambda}_{1i}\right)  \\ & +2\boldsymbol{\sigma}_{nt}\left(\mathbf{u}_i\right)\boldsymbol{\epsilon}_{nt}\left(\boldsymbol{\lambda}_{1i}\right) + \boldsymbol{\sigma}_{tt}\left(\mathbf{u}_i\right)\boldsymbol{\epsilon}_{tt}\left(\boldsymbol{\lambda}_{1i}\right).
\end{aligned}
\end{equation}

The transmission conditions ensure that $\boldsymbol{\sigma}_{nn}(\mathbf{u}_1) = \boldsymbol{\sigma}_{nn}(\mathbf{u}_2)$ and $\boldsymbol{\sigma}_{nt}(\mathbf{u}_1) = \boldsymbol{\sigma}_{nt}(\mathbf{u}_2)$.
Furthermore, the continuity of the solution field at the interface, including both the displacement $\mathbf{u}$ and the Lagrange multiplier $\boldsymbol{\lambda}$, dictates that $\boldsymbol{\epsilon}_{tt}(\boldsymbol{\lambda}_{11}) = \boldsymbol{\epsilon}_{tt}(\boldsymbol{\lambda}_{12})$. Therefore, the first segment can be summarized as
\begin{equation}
\label{eq:L1}
\begin{aligned}
         L'_1 = & \int_{\Sigma} \boldsymbol{\sigma}_{nn}(\mathbf{u})\left[\boldsymbol{\epsilon}_{nn}(\boldsymbol{\lambda}_{1i})\right] + 2\boldsymbol{\sigma}_{nt}(\mathbf{u})\left[\boldsymbol{\epsilon}_{nt}(\boldsymbol{\lambda}_{1i})\right] \\ &+ \boldsymbol{\epsilon}_{tt}(\boldsymbol{\lambda}_{1i})\left[(\boldsymbol{\sigma}_{tt}(\mathbf{u}_i) \right] \cdot \mathbf{n}\,d\Gamma
\end{aligned}
\end{equation}
where  $\boldsymbol{\sigma}_{nn}(\mathbf{u})$, $\boldsymbol{\sigma}_{nt}(\mathbf{u})$ and $\boldsymbol{\epsilon}_{tt}(\boldsymbol{\lambda})$ denote the continuous quantities at the interface, and the jump of a quantity at the material interface is expressed as $\left[~\cdot~\right] = \left[~\cdot~\right]_1 - \left[~\cdot~\right]_2$.  
We prescribe the boundaries $\Gamma_D$ and $\Gamma_N$ to be constant during the optimization process. 
Consequently, our derivations will consider only the material interface $\Sigma$.
Using the surface expression of Lemma \ref{lem:lemma1}, the shape derivative of the first integral at the material interface is expressed as
\begin{equation}
\label{eq:trans1}
\begin{aligned}
         L'_2 & = \frac{1}{2} \int_{\Sigma} \left(\frac{\partial}{\partial \mathbf{n}} + H \right) \\ 
         & \left[\left(\mathbf{u}_1 - \mathbf{u}_2\right)\left(\boldsymbol{\sigma}(\boldsymbol{\lambda}_{11})\cdot \mathbf{n}_1 - \boldsymbol{\sigma}(\boldsymbol{\lambda}_{12})\cdot \mathbf{n}_2\right) \right] \, d\Gamma \\ 
        & = -\frac{1}{2} \int_{\Sigma} \left(\frac{\partial}{\partial \mathbf{n}} + H \right) \\ 
        & \left[\left(\mathbf{u}_1 - \mathbf{u}_2\right)\left(\boldsymbol{\sigma}(\boldsymbol{\lambda}_{11})\cdot \mathbf{n} + \boldsymbol{\sigma}(\boldsymbol{\lambda}_{12})\cdot \mathbf{n}\right) \right] \, d\Gamma.
\end{aligned}
\end{equation}
Since $\mathbf{u}_1 = \mathbf{u}_2$ on $\Sigma$, the term that contains the curvature vanishes. 
Substituting the first identity into \prettyref{eq:trans1} gives
\begin{equation}
\begin{split}
\label{eq:L2}
     L'_2 &=  -\int_{\Sigma}\Bigl(\frac{1}{2}\left(\boldsymbol{\sigma}_{nn}(\boldsymbol{\lambda}_{11}) + \boldsymbol{\sigma}_{nn}(\boldsymbol{\lambda}_{12})\right)\left[\boldsymbol{\epsilon}_{nn}(\mathbf{u})\right]  \\&+ \left(\boldsymbol{\sigma}_{nt}(\boldsymbol{\lambda}_{11})  + \boldsymbol{\sigma}_{nn}(\boldsymbol{\lambda}_{12})\right)\left[\boldsymbol{\epsilon}_{nt}(\mathbf{u})\right]\Bigr)\cdot \mathbf{n}\,d\Gamma.
\end{split}
\end{equation}
Similarly, the second and the third integrals on the material interface yield
\begin{equation}
\label{eq:L34}
\begin{aligned}
    L'_3 &=  -\int_{\Sigma}\Bigl(\frac{1}{2}\left(\boldsymbol{\sigma}_{nn}(\mathbf{u}_1) + \boldsymbol{\sigma}_{nn}(\mathbf{u}_2)\right)\left[\boldsymbol{\epsilon}_{nn}(\boldsymbol{\lambda}_{1i})\right]\\& + \left(\boldsymbol{\sigma}_{nt}(\mathbf{u}_1) + \boldsymbol{\sigma}_{nn}(\mathbf{u}_2)\right)\left[\boldsymbol{\epsilon}_{nt}(\boldsymbol{\lambda}_{1i})\right]\Bigr)\cdot \mathbf{n} d\Gamma \\
    &= -\int_{\Sigma}\Bigl(\boldsymbol{\sigma}_{nn}(\mathbf{u})\left[\boldsymbol{\epsilon}_{nn}(\boldsymbol{\lambda}_{1i})\right] + 2\boldsymbol{\sigma}_{nt}(\mathbf{u})\left[\boldsymbol{\epsilon}_{nt}(\boldsymbol{\lambda}_{1i})\right]\Bigr)\cdot \mathbf{n}\,d\Gamma \quad ,\\
     L'_4 &= \int_{\Sigma} \Bigl(\frac{1}{2}\left(\boldsymbol{\sigma}_{nn}(\mathbf{W}_1) + \boldsymbol{\sigma}_{nn}(\mathbf{W}_2)\right)\left[\boldsymbol{\epsilon}_{nn}(\mathbf{u})\right] \\& + \left(\boldsymbol{\sigma}_{nt}(\mathbf{W}_1) + \boldsymbol{\sigma}_{nn}(\mathbf{W}_2)\right)\left[\boldsymbol{\epsilon}_{nt}(\mathbf{u})\right]\Bigr) \cdot \mathbf{n}\,d\Gamma \quad .
    \end{aligned}
\end{equation}

The final expression of the sensitivity of the objective is a simple summation of Eqs.~\eqref{eq:L1}, \eqref{eq:L2} and \eqref{eq:L34}.
Because compliance is a self-adjoint functional, its final shape derivative expression can be significantly simplified, as shown in \cite{allaire2014multi} 

\begin{equation}
\label{eq:comp}
    \begin{aligned}
     L' & = \int_\Gamma D\cdot \mathbf{n} d\Gamma, \\
     D & = -\boldsymbol{\sigma}\left(\mathbf{u}\right)_{nn}[\boldsymbol{\epsilon}\left(\mathbf{u}\right)_{nn}] -2\boldsymbol{\sigma}\left(\mathbf{u}\right)_{nt}[\boldsymbol{\epsilon}\left(\mathbf{u}\right)_{nt}] \\
     & +[\boldsymbol{\sigma}\left(\mathbf{u}\right)_{tt}]\boldsymbol{\epsilon}\left(\mathbf{u}\right)_{tt} \quad .
    \end{aligned}
\end{equation}

The final expression for the SA of both compliance and stress functionals is defined as a boundary integral.
The explicit, smooth and precise boundary representation that characterizes our framework unlocks new opportunities to maximize the potential of this method, which, to the best of the authors' knowledge, has not been previously fully utilized and investigated.

\subsubsection{Sensitivity analysis on a reduced discretized
domain -- parameterized shape SA}
\label{sec:Para}
C\'ea's method as formulated in Section \ref{sec:deriv_Cea} is much more efficient than the discrete formulation of Section \ref{sec:FirstSA} in terms of computational effort.
Nonetheless, as will be discussed in Section~\ref{sec:CeaInsights}, it suffers from two main drawbacks:
1) Lower accuracy compared to the discretized formulation; 
2) Challenge in applying to functionals involving interfacial stresses.
To overcome the second limitation, \citet{liu2020multi} suggested to evaluate the interfacial stresses as a narrow-band domain integral in the vicinity of the material interface.
In a similar case, \citet{feppon2020topology} reformulated the lift functional as a volume integral to maximize it.

Herein we suggest an alternative formulation of the design sensitivities that does not require any modification to the interfacial stress functional -- i.e., it can be evaluated and derived precisely.
The formulation integrates features from the previous two types of sensitivity analyses.
On the one hand, we exploit the explicit and parameterized physical representation, including that of the interface, hence the functional and the sensitivities with respect to the control points can be computed as in Section \ref{sec:FirstSA}.
On the other hand, we are interested only in the design sensitivities that may govern the shape update, as in classical shape optimization.  
This means that we concentrate solely on control points that parameterize the interface and on those that have a direct impact on the evaluation of the response functional.

\begin{figure}[tbp] 
\centering
\subfloat[]{\label{fig:CompSet}
\resizebox{0.49\textwidth}{!}{
\input{Figures/OptDes/ProblemSet1}}} \\
\subfloat[]{\label{fig:CompDes} 
\includegraphics[width=0.41\textwidth]{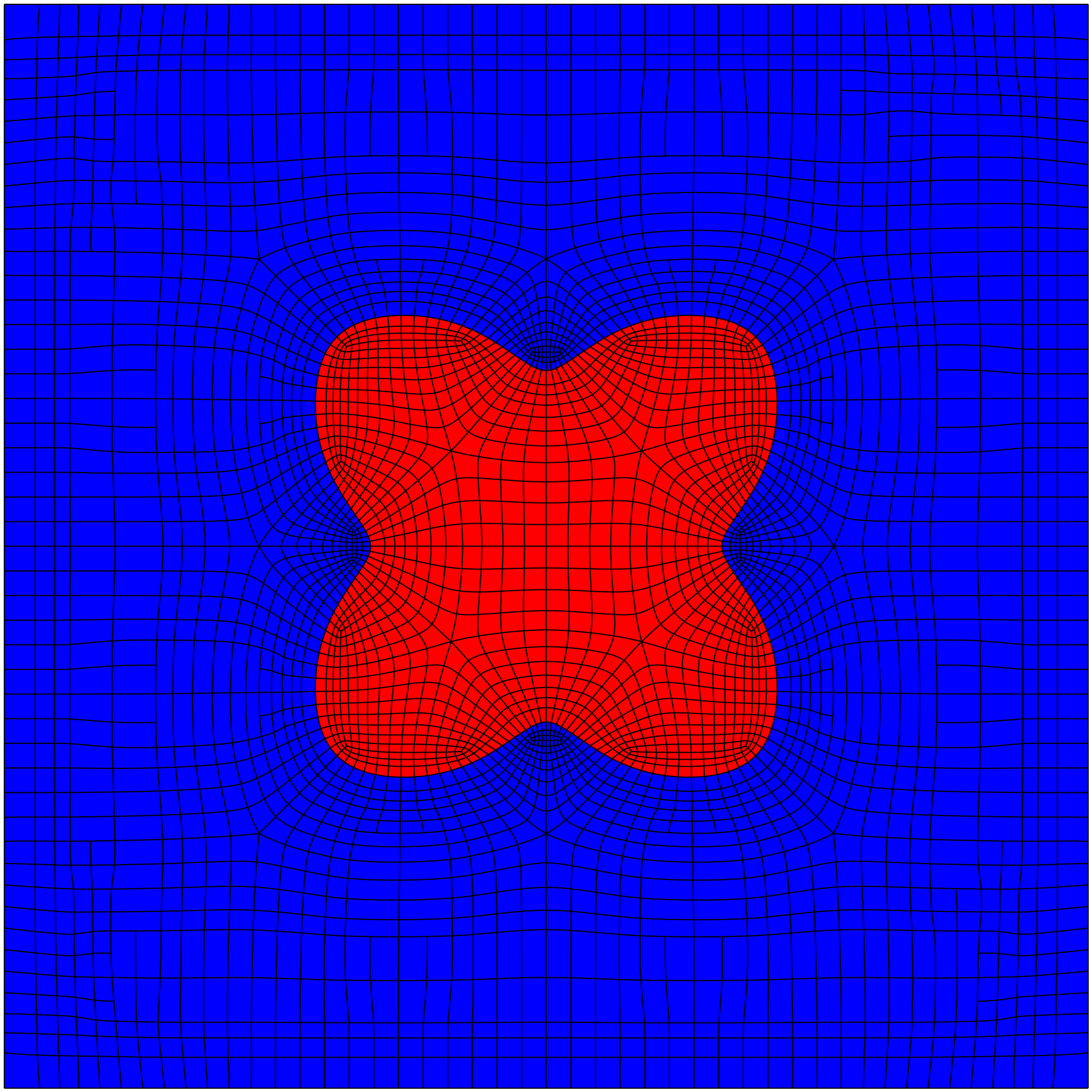}}
\caption{The setup used for comparison between the three types of sensitivity analyses. 
Panels (a) and (b) show the setup and the design, respectively. 
A stiff inclusion (red) is embedded in a soft matrix (blue), subjected to a tensile load on one edge.}
\label{fig:ComparStruc}
\end{figure}

Sensitivity analysis of compliance is computed similarly to \prettyref{eq:SA_comp}, but the intermediate design variables $\mathbf{P}_M$ are only the control points that define the material interface. 
This is a viable choice because when considering all control points as intermediate design variables (as in Section \ref{sec:FirstSA}), only the interfacial control points have significant sensitivity values. 
To illustrate this, we consider a $10\times10$ design domain composed of a flower-shaped stiff inclusion ($E_{\text{inc}} = 1000)$ surrounded by a softer matrix material with ($E_{\text{mat}} = 200$), as displayed in \prettyref{fig:ComparStruc}.

\begin{figure}[tbp] 
\centering
\subfloat[]{\label{fig:NonZeroComp} 
\includegraphics[width=0.48\textwidth]{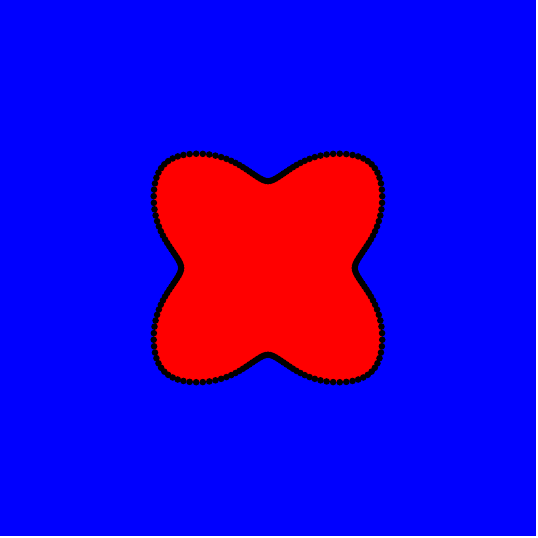}} \\
\subfloat[]{\label{fig:NonZeroSig}
\includegraphics[width=0.48\textwidth]{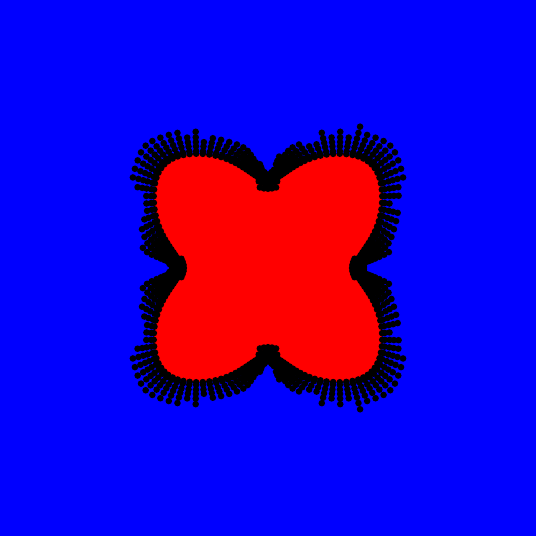}}
\caption{Sensitivity analysis w.r.t.~intermediate design variables using a discretized formulation. 
The values are normalized, and black dots appear only where the derivative is larger that $10^{-3}$. Panels (a) and (b) show sensitivities of compliance and interfacial stresses, respectively.}
\label{fig:NonZero}
\end{figure}

The design derivatives are computed on the entire discretized domain and are normalized with respect to the maximal value. 
Results show that significant derivatives exist only on the interface, while the normalized sensitivities of other control points are lower than $10^{-3}$, as shown in \prettyref{fig:NonZeroComp}. 
This result agrees with the interpretation of the continuum structure:
any perturbation (movement of a control point) that does not affect the shape of the interface and the topology of the structure, has no effect on the structural response, and hence yields the same compliance. 
It should be noted that the sensitivities with respect to control points outside the interface are insignificant, but do not completely vanish. 
This is because we derive a discretized model, where the stiffness matrix is directly influenced by the positions of the control points, hence the \textit{numerical evaluation} of compliance changes slightly with the movement of control points.

As for the stress functional, we show that the sensitivity analysis is not as local as for compliance. 
It is insufficient to consider control points on the interface only, and the computation should encompass a certain range of control points in the vicinity of the points where stress is evaluated. 
To illustrate this, \prettyref{fig:NonZeroSig} shows the normalized design sensitivities of the functional 
$f^{\RNum{3}}_{\sigma_i}$, computed as in \prettyref{eq:Pnorm3}. 
Without loss of generality, we choose to evaluate the interfacial stresses on the matrix (blue) side.
The results show that not only interface control points, but rather a narrow band of points, have significant sensitivities.  
This can be explained by considering both parts of \prettyref{eq:Adjoint}. Since each patch is defined by a set of control points -- usually 16 for a bi-cubic patch -- the first (explicit) part that correlates to $\frac{\partial \boldsymbol{\sigma}_{ij}}{\partial \mathbf{P}_M}$ is non-zero for the whole set, including control points away from the interface. 
The same dependency also applies for the derivative with respect to the displacements $\frac{\partial \boldsymbol{\sigma}_{ij}}{\partial \mathbf{u}}$, only that this dependency requires a wider range of control points due to the global nature of $\frac{d\mathbf{u}}{d\mathbf{P}_M}$.
In other words, changing the location of any control point will change the evaluation of the displacement field and henceforth affect the stress evaluation at the hosting patch. 
Mathematically, this can be explained as follows: 
even though the right hand side of \prettyref{eq:AdEq} is non-zero for an exclusive set of control points, the adjoint vector is solved for the whole domain. 
Hence, the second part of \prettyref{eq:Adjoint} is non-zero for a wider range of control points.

Fortunately, the impact of each design variable diminishes as the distance from it increases. 
Given that cubic B-splines are employed as the basis functions, we define a 4$\times$4 grid of control points as the `influence range' of each control point. 
Therefore, for any stress evaluation point, the complete set of relevant control points is composed of two groups: 
1) Control points of the element (usually 16), referred to as hosting control points; 
and 2) The $4\times4$ grid surrounding each of these hosting control points.

Clearly, if the aim is to control stresses in the whole domain -- e.g., by evaluating at the center of each patch as in the stress functional $f^{\RNum{1}}_{\sigma_i}$ -- then all control points will be included, resulting in the same procedure as described in Section \ref{sec:FirstSA}.
However, to control stresses at the interface with the functional $f^{\RNum{3}}_{\sigma_i}$ of \prettyref{eq:Pnorm3}, the set of control points needed for sensitivity analysis is reduced significantly. 
Therefore, the formulation based on a reduced discretized domain offers a valuable trade-off.
It allows to compute consistent derivatives of interfacial stresses, a task that has not been accomplished so far using C\'ea's method. 
At the same time, it does not suffer from the high computational burden of the discretized derivation of the whole domain. 
Ultimately, it resembles \emph{parameterized shape SA} -- the movement of the boundary is governed by the control points of the boundary, but consistent derivation requires to account also for other control points that affect the discretization. 
 
\section{Comparisons and insights}\label{sec:insights}

This section is dedicated to investigations and comparisons of the three types of sensitivity analyses.
For consistency, all examinations will be carried out on the same structural setup, presented in \prettyref{fig:ComparStruc}.
A stiff inclusion ($E_{\text{inc}} = 1000$) is embedded in a softer matrix ($E_{\text{mat}} = 200$), subjected to a tensile load on one edge. 
Both materials have the same Poisson's ratio of $\nu = 0.3$.

\begin{table}
    \centering
    \normalsize
    \begin{tabular}{c|c c c}
    \hline
        Stress & Discretized & C\'ea's & Parameterized \\
        functional & domain SA &  method & shape opt. \\
        \hline
        $f^{\RNum{1}}_{\sigma_i}$ & \checkmark & &\\
        \hline
        $f^{\RNum{2}}_{\sigma_i}$ &  & \checkmark & \\
        \hline
        $f^{\RNum{3}}_{\sigma_i}$ & \checkmark & & \checkmark\\
        \hline\hline
    \end{tabular}
    \caption{Summary of the scope and limitations of the three types of SA, with respect to the three stress functionals.}
    \label{tab:SUm}
\end{table}

While all three types of sensitivity analyses share a common definition of the compliance functional, the definition of the stress functional differs according to the scope and relevance of each type. 
The various definitions are outlined in \prettyref{tab:SUm}. 
Note that the discrete procedure of Section \ref{sec:FirstSA} can consider either $f^{\RNum{1}}_{\sigma_i}$ or $f^{\RNum{3}}_{\sigma_i}$, or a combination of the two.
C\'ea's approach however, is formulated consistently only for $f^{\RNum{2}}_{\sigma_i}$ whereas the parameterized shape SA is beneficial only when considering $f^{\RNum{3}}_{\sigma_i}$. 
To facilitate the comparison, and unless stated otherwise, all mentions of von Mises stresses -- both on the interface and within the domain -- refer to the matrix (blue) material. 
Consequently, the subscript indicating the phase number, $i$, is omitted, bearing in mind that the conclusions are applicable to both materials equally.

\textbf{Remark}: Since the integration in \prettyref{eq:Pnorm2} is carried out numerically via Gauss integration -- a summation of discrete points each multiplied by a weighting factor -- the discrete procedure can be applied, in principle, also to $f^{\RNum{2}}_{\sigma_i}$.
However, we choose not to include $f^{\RNum{2}}_{\sigma_i}$ within the scope of the discretized SA, because the functional itself is continuous, and Gauss integration is merely an acceptable approximation for implementing a ``differentiate-then-discretize'' approach.

This section is structured as follows: first, a verification of the discretized domain SA is presented. 
Subsequently, we explore the accuracy of C\'ea's method and discuss the implications of the choice to ``differentiate-then-discretize''.
Finally, we discuss the convergence of the parameterized shape SA, and compare its performance to the two other types.

\subsection{Verification of the discretized domain SA}
\begin{figure}[htbp]
    \centering
    \includegraphics[width=\linewidth]{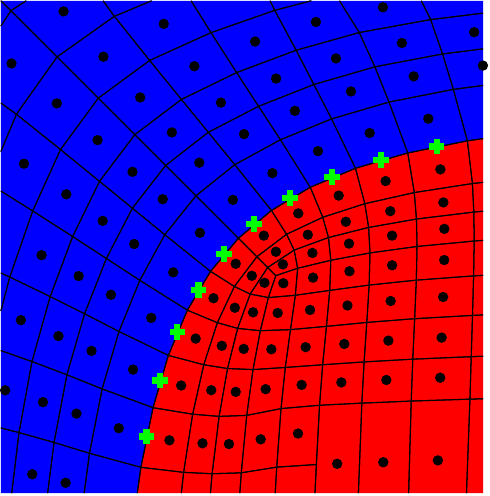}
    \caption{Stress computation points at the center of each patch (black dots) and on the material interface (green crosses), for quantifying the functionals $f^{\RNum{1}}_{\sigma}$ and $f^{\RNum{3}}_{\sigma}$.}
    \label{fig:StressPoints}
\end{figure}

\begin{figure}[h!]
    \centering
    \subfloat[]{\label{fig:VD_Comp}
    \resizebox{0.5\textwidth}{!}{
    \input{Figures/SA/VOL_FD_Comp}}} \\
    \subfloat[]{\label{fig:VD_Stress}
    \resizebox{0.5\textwidth}{!}{
    \input{Figures/SA/VOL_FD_STressNEW}}}
    \caption{The analytical sensitivity analysis using the full discretized domain versus numerical (finite difference) derivatives, depicted as blue circles. 
    Panels (a) and (b) shows the results for compliance and von Mises stresses, respectively. 
    The total number of design variables is 169, with 77 non-zero derivatives for both functionals. The mean relative error for compliance is $1.3\cdot10^{-5}$ and the mean relative error for stresses is $2.0\cdot10^{-4}$.}
    \label{fig:FD_VOL}
\end{figure}
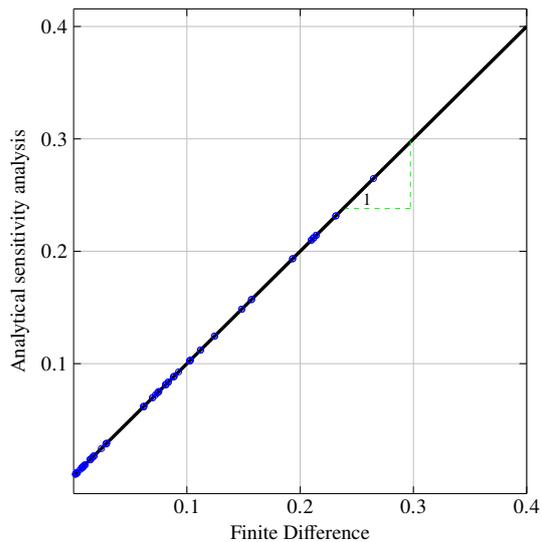
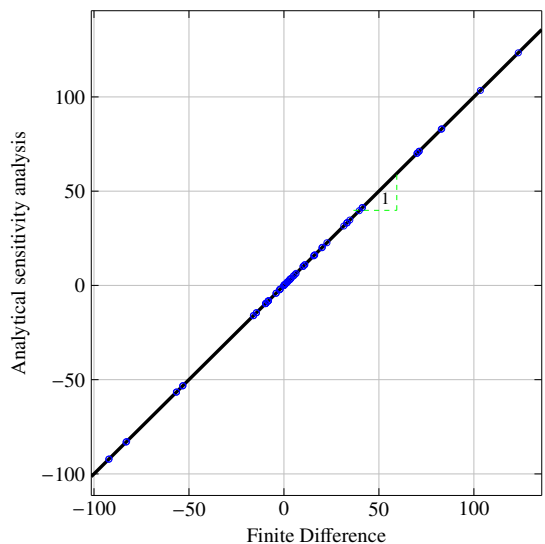

In this section, we provide a verification of the discretized domain SA. 
To this end, the analytical sensitivity analysis is compared to numerical derivatives that are computed by finite differences. 
The structural setup is the same as the one given in \prettyref{fig:ComparStruc}.
The stress functional is defined as a summation of both $f^{\RNum{1}}_{\sigma}$ and $f^{\RNum{3}}_{\sigma}$, namely, discrete computational points at the center of each patch, as well as at the center of the interface segment, as shown in \prettyref{fig:StressPoints}. 
The results are displayed in \prettyref{fig:FD_VOL} and show excellent agreement for both functionals -- compliance and stresses in the matrix.
This confirms the correctness of our derivations and implementation. 
The achieved accuracy is not surprising since we ``discretize-then-differentiate'', so consistency is preserved when we derive the same discrete model that was used for the analysis. 
The excellent agreement between the discretized domain SA and numerical derivatives -- with a mean relative error of $1.3\cdot10^{-5}$ for compliance and $2.0\cdot10^{-4}$ for von Mises stresses -- makes it a perfect candidate in terms of accuracy.
Hence it is used hereafter as a reference for comparisons to the other types of SA. 

\subsection{Accuracy of SA using C\'ea's method}
\label{sec:CeaInsights}
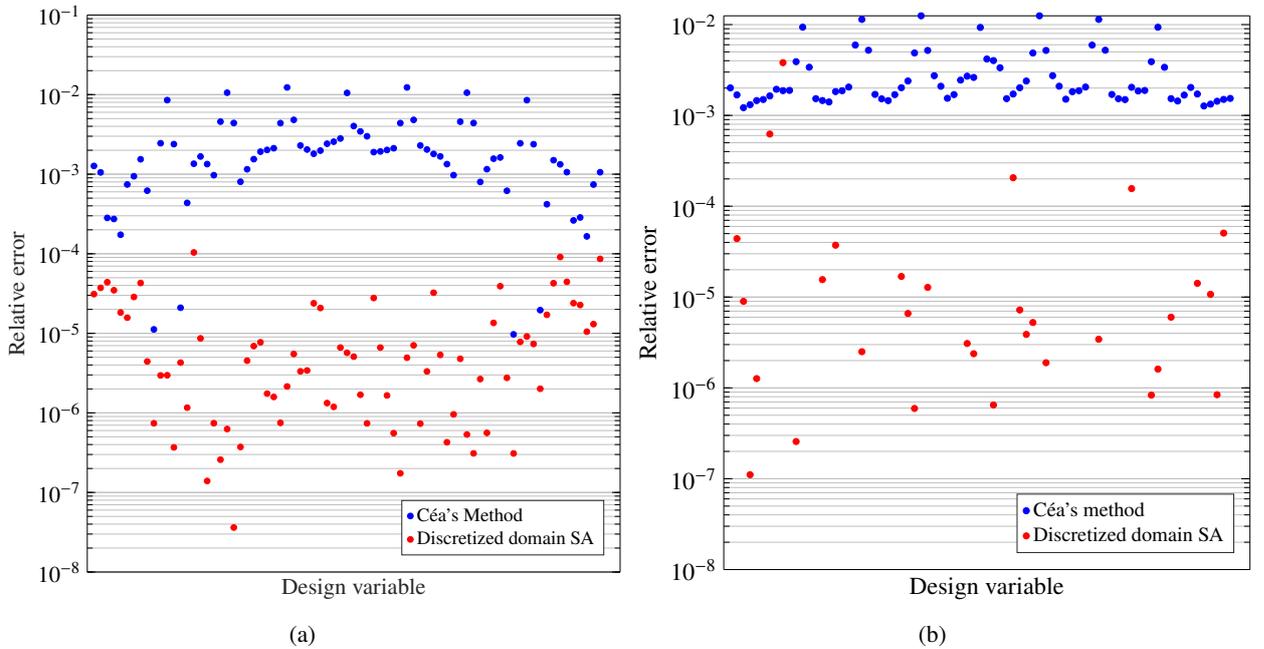
\begin{figure*}[htbp] 
\centering
\subfloat[]{\label{fig:ComparisonJ}
\resizebox{0.5\textwidth}{!}{
\input{Figures/SA/Compliance_Comp}}}
\subfloat[]{\label{fig:ComparisonS}
\resizebox{0.5\textwidth}{!}{
\input{Figures/SA/Comp_Stress_New}}}
\caption{Comparison between ``discretize-then-differentiate'' and ``differentiate-then-discretize''. Panels (a) and (b) show the relative errors of design sensitivities, for compliance and domain stresses, respectively.}
\label{fig:comparison}
\end{figure*}

Theoretically, reversing the sequence of discretization and differentiation should lead to similar results, but this equivalence breaks down when the response functional is evaluated using a discretized model \citep{hiptmair2015comparison}. 
In such a case, the discretized domain SA -- representing a ``discretize-then-differentiate'' scheme -- is more consistent, since the derivation and the analysis are performed on the same model. 
Hence, it is not surprising that the discretized domain SA offers better accuracy than C\'ea's method, as can be seen in \prettyref{fig:comparison}.
The relative error by comparison to numerical finite differences is higher for C\'ea's method for both compliance and domain stresses (computed via $f^{\RNum{1}}_{\sigma}$ for the discretized domain SA and via $f^{\RNum{2}}_{\sigma}$ for C\'ea's method.)

Since the lower accuracy of C\'ea's method is a direct result of the choice to ``differentiate-then-discretize'', it is anticipated that the derivatives will converge with mesh refinement. 
Indeed, such a conclusion can be drawn from \prettyref{tab:refinement}, where the relative errors of C\'ea's method are presented, for the same design with different refinement levels. 
Clearly, the drawbacks of ``differentiate-then-discretize'' can be overcome by mesh refinement, at the expense of computational cost. 

\begin{table}[htbp]
    \centering
    \normalsize
    \begin{tabular}{c| c c}
    \hline
        Number of elements & $f_c = \mathbf{f}^T\mathbf{u}$ & $f^{\RNum{2}}_{\sigma}$ \\
        \hline
        3,690 & 0.159 & 0.318 \\
        4,036 & 0.158 & 0.3126 \\
        28,498 & 0.0328 & 0.0197 \\
        71,428  & 0.0316 &  0.0135 \\
        \hline\hline
    \end{tabular}
    \caption{Mesh convergence of C\'ea's method. The mean relative error (in \%) of the design sensitivities, computed on different refinement levels.}
    \label{tab:refinement}
\end{table}

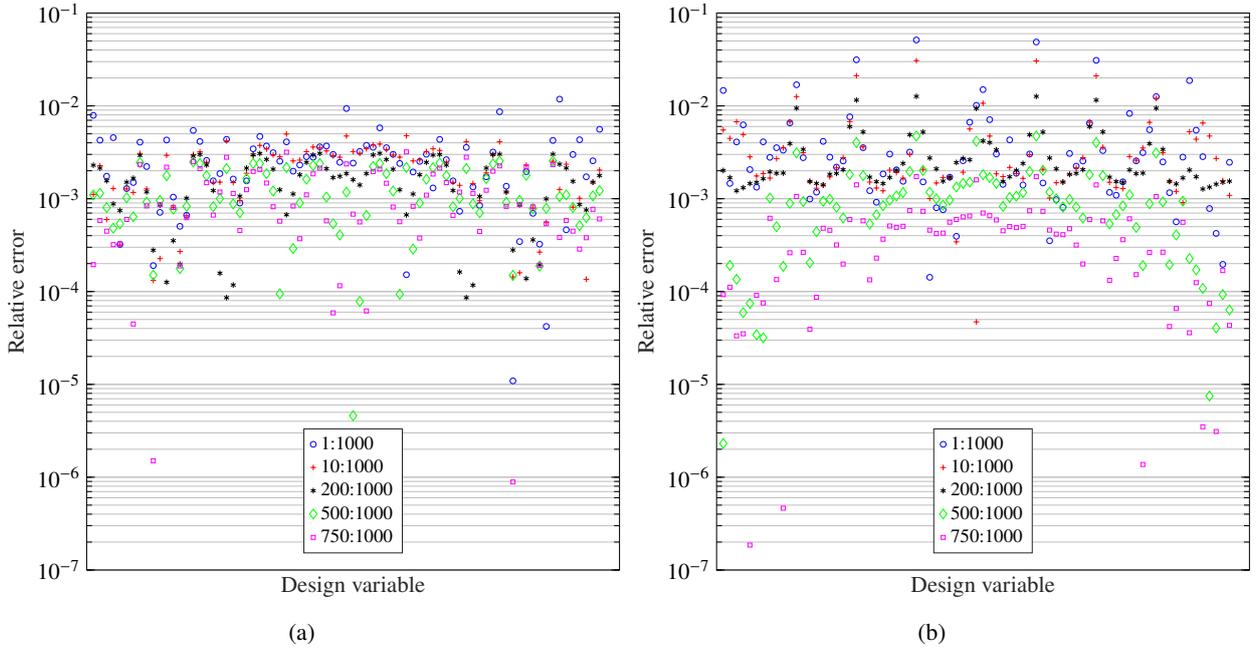
\begin{figure*}[htbp] 
\centering
\subfloat[]{\label{fig:diffE_J}
\resizebox{0.5\textwidth}{!}{
\input{Figures/SA/Compliance_diffE}}}
\subfloat[]{\label{fig:diffE_S}
\resizebox{0.5\textwidth}{!}{
\input{Figures/SA/Stress_diffE}}}
\caption{The impact of stiffness contrast between phases on the accuracy of design sensitivities using C\'ea's method, for compliance in panel (a) and von Mises stress in panel (b).
Higher errors are observed as the contrast is increased, with the effect slightly enhanced for the stress functional.}
\label{fig:diffE}
\end{figure*}

Furthermore, our findings indicate that the precision of C\'ea's method is affected by the contrast between the two phases. 
This is illustrated in \prettyref{fig:diffE}, where the relative error of  C\'ea's method is computed for different ratios between the elasticity moduli of the two phases.
We observe that a lower contrast between the properties of the two materials leads to higher accuracy. 
This can be explained by the maintenance of the transmission conditions: for high-contrast structures, the jump at the interface is larger, and such conditions are more difficult to uphold numerically. 
In addition, a high stress concentration might appear at the interface for high-contrast structures.
These factors aggravate the numerical difficulties in computing accurate shape derivatives, where the strains and stresses at the interface are an explicit part of the expression, see Section \ref{sec:deriv_Cea}.

An intriguing finding is the similar accuracy of C\'ea's method when applied to either compliance or stress. 
One might reasonably anticipate greater accuracy in the SA of compliance, due to its inherent lower mesh-dependency. 
However, the results indicate only a marginally higher accuracy, which is not as significant as expected. 
This can be attributed to the integration across the entire domain of the matrix material, which reduces the effect of the discretization.
In particular, it diminishes the impact the discretization has on the perturbation of the interface.
This observation will be explored and elaborated upon in the following sections.

\subsection{Accuracy of parameterized shape SA}
\label{sec:APSO}
In this section, we investigate the accuracy of the parameterized shape SA -- that is formulated based on the discretized domain, but uses a reduced set of control points in a predefined support range, in the vicinity of the moving interface or boundary. 
We demonstrate the effect of the support range and compare the accuracy of the parameterized shape SA to the two previous types of differentiation.

\begin{figure}[tbp] 
\centering
\subfloat[]{\label{fig:SA_Points} 
\includegraphics[width=0.5\textwidth]{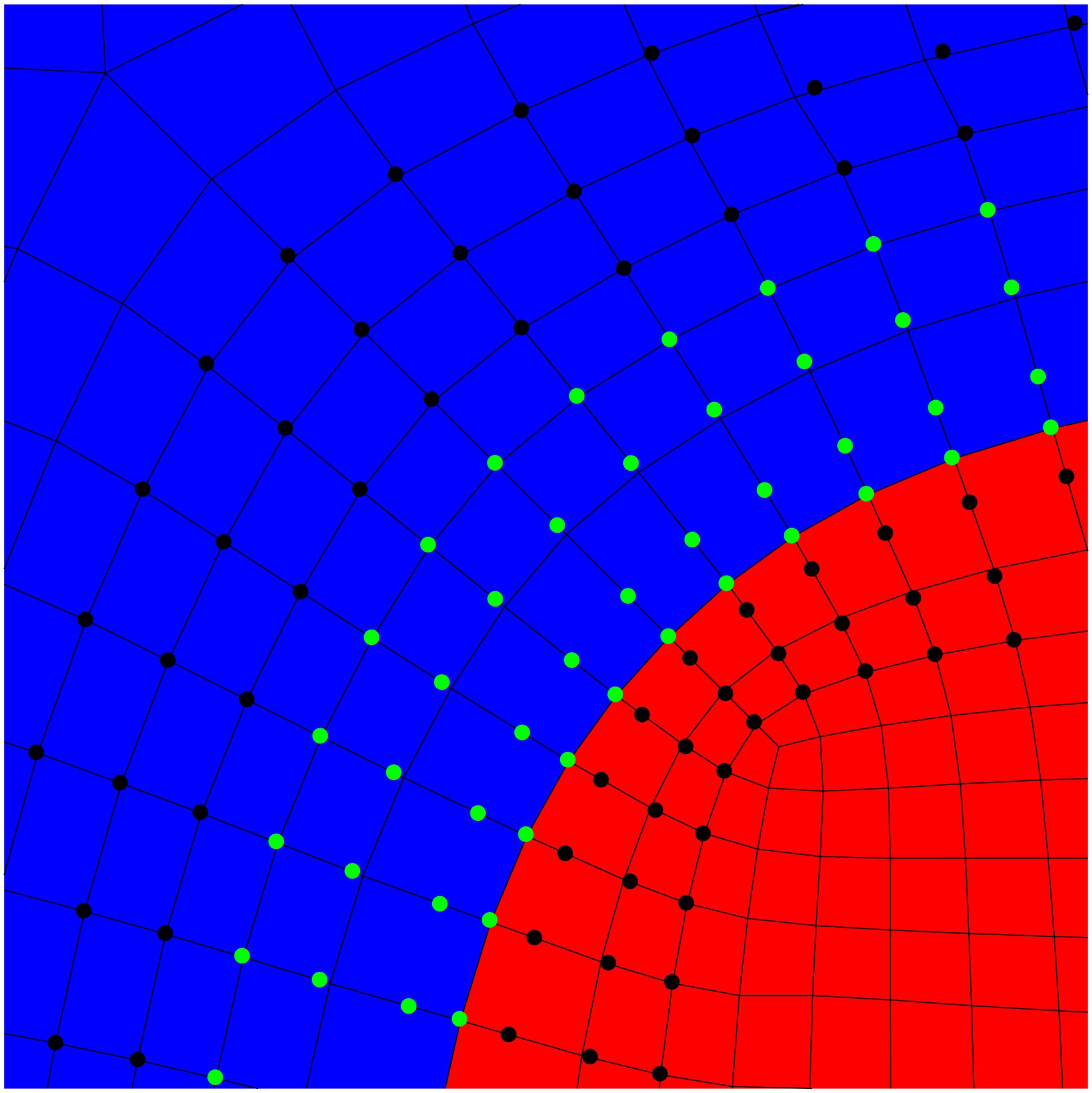}} \\
\subfloat[]{\label{fig:SA_conv}
\resizebox{0.5\textwidth}{!}{
\input{Figures/SA/Convg_AddP}}}
\caption{The effect of the set of control points used for parameterized shape SA. Panel (a) shows the complete collection of control points, where hosting control points are indicated by green dots and black dots represent control points within the influence range of the hosting points.
Panel (b) displays the relative error of the sensitivity analysis of the functional $f^{\RNum{3}}_{\sigma}$ compared to finite differences, computed for various support ranges, from zero to four layers.}
\label{fig:SApoints}
\end{figure}

An example of the complete collection of control points -- hosting control points and control point within the influence ranges -- is shown in \prettyref{fig:SA_Points}. The green dots represent hosting control points and the black dots denote the control point within the support range of the hosting points. The term \textit{layers} is used to describe the number of rings of black control points included within the support range. To illustrate the impact of the support range, we evaluate the relative error of the proposed method in comparison to finite differences across various number of layers. 
The relative errors of the design sensitivities of the stress functional $f^{\RNum{3}}_{\sigma}$ are
depicted in \prettyref{fig:SA_conv}.
As expected, the results indicate that a wider range improves the accuracy of the design sensitivities.
Moreover, it is shown that high accuracy can be obtained without considering the whole domain -- meaning that computational efficiency can be improved compared to the full domain approach.

\begin{figure}
    \centering
    \includegraphics[width=0.95\linewidth]{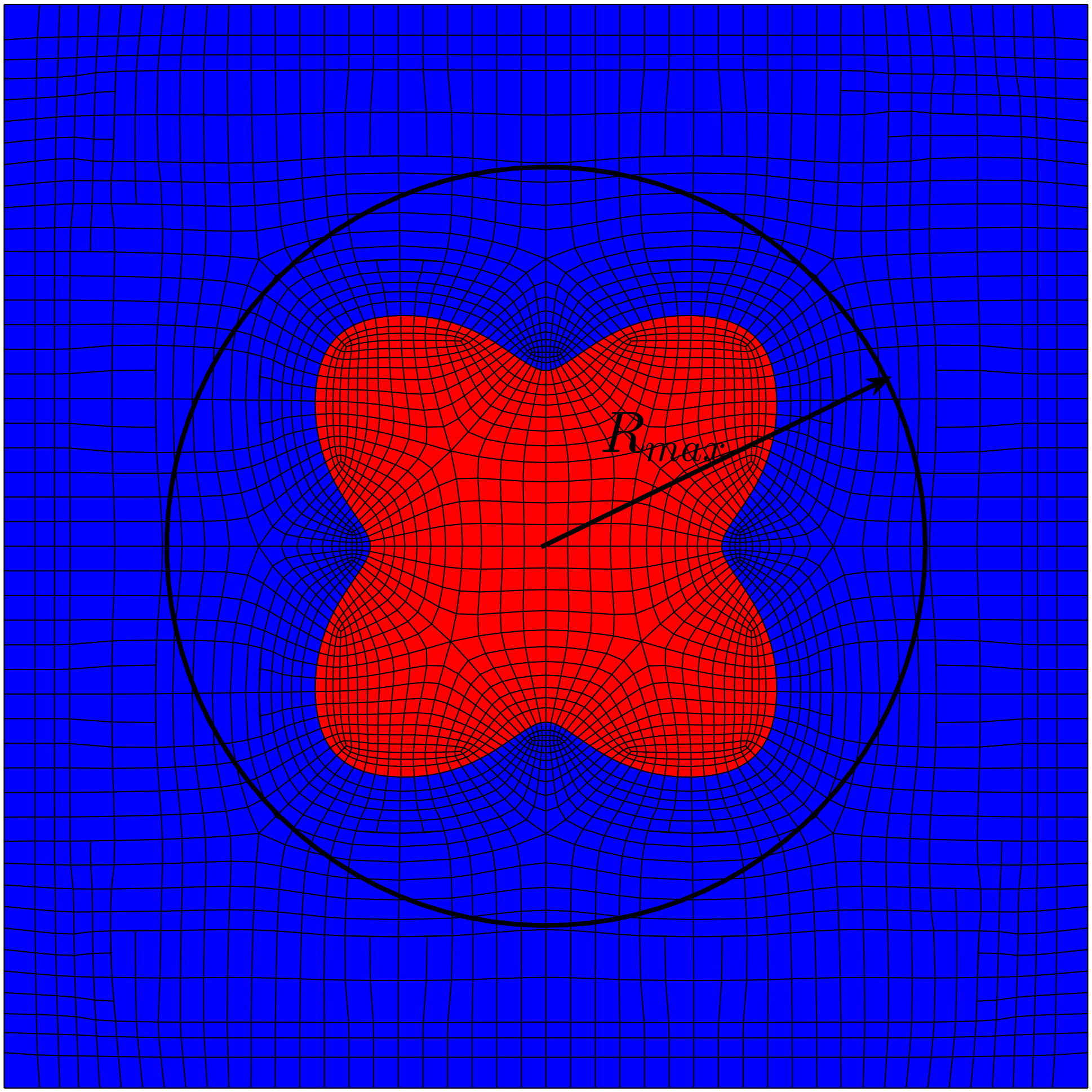}
    \caption{The definition of $\Omega_i$ in \prettyref{eq:Pnorm2} based $R_{max}$: the integration domain includes only the elements whose centroid is inside the black circle.}
    \label{fig:Rmax}
\end{figure}

The results of \prettyref{fig:SApoints} raise questions about the accuracy of C\'ea's method. 
We use C\'ea's method to optimize domain stresses, which include also stresses on elements that touch the interface. 
So, if interfacial stresses require a narrow band of control points to be considered as shown above, why is it acceptable to rely only on boundary control points for computing the derivative in C\'ea's method? 
A possible answer is that the stresses in elements near the interface are not the dominant part in the stress functional $f^{\RNum{2}}_{\sigma}$ of \prettyref{eq:Pnorm2}, and hence the accuracy of C\'ea's method is not strongly affected. 
To investigate this, we examine the accuracy of C\'ea's method for various definitions of $\Omega_i$ in \prettyref{eq:Pnorm2}. 
To this end, we define $R_e$ as the distance between the centroid of element $e$ and the center of the entire domain. 
Then, instead of considering all elements, we consider only the elements whose distance $R_e$ is smaller than a certain value $R_{max}$. 
This implies that the definition of $\Omega_i$ changes according to the value of $R_{max}$: 
for smaller values of $R_{max}$, the boundaries of the domain $\Omega_i$ are closer to the interface, and the weight of the interfacial stresses in \prettyref{eq:Pnorm2} increases accordingly.
For an illustration, see \prettyref{fig:Rmax}: the integration domain in \prettyref{eq:Pnorm2} is defined as all blue elements that are inside the black circle.
The results presented in \prettyref{tab:DiffRadius} show that the smaller is $R_{max}$, the higher is the relative error of C\'ea's method when differentiating domain stresses.
In other words, considering only the movement of the boundary when computing design sensitivities of stresses near the boundary, may lead to significant inaccuracies.

\begin{table}
    \centering
    \normalsize
    \begin{tabular}{c|c}
    \hline
        Rmax  & Relative error \\ 
        \hline
        3 & 4.25 \\
        3.5 & 0.568 \\
        4 & 0.189\\
        4.5 & 0.056\\
        5 & 0.0147 \\
        5$\sqrt{2}$ & 0.0032\\ 
        \hline\hline
    \end{tabular}
    \caption{The relative error of C\'ea's method when differentiating domain stresses, computed for different definitions of the domain $\Omega_i$ in \prettyref{eq:Pnorm2}.}
    \label{tab:DiffRadius}
\end{table}

\begin{figure*}[t!] 
\centering
\subfloat[]{\label{fig:BigCompar_Compliance}
\resizebox{0.5\textwidth}{!}{
\input{Figures/SA/Compare3_compliance}}}
\subfloat[]{\label{fig:BigCompar_Stres}
\resizebox{0.5\textwidth}{!}{
\input{Figures/SA/Stress_ParaVsDom}}}
\caption{The relative accuracy of the SA formulated as parameterized shape derivatives. Panel (a) shows the relative error for compliance, computed using the three methods. Panel (b) shows the relative error for interfacial stresses, computed using the full domain and the parameterized shape SA.
}
\label{fig:BigCompar}
\end{figure*}
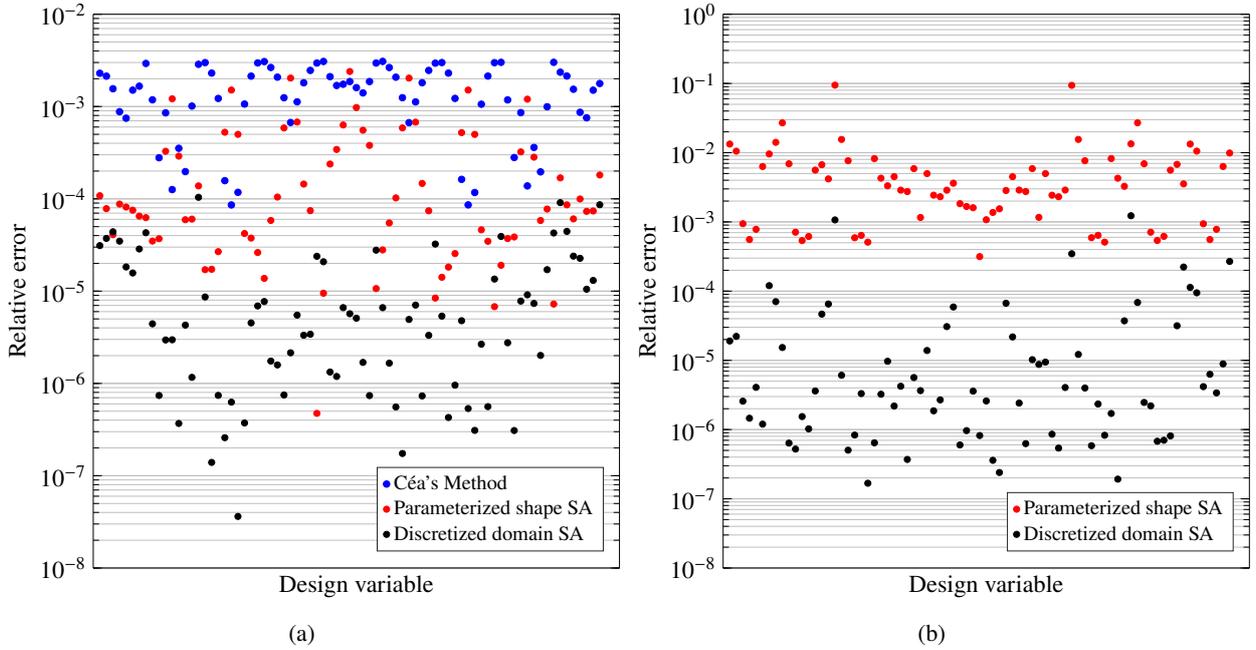

Finally, a thorough comparison between the three formulations when computing the sensitivities of compliance is shown in \prettyref{fig:BigCompar_Compliance}. 
It is evident that the discretized domain SA achieves the highest accuracy, because it follows the consistent discretize-then-differentiate approach, whereas C\'ea's method does the opposite. 
As anticipated, the parameterized shape SA ranks between the two, as it also relies on differentiating the discretized solution, but it utilizes a limited support range of control points in the vicinity of the interface. 
This reduces the accuracy compared to SA on the full discretized domain. 
This effect is evident also when evaluating the accuracy of SA of interfacial stresses measured by $f^{\RNum{3}}_{\sigma_i}$, as shown in \prettyref{fig:BigCompar_Stres}. 
As expected, the discretized domain SA shows higher accuracy than the parameterized shape SA, but the latter reaches sufficient accuracy using much less computational cost. Herein, the support range comprised of 3 layers.  

\section{Design examples}\label{sec:examples}

In this section, we discuss several results of optimization runs that utilize the three types of sensitivity analyses.
All results were obtained using the Method of Moving Asymptotes -- MMA \citep{svanberg1987method}, implemented in the authors' in-house code written in C. 
Unless stated otherwise, the move limit in MMA is set to 0.1 for the first 10 iterations, and subsequently is reduced to 0.05 to ensure smooth convergence. 

\subsection{Compliance minimization}
\label{sec:compmin}
In the first example, our objective is to maximize the rigidity of the  structure in \prettyref{fig:CompSet} without considering a stress limit, meaning $\omega = 0$.
To maintain symmetry conditions, all nodes along the left edge can move vertically (horizontal displacement is restricted), except for the node at the height of $\frac{L_y}{2}$, which is also restricted vertically to preserve the structure's stability.
The elasticity moduli of the inclusion and the matrix are 1000 and 200, respectively, and Poisson's ratio is set to $\nu = 0.3$ for both materials. 
The maximum volume of the stiff inclusion is limited to 15\% of the total volume.
We note that the design is restricted to a single inclusion inside a square matrix, where the main concern is finding the optimal shape of the inclusion. 
We note that topology optimization can be achieved with the same sensitivity analysis by starting the optimization with multiple inclusions.
This class of problems is left out of the current scope, to focus on investigating the formulations of design sensitivities and how they affect the optimization.

\begin{figure}[htbp] 
\subfloat[]{\label{fig:OptimalComp} 
\includegraphics[width=0.48\textwidth]{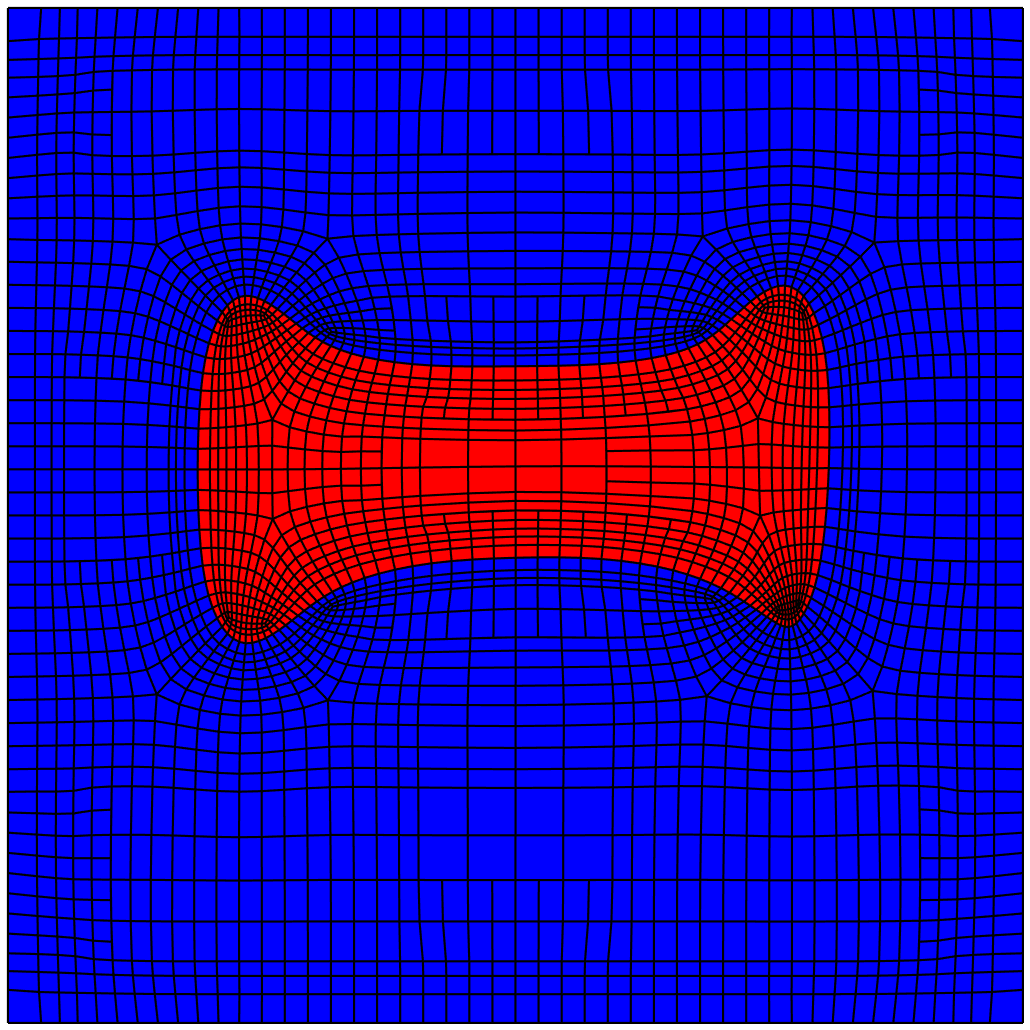}} \\
\subfloat[]{\label{fig:ConvGraph}
\resizebox{0.48\textwidth}{!}{
\input{Figures/OptDes/ComplianceFirstRun_Conv}}}
\caption{Results for minimizing compliance. Panel (a) displays the optimized design obtained with all three types of SA and panel (b) shows the compliance versus the iteration number.
}
\label{fig:FirstComp}
\end{figure}

The optimized designs obtained with all three types of sensitivities are practically identical.
One sample is presented in \prettyref{fig:OptimalComp} together with the convergence plots in \prettyref{fig:ConvGraph}.
Clearly, all three runs reach a very similar objective value.
The final values of the compliance for C\'ea's method, discretized domain and parameterized shape SA are $41.12$, $41.07$ and $41.02$, respectively.
The maximum number of iterations for each run is limited to $150$. 
We did not impose another stopping criterion in order to avoid an early termination of the optimization process of a certain run. 
We observe that the number of iterations needed for practical convergence is similar, with all three simulations reaching the minimal compliance after roughly 60 iterations.

\subsection{Compliance and stress minimization}
\label{sec:CompStress}
Once a stress functional is added to the objective function, the differences between the various formulations of the design sensitivities affect the outcome of optimization.
To demonstrate this effect, the weight factor is set to $\omega = 0.5$. 
Without loss of generality, in the following examples the stress functional is evaluated on the stiff inclusion only.
For all subsequent examples, the normalization factor is calculated as $80\%$ of the stress functional obtained from the design optimized for compliance.

\begin{figure*}[h!] 
\centering
\subfloat[$f_c$ = 44.14, $f^{\RNum{3}}_{\sigma_2}$ = 20.13, $f = 0.81$]{\label{fig:FDIStrsB} 
\includegraphics[width=0.5\textwidth]{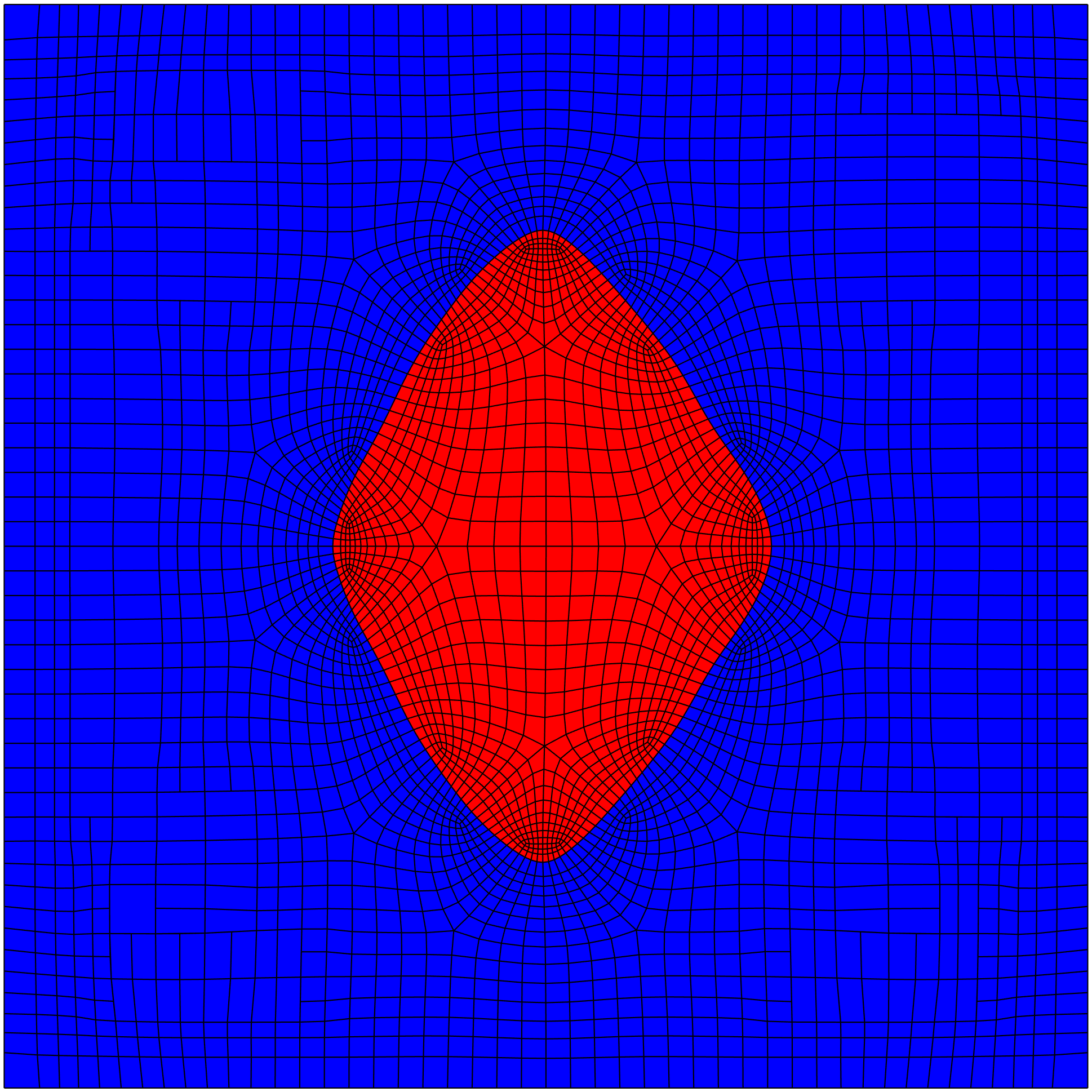}}
\subfloat[$f_c$ = 44.14, $f^{\RNum{3}}_{\sigma_2}$ = 20.13, $f = 0.81$]{\label{fig:ParamStrsB} 
\includegraphics[width=0.5\textwidth]{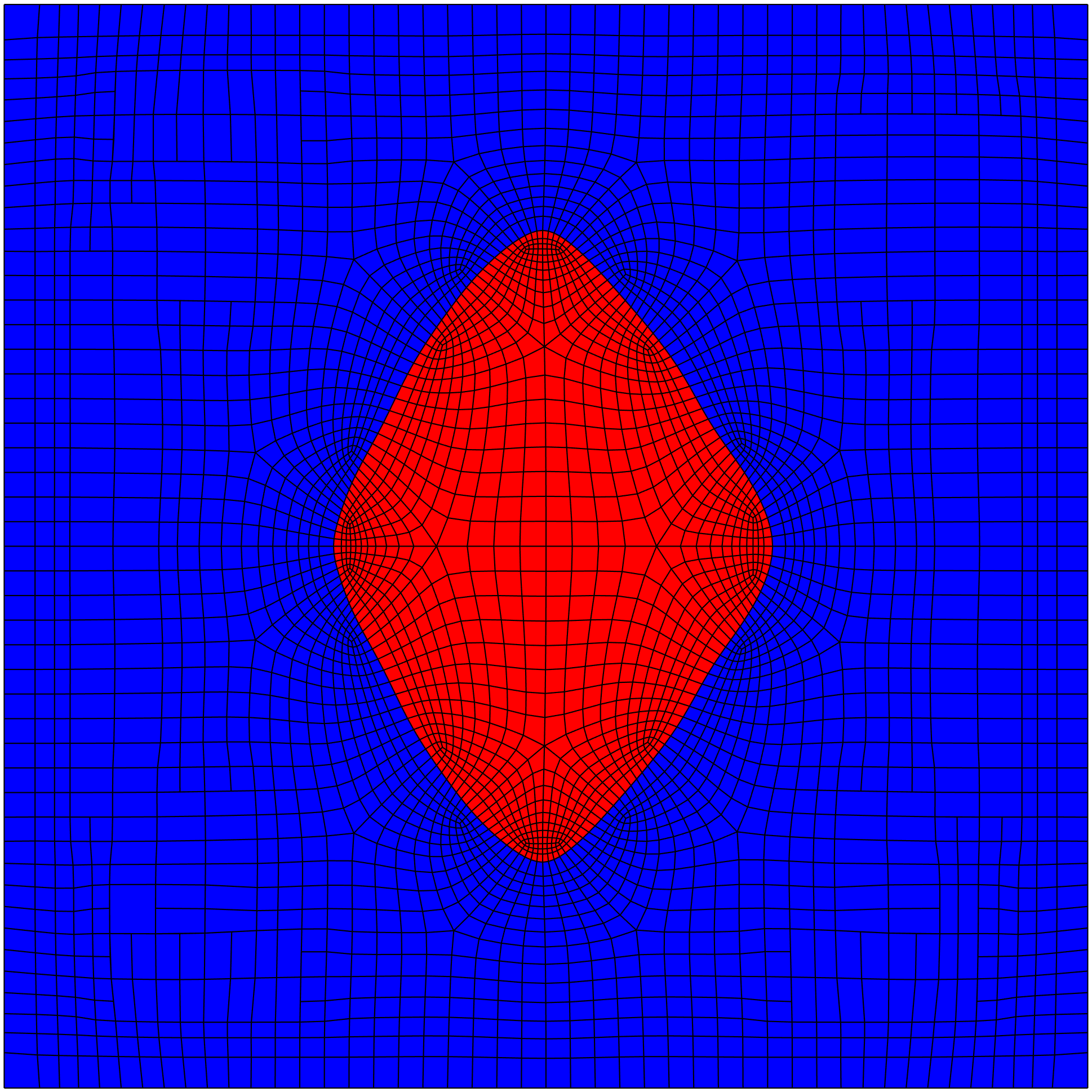}}
\caption{Compliance and interfacial stress minimization. Panels (a) and (b) display the optimized designs obtained with the discretized domain SA and parameterized shape SA, respectively.}
\label{fig:CompStress}
\end{figure*}

We start by including the stresses at the interface, namely the stress functional $f^{\RNum{3}}_{\sigma_2}$, in the objective function. 
The normalization factor for both the discretized domain SA and the parameterized shape SA is specified in \prettyref{tab:stresses}. 
We note that despite the notable similarity in the designs obtained for compliance with both types of SA, there is a minor variation in the stress field, which results in slightly different normalization factors.
Both optimization runs result in very similar optimized designs, as illustrated in \prettyref{fig:CompStress}. 
The compliance values are the same ($J = 44.14$) and so are the values of the von Mises interfacial stress functionals ($f^{\RNum{3}}_{\sigma_2} = 20.13$).
The true maximal von Mises stresses are nearly identical, $12.63$ for the discretized domain and $12.59$ for the parameterized shape.  
These results show that computing the sensitivities of interfacial stresses using a narrow region near the interface is a viable approach, since the contribution of more distant control points is small. 

\begin{table}
    \centering
    \normalsize
    \begin{tabular}{c | c c}
    & Discretized & Parameterized \\
    & domain SA & shape SA \\
    \hline
        Normalization value & $f*^{\RNum{3}}_{\sigma_2}$ = 28.7 & $f*^{\RNum{3}}_{\sigma_2}$ =29 \\
        Optimized stress & $f^{\RNum{3}}_{\sigma_2}$ = 20.13 & $f^{\RNum{3}}_{\sigma_2}$ = 20.13 \\
        True maximum stress  & 12.63 & 12.59  \\
        \hline\hline
    \end{tabular}
    \caption{Results for minimizing the compliance and the interfacial stresses.
    All results shown are for the stiff inclusion only, and related to the optimized designs shown in \prettyref{fig:CompStress}.}
    \label{tab:stresses}
\end{table}

\begin{figure*}[h!] 
\centering
\subfloat[$f_c$ = 44.08, $f^{\RNum{1}}_{\sigma_2}$ = 26.42, $f = 0.88$]{\label{fig:FDDStrsB} 
\includegraphics[width=0.5\textwidth]{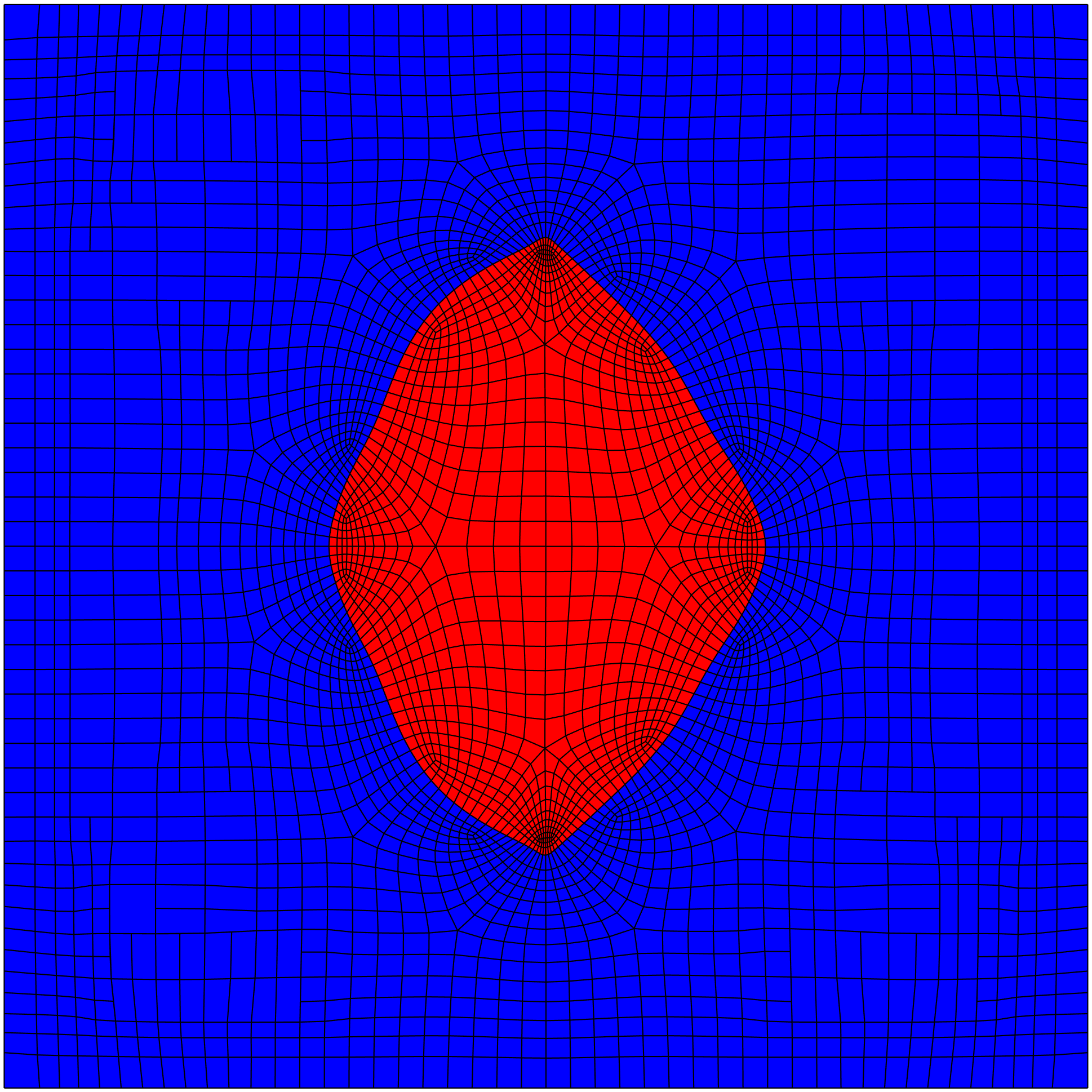}}
\subfloat[$f_c$ = 41.73, $f^{\RNum{2}}_{\sigma_2}$ = 21.40, $f = 1.01$ ]{\label{fig:CEAStrsB} 
\includegraphics[width=0.5\textwidth]{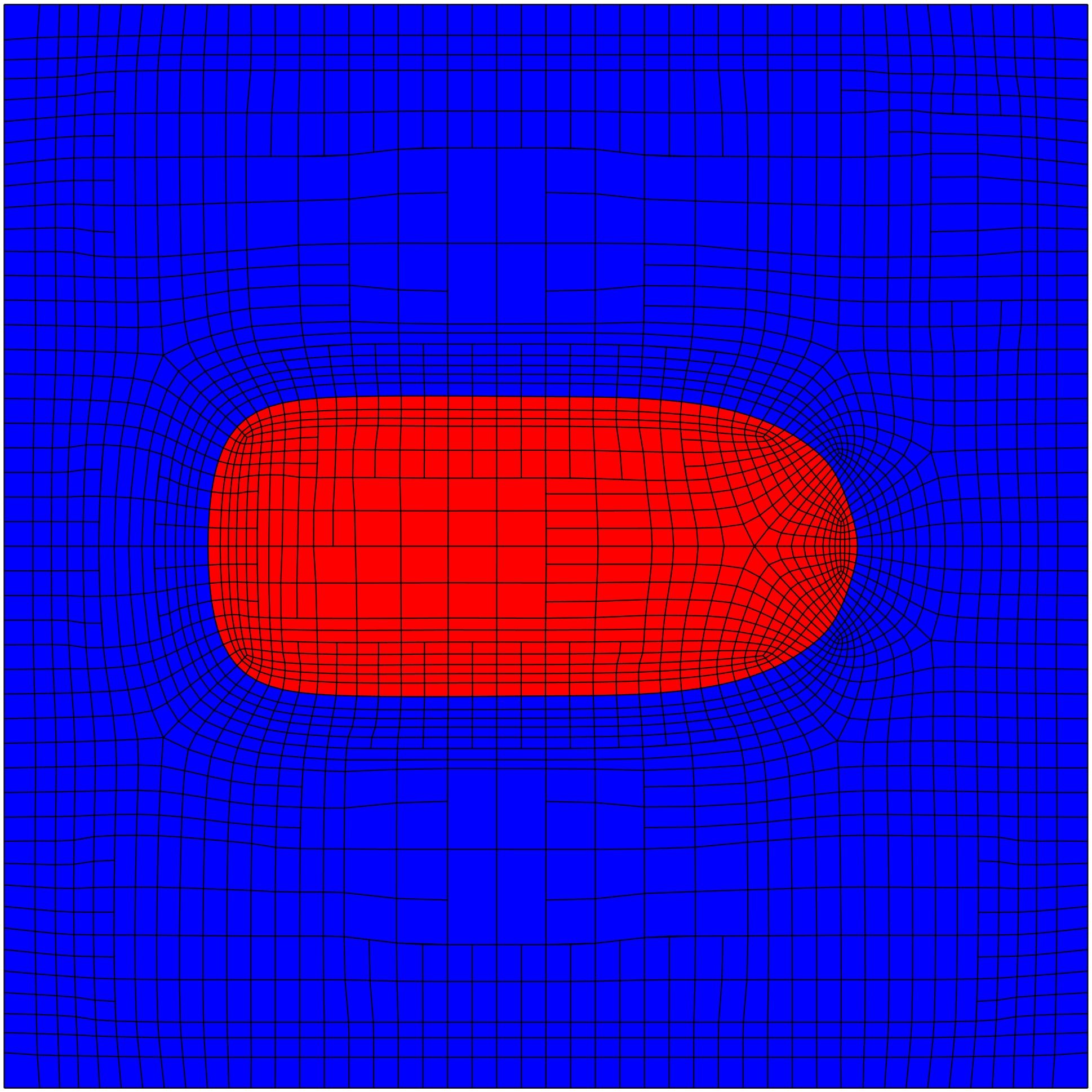}}
\caption{Compliance and domain stress minimization. Panels (a) and (b) display the optimized designs obtained via the discretized domain SA and C\'ea's method, respectively.}
\label{fig:CompStress2}
\end{figure*}

A different outcome is reached when minimizing the stresses throughout the domain of the inclusion, using either the discretized domain SA with $f^{\RNum{1}}_{\sigma_2}$ or C\'ea's method with $f^{\RNum{2}}_{\sigma_2}$. 
The normalization value for each run is given in \prettyref{tab:stresses2}.
The run with the discretized domain SA yields a design that resembles the designs obtained for interfacial stresses -- presumably because the latter dominate the stress field in the inclusion -- with $f_c = 44.08$, $f^{\RNum{1}}_{\sigma_2} = 26.42$, and a true maximal von Mises stress of $12.38$. 
Conversely, C\'ea's method converged to a different design, with $f_c = 41.73$, $f^{\RNum{2}}_{\sigma_2} = 21.40$ and a true maximal von Mises stress of $17.72$. 
The objective value of the optimized design, calculated using \prettyref{eq:formulation2}, also shown in \prettyref{fig:CompStress2}, indicates that C\'ea's approach converged to an inferior design in terms of stresses, that prioritized the compliance part over the stress part.
As inferred from \prettyref{tab:stresses2}, with  C\'ea's method the optimized stress is larger than the normalization value, exhibiting a ratio of 1.14, while the discretized SA lead to a ratio of 0.84.
This result is likely due to the lower accuracy of stress derivatives, that is aggravated if the stresses near the interface dominate the computation of the stresses in the domain.

\begin{figure*}[h!] 
\centering
{\label{fig:FDS} 
\includegraphics[width=0.45\textwidth]{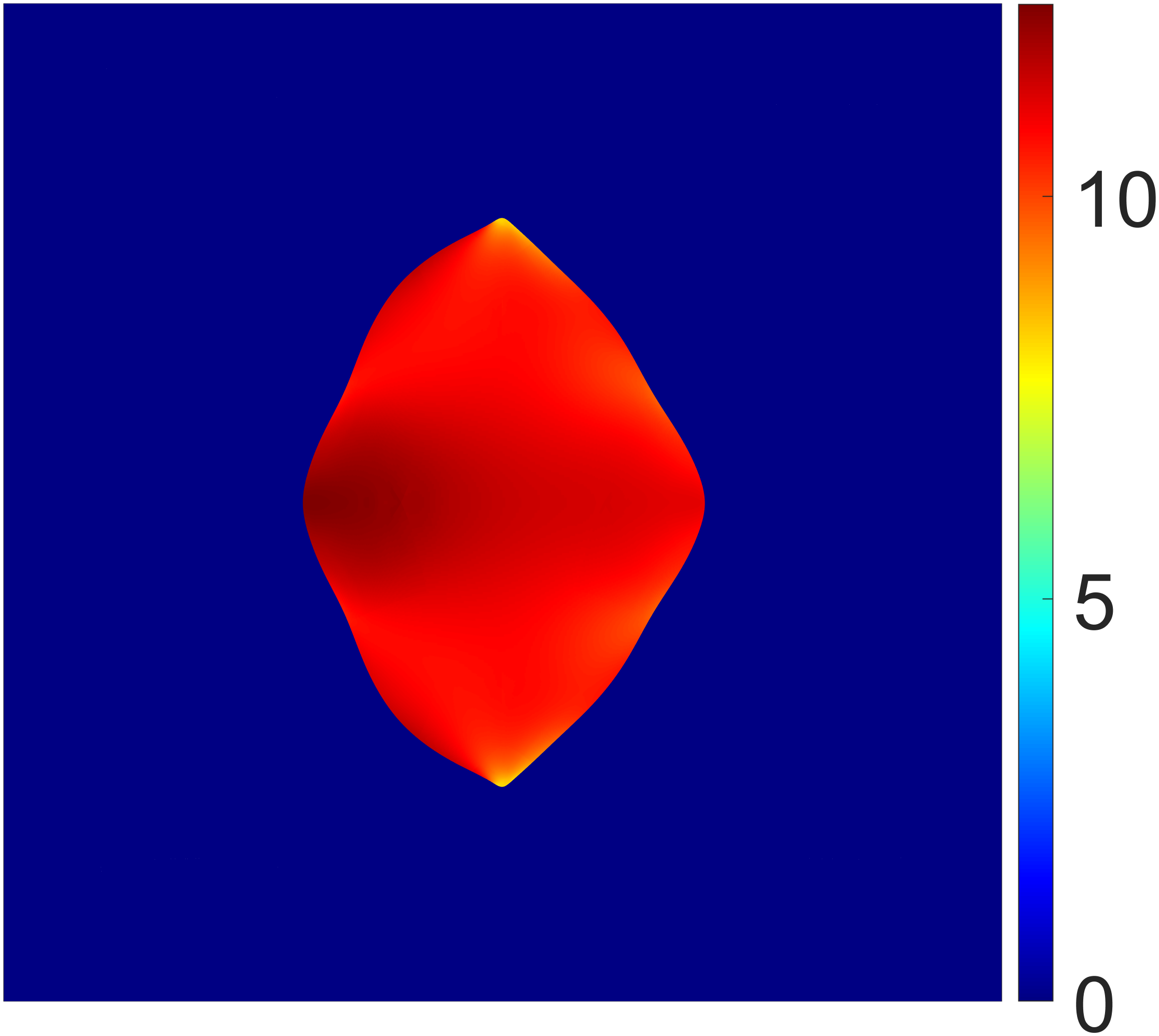}}
{\label{fig:CEAS} 
\includegraphics[width=0.45\textwidth]{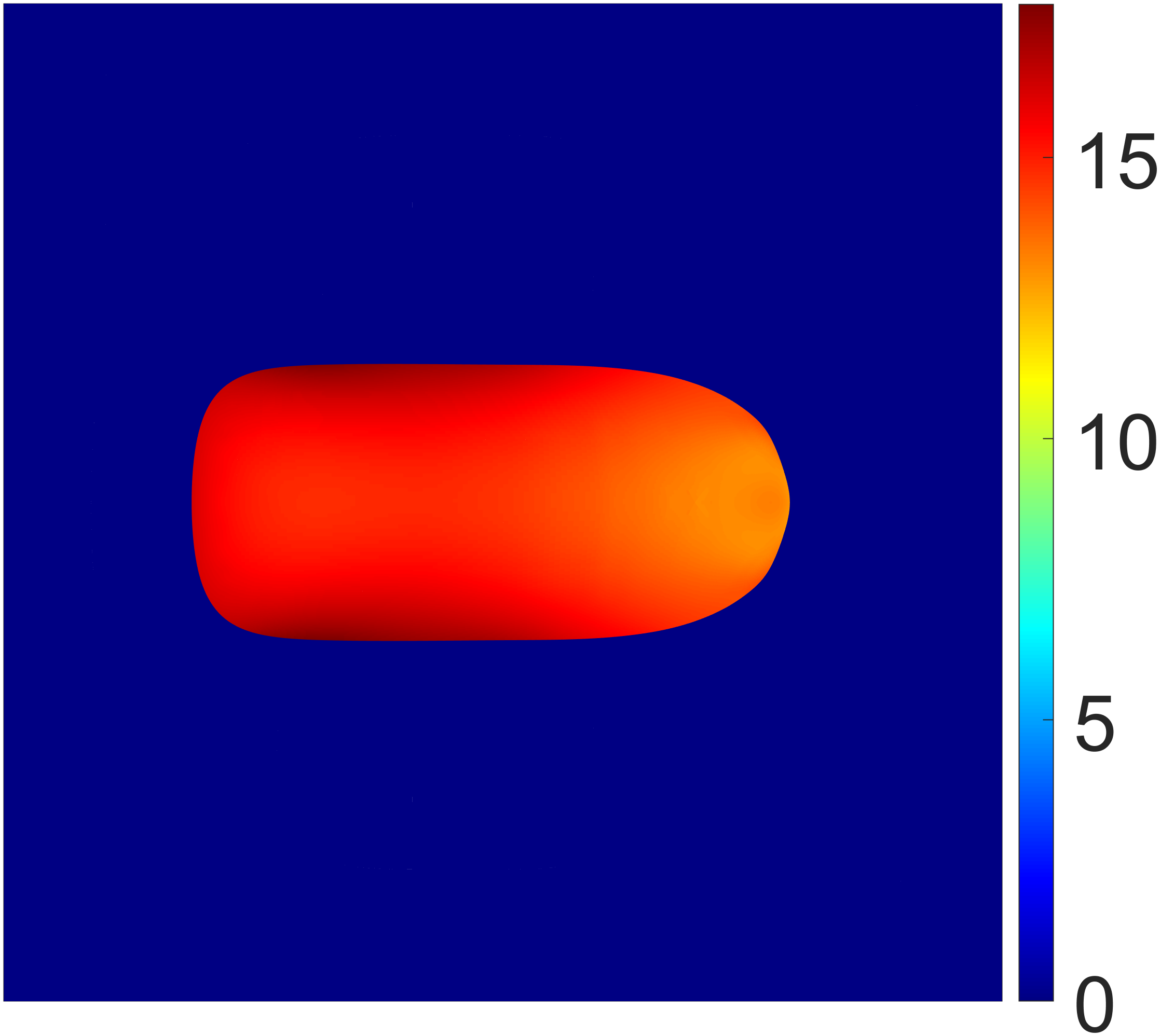}}
\caption{Compliance and domain stress minimization. Panels (a) and (b) display the von Mises stress field of the optimized designs obtained via the discretized domain SA and C\'ea's method, respectively. For clarity, the stress field within the matrix is omitted to highlight the stress distribution within the inclusion.}
\label{fig:StrField}
\end{figure*}

To clarify the proposed explanation, we present the von Mises stress field for the latter two optimized designs in \prettyref{fig:StrField}. Notably, the maximal stresses are located at the interface in both cases. This suggests that minimizing the stress field necessitates a tool capable of reducing interfacial stresses, which, as previously discussed in Section \ref{sec:APSO} and corroborated by the findings in \prettyref{tab:DiffRadius}, is limited in C\'ea's method.

\begin{table}
    \centering
    \normalsize
    \begin{tabular}{c | c c}
    & Discretized & C\'ea's\\ 
    & domain SA & method \\
    \hline
        Normalization value & $f*^{\RNum{1}}_{\sigma_2}$ = 31.30 & $f*^{\RNum{2}}_{\sigma_2}$ = 18.70 \\
        Optimized stress & $f^{\RNum{1}}_{\sigma_2}$ = 26.42 & $f^{\RNum{2}}_{\sigma_2}$ = 21.40 \\
        True maximum stress  & 12.38 & 17.72  \\
        \hline\hline
    \end{tabular}
    \caption{Results for minimizing the compliance and the domain stresses. All results shown are for the stiff inclusion only, and related to the optimized designs shown in \prettyref{fig:CompStress2}.}
    \label{tab:stresses2}
\end{table}

\section{Conclusion}\label{sec:conclusion}
We investigated and compared various formulations for computing shape derivatives in bi-material level-set optimization with a precise interface.
The overarching goal was to formulate a consistent and efficient procedure for sensitivity analysis of stress functionals, that can be used subsequently for shape and topology optimization of multi-material structures, with precise modeling of boundaries and interfaces. 
Untrimming techniques and IGA on unstructured meshes were used in this study to simulate the response according to the exact boundaries as they evolve during optimization.

For minimizing stresses in a volumetric domain, we compared a ``discretize-then-differentiate'' approach versus a ``differentiate-then-discretize'' approach, the latter realized using C\'ea's Lagrangian method.
We demonstrated several characteristics of C\'ea's method: 1) Inferior accuracy compared to the consistent discrete formulation, as expected; 2) Possibility to improve the accuracy by mesh refinement; 3) Dependence on the stiffness contrast between material phases, that affects the numerical accuracy of the transmission conditions; and 4) Inaccuracy when stresses near the interface dominate the stress field within the volumetric domain. 
When aiming to minimize the domain stresses in a stiff inclusion, C\'ea's method attained an inferior local minimum -- presumably because of the lower accuracy of stress derivatives in general, that was aggravated because stresses near the interface dominated the computation of stresses in
the entire domain.
C\'ea's method is much more efficient because it relies on boundary integrals rather then volumetric ones, hence it could be preferred for compliance and for stress functionals that are not dominated by stresses near the material interface. 

For minimizing stresses precisely at the interface between two materials, the complete ``discretize-then-differentiate'' approach is inefficient, because it considers the movement of all points in the design domain. 
At the same time, C\'ea's method is not a viable option, for two reasons: 1) So far, there is no rigorous mathematical derivation of the method for stress functionals on the interface; and 2) Based on the observations above, such derivations will most likely suffer from numerical inaccuracy.
A suitable compromise was presented in the form of a reduced discretized approach, where only interface control points and a selective subset of adjacent control points are used as intermediate design variables -- resembling a parameterized shape SA.
When aiming to minimize interfacial stresses between a stiff inclusion and a soft matrix, the parameterized shape SA yielded practically the same result as the complete discretized approach. 
Consequently, the parameterized shape SA will be preferred in future studies for controlling interfacial stresses, as it provides a superior trade-off between numerical consistency and computational efficiency.  

\section*{Conflict of interest}
The authors declare that they have no conflicts of interest or personal relationships that could have appeared to influence the work reported in this paper.

\section*{Replication of results}
The results and methodology discussed have been achieved using the authors' in-house code developed in C. The article contains all essential data configurations required for replicating the results. Readers can obtain the result files by contacting the corresponding author.

\section*{Acknowledgment}
This research was funded by the Israel Science Foundation , grant number 2594/21. The corresponding author wish to thank Neubauer doctoral fellowship fund for minority students for the generous financial support.
\begin{appendices}

\section{The derivation of the Lagrangian}\label{Appendix_Lag}

In this appendix, we present the full formulation of \prettyref{eq:UpdLag}. To this end, we establish the connections between the Lagrange multipliers by enforcing $\frac{\partial L}{\partial \mathbf{u}_i} = 0$. For clarity, the derivation is divided into individual domains, each presented separately.
First, the derivative of $\hat{J}$ with respect to $u_i$ can be written as:
\begin{equation}
\label{eq:StressDer}
    \begin{aligned}
    & \int_{\Omega_i} \frac{\partial \hat{J}}{\partial \mathbf{u}}\boldsymbol{\boldsymbol{\delta}}_u d\Omega = \int_{\Omega_i} \frac{\partial \hat{J}}{\partial\boldsymbol{\sigma}_{ij}}\frac{\partial\boldsymbol{\sigma}_{ij}}{\partial \mathbf{u}}\boldsymbol{\delta}_{u_i} d\Omega = \int_{\Omega_i} \boldsymbol{\chi}_i \mathbf{C} \boldsymbol{\delta}_{\epsilon_i} d\Omega\quad, 
    \end{aligned}
\end{equation}
where $\boldsymbol{\chi}_i =  \frac{\partial \hat{J}}{\partial\boldsymbol{\sigma}_{ij}}$ and $\frac{\boldsymbol{\sigma}_{ij}}{\partial u}\boldsymbol{\delta}_{u_i} = \mathbf{C} \boldsymbol{\delta}_{\epsilon_i}$. Exploiting Lemma \ref{lemma:Green}, \prettyref{eq:StressDer} can be rewritten as
\begin{equation}
\label{eq:StressDer2}
    \begin{aligned}
            & \int_{\Omega_i} \boldsymbol{\chi}_i \mathbf{C} \boldsymbol{\delta}_{\epsilon_i} d\Omega = \int_\Gamma \boldsymbol{\chi}_i \mathbf{C}\cdot \mathbf{n}_i \boldsymbol{\delta}_{u_i} d\Gamma - \int_\Omega div(\boldsymbol{\chi}_i \mathbf{C}) \boldsymbol{\delta}_{u_i} d\Omega \\
            & = \int_\Gamma \boldsymbol{\sigma}(\mathbf{W}_i)\cdot \mathbf{n}_i \boldsymbol{\delta}_{u_i}d\Gamma - \int_\Omega div\left(\boldsymbol{\sigma}(\mathbf{W}_i)\right)\boldsymbol{\delta}_{u_i} d\Omega\quad,
    \end{aligned}
\end{equation}
where $\mathbf{W}_i = \boldsymbol{\chi}_i\mathbf{C}$ is a non-dimensional vector, considered as an \textit{effective} stress. Henceforth, the integrand of the surface integral in \prettyref{eq:StressDer2} will be denoted by $\partial \hat{J}_{i,\text{surf}}$, and the integrand of the domain integral by $\partial \hat{J}_{i,vol.}$. 

The domain integral is summed to the derivative of the state equation w.r.t the displacement, which yields:

\begin{equation}
\begin{aligned}
            \frac{\partial L}{\partial \mathbf{u}_i} \bigg|_{\Omega^i} = div\left(\boldsymbol{\sigma}_i\left(\boldsymbol{\lambda}_{1i}\right)\right) - \partial \hat{J}_{i,vol.} =  0 & \:\: \text{in} \:\:\Omega_i\quad,
\end{aligned}
\end{equation}
which defines the adjoint equation.
We follow a similar approach for the Neumann boundary integration, where varying the trace of $\phi$ on $\Gamma_N^i$ yields
\begin{equation}
\label{eq:N_P}
    \begin{aligned}
            -\boldsymbol{\sigma}(\boldsymbol{\lambda}_{1i})\cdot \mathbf{n}_i + \partial \hat{J}_{i,\text{surf}} =  0 && \text{on} \quad\Gamma_N^i\quad,
    \end{aligned}
\end{equation}
and varying the normal stress $\boldsymbol{\sigma}(\phi)\mathbf{n}$ on $\Gamma_N^i$ yields
\begin{equation}
\label{eq:N_S}
\begin{aligned}
    \boldsymbol{\sigma}(\boldsymbol{\lambda}_{1i})\cdot \mathbf{n}_i + \boldsymbol{\sigma}(\boldsymbol{\lambda}_{2i})\cdot \mathbf{n}_i =  0 && \text{on} \quad\Gamma_N^i\quad,
\end{aligned}
\end{equation}
where \prettyref{eq:N_P} is a boundary condition of the adjoint problem, and \prettyref{eq:N_S} yields $\boldsymbol{\lambda}_{1i} = -\boldsymbol{\lambda}_{2i}$, on $\Gamma_N^i$.

Similarly, varying the trace of $\phi$ and the normal stress $\boldsymbol{\sigma}(\phi)\mathbf{n}$ on $\Gamma_D^i$, respectively, yields:

\begin{equation}
\begin{aligned}
        & \boldsymbol{\lambda}_{3i} - \boldsymbol{\sigma}(\boldsymbol{\lambda}_{1i}) + \partial \hat{J}_{i,\text{surf}} =  0 && \text{on} \quad\Gamma_D^i, \\
        & \boldsymbol{\sigma}(\boldsymbol{\lambda}_{1i})\cdot \mathbf{n}_i = 0 && \text{on} \quad\Gamma_D^i,
\end{aligned} 
\end{equation}
which yields
\begin{equation}
\begin{aligned}
        & \boldsymbol{\lambda}_{3i} = \boldsymbol{\sigma}(\boldsymbol{\lambda}_{1i})\cdot \mathbf{n}_i - \partial \hat{J}_{i,\text{surf}}  && \text{on} \quad\Gamma_D^i, \\
        & \boldsymbol{\lambda}_{1i} = 0 && \text{on} \quad\Gamma_D^i,
\end{aligned} 
\end{equation}
where the first equation is to be substituted into \prettyref{eq:Lagrangian} and the second equation is the Dirichlet boundary condition for the adjoint problem.
As for the material interface $\Sigma$, varying the trace of $\phi$ and the normal stress $\boldsymbol{\sigma}(\phi)\mathbf{n}$, for the derivative with respect to $\mathbf{u}_1$,
\begin{equation}
\label{eq:Sigma_1}
    \begin{aligned}
        & \boldsymbol{\lambda}_4 - \boldsymbol{\sigma}(\boldsymbol{\lambda}_{11})\cdot \mathbf{n}_1 + \partial \hat{J}_{1,\text{surf}} = 0 && \text{on} \quad\Sigma, \\
        & \boldsymbol{\sigma}(\boldsymbol{\lambda}_5)\cdot \mathbf{n}_1 + \boldsymbol{\sigma}(\boldsymbol{\lambda}_{11})\cdot \mathbf{n}_{1} = 0 && \text{on} \quad\Sigma,
    \end{aligned}
\end{equation}
and for the derivative with respect to $\mathbf{u}_2$,
\begin{equation}
\label{eq:Sigma_2}
    \begin{aligned}
        & -\boldsymbol{\lambda}_4 - \boldsymbol{\sigma}(\boldsymbol{\lambda}_{12})\cdot \mathbf{n}_2 + \partial \hat{J}_{2,\text{surf}} = 0 && \text{on} \quad\Sigma, \\
        & \boldsymbol{\sigma}(\boldsymbol{\lambda}_5)\cdot \mathbf{n}_2 + \boldsymbol{\sigma}(\boldsymbol{\lambda}_{12})\cdot \mathbf{n}_{2} = 0 && \text{on} \quad\Sigma .
    \end{aligned}
\end{equation}
Isolating $\boldsymbol{\lambda}_4$ from \prettyref{eq:Sigma_1} and \prettyref{eq:Sigma_2}, and summing both results yields
\begin{equation}
    \begin{aligned}
        \boldsymbol{\lambda}_4 = \frac{1}{2}\left(\boldsymbol{\sigma}(\boldsymbol{\lambda}_{11})\cdot \mathbf{n}_1 - \boldsymbol{\sigma}(\boldsymbol{\lambda}_{12})\cdot \mathbf{n}_2\right) + \frac{1}{2} \left(-\partial \hat{J}_{1,\text{surf}} + \partial \hat{J}_{2,\text{surf}}  \right)
    \end{aligned}
\end{equation}
and equating both results also gives
\begin{equation}
    \boldsymbol{\sigma}(\boldsymbol{\lambda}_{11})\cdot \mathbf{n}_1 + \boldsymbol{\sigma}(\boldsymbol{\lambda}_{12})\cdot \mathbf{n}_2 = \partial \hat{J}_{1,\text{surf}} + \partial \hat{J}_{2,\text{surf}} .
\end{equation}
Also, isolating $\boldsymbol{\lambda}_5$ gives $\boldsymbol{\lambda}_5 = -\boldsymbol{\lambda}_{11}= -\boldsymbol{\lambda}_{12}$ on $\Sigma$, which also implies that $\boldsymbol{\lambda}_{11}= \boldsymbol{\lambda}_{12}$, and $\boldsymbol{\lambda}_5 = -\frac{1}{2}\left(\boldsymbol{\lambda}_{11} + \boldsymbol{\lambda}_{12} \right)$ on $\Sigma$.

To sum-up, substituting $\boldsymbol{\lambda}_{2i},\boldsymbol{\lambda}_{3i},\boldsymbol{\lambda}_4$ and $\boldsymbol{\lambda}_5$ into \prettyref{eq:Lagrangian} yields
\begin{equation}
        \begin{aligned}
        L = & \hat{J} +  \sum_{i=1}^2 \int_{\Omega^i}\left(-\boldsymbol{\sigma}\left(\mathbf{u}_i\right)\boldsymbol{\epsilon}\left(\boldsymbol{\lambda}_{1i}\right) + \mathbf{f}_i \right) d\Omega \\
        & +\sum_{i=1}^2 \int_{\Gamma^i_{N}}\boldsymbol{\lambda}_{1i} \mathbf{g}_i   \;d\Gamma \\ & + \sum_{i=1}^2  \int_{\Gamma^i_{D}}\left(\boldsymbol{\sigma}(\boldsymbol{\lambda}_{1i})\cdot \mathbf{n}_i - \partial \hat{J}_{i,\text{surf}} \right) \mathbf{u}_i + \boldsymbol{\lambda}_{1i}\boldsymbol{\sigma}\left(\mathbf{u}_i\right)\cdot \mathbf{n}_i \;d\Gamma \\
        & +\frac{1}{2} \int_{\Sigma} \left(\mathbf{u}_1 - \mathbf{u}_2\right)\left(\boldsymbol{\sigma}(\boldsymbol{\lambda}_{11})\cdot \mathbf{n}_1 - \boldsymbol{\sigma}(\boldsymbol{\lambda}_{12})\cdot \mathbf{n}_2\right) d\Gamma \\
         & + \frac{1}{2} \int_{\Sigma} \left(\boldsymbol{\lambda}_{11} - \boldsymbol{\lambda}_{12} \right)
        \cdot\left(\boldsymbol{\sigma}_1\left(\mathbf{u}_1\right)\cdot\mathbf{n}_1 - \boldsymbol{\sigma}_2\left(\mathbf{u}_2\right)\cdot\mathbf{n}_2 \right) d\Gamma \\
        & +\frac{1}{2} \int_{\Sigma} \left(\mathbf{u}_1 - \mathbf{u}_2\right)\left(-\partial \hat{J}_{1,\text{surf}} + \partial \hat{J}_{2,\text{surf}}  \right) d\Gamma,
    \end{aligned}
\end{equation}
\textbf{Remark}: Minimizing the stress of a specific phase, e.g.~phase 1, implies that $\frac{\partial \hat{J}}{\partial \boldsymbol{\sigma}_{2j}} = 0$. 
As a result, when examining a specific material, one of the $\partial\hat{J}_{i,\text{surf}}$ invariably falls, as elaborated in Sections \ref{sec:insights} and \ref{sec:examples}.



\end{appendices}


\bibliography{sn-bibliography}

\input{sn-article.bbl}

\end{document}

%% file: Figures/OptDes/ProblemSet1.tex
\begin{tikzpicture}

\point{a1}{0}{0.05}; \point{a2}{0}{1.45}; \point{a3}{0}{2.9}; \point{a4}{0}{4.3}; \point{a5}{0}{5.7};
\point{b1}{6.5}{0.05}; \point{b2}{6.5}{5.7}; 
\draw[black, fill=gray!20,  thick] (0,0.05) rectangle (5.7,5.7);


\support{2}{a1}[270];  \support{1}{a3}[270];  \support{2}{a5}[270];
\support{2}{a2}[270];  \support{2}{a4}[270];
\lineload{1}{b1}{b2}[0.5][0.5][.2];
\node[draw] at (2.85,2.8) {\large Design domain};

\draw[stealth-stealth] (0.51,0.05) -- (0.51,5.7)
  node[sloped, pos=0.5,above=0.0005mm]{\scriptsize $L_y = 10$};
\draw[stealth-stealth] (0.51,0.05) -- (0.51,5.7)
  node[sloped, pos=0.5,below=0.01mm]{\scriptsize 10 patches, 13 control points};
\draw[stealth-stealth] (0,0.6) -- (5.7,0.6)
  node[sloped, pos=0.5,above=0.0005mm]{\scriptsize $L_x = 10$};
\draw[stealth-stealth] (0,0.6) -- (5.7,0.6)
  node[sloped, pos=0.5,below=0.0005mm]{\scriptsize 10 patches, 13 control points};

\end{tikzpicture}

%% file: Figures/SA/VOL_FD_Comp.tex
%
%
\begin{tikzpicture}

\begin{axis}[%
width=3.66in,
height=3.946in,
at={(0.779in,0.679in)},
scale only axis,
separate axis lines,
every outer x axis line/.append style={black},
every x tick label/.append style={font=\color{black}},
every x tick/.append style={black},
xmin=0,
xmax=0.4,
xtick={0.1, 0.2, 0.3, 0.4},
xticklabel style={font=\large},
xlabel={\large{Finite Difference}},
every outer y axis line/.append style={black},
every y tick label/.append style={font=\color{black}},
every y tick/.append style={black},
ymin=-0.0156356739065576,
ymax=0.415635673906558,
ytick={0.1, 0.2, 0.3, 0.4},
ylabel={\large{Analytical sensitivity analysis}},
yticklabel style={font=\large},
axis background/.style={fill=white},
xmajorgrids,
ymajorgrids
]
\addplot [color=black, line width=2.0pt, forget plot]
  table[row sep=crcr]{%
0	0\\
0.0444444444444444	0.0444444444444444\\
0.0888888888888889	0.0888888888888889\\
0.133333333333333	0.133333333333333\\
0.177777777777778	0.177777777777778\\
0.222222222222222	0.222222222222222\\
0.266666666666667	0.266666666666667\\
0.311111111111111	0.311111111111111\\
0.355555555555556	0.355555555555556\\
0.4	0.4\\
};
\addplot[only marks, mark=o, mark options={}, mark size=1.7678pt, draw=blue, forget plot] table[row sep=crcr]{%
x	y\\
0.001675232397247	0.00167518023967\\
0.00808000777397	0.00807970652451\\
0.007191438555765	0.00719112250299\\
0.003024148043096	0.003024042786308\\
0.008667868200973	0.008668026632848\\
0.009996405481161	0.009996248224208\\
0.002087737982492	0.002087797750932\\
0.003210956833755	0.003210819019365\\
0.028617868963465	0.028617995660066\\
0.075105937469289	0.075105993060587\\
0.061662959183195	0.061662776772923\\
0.06197957702625	0.061979761244822\\
0.06976890176702	0.069768876070146\\
0.083565396380436	0.083565038261784\\
0.029100377219038	0.029100343354916\\
0.002880852889575	0.002880553553565\\
0.016557024196118	0.016556881115698\\
0.112020188680617	0.11202020429518\\
0.214228158768037	0.214227999800995\\
0.148430544300027	0.148430582638231\\
0.209857766719779	0.209857635162181\\
0.1570618692881	0.157061874978345\\
0.212006725064384	0.212006804130103\\
0.103116938987569	0.103116471998205\\
0.014484839994111	0.01448473987586\\
0.018065378526444	0.018065239172298\\
0.124454018646247	0.124454236087023\\
0.231457107702226	0.231457474947022\\
0.088671924913797	0.0886718584561\\
0.081561644549311	0.081561469360441\\
0.081162127685275	0.081161681462744\\
0.193360712330559	0.193360068572309\\
0.102346845665124	0.102346496423215\\
0.014731007524915	0.014730656199948\\
0.007427352954892	0.007427198311396\\
0.092684700803147	0.092684823950002\\
0.264686065065689	0.264686380072891\\
0.088046462565217	0.08804587897611\\
0.024522115893433	0.024522255390098\\
0.072165228459653	0.072164861097748\\
0.211454491250151	0.211454848690706\\
0.073940412903539	0.073940467507533\\
0.005925215873503	0.005925051186633\\
0.018065058782213	0.018065178445322\\
0.124454075489666	0.124453869117853\\
0.231457036647953	0.231456908041054\\
0.088671676223839	0.088671660783792\\
0.081561722709012	0.081561320269293\\
0.081162212950403	0.081161639793167\\
0.193360236266926	0.193360094643813\\
0.102346852770552	0.102346512894265\\
0.014731135422608	0.014730658610403\\
0.016556903403853	0.016556814683815\\
0.112019797882112	0.112019749983645\\
0.214226929529104	0.214227134590594\\
0.148430842727976	0.148430133165399\\
0.209857354604992	0.209857242329103\\
0.157061812444681	0.157061763770453\\
0.212007506661394	0.21200694219328\\
0.10311661213791	0.103116554330931\\
0.014484555777017	0.014484751566023\\
0.003210679722088	0.003210805352662\\
0.028617790803764	0.028617869729105\\
0.075105688779331	0.075105665576176\\
0.061663051553751	0.06166257055762\\
0.061980209409285	0.0619796440339\\
0.069769349408944	0.069768836123634\\
0.083565261377316	0.083565093406244\\
0.029100867493526	0.029100370434366\\
0.002880433669361	0.002880556510245\\
0.001675324767803	0.001675172405965\\
0.008080029090252	0.008079670482893\\
0.007191268025508	0.007191095842087\\
0.003024105410532	0.003024036787689\\
0.00866793214982	0.008668023132211\\
0.009996384164879	0.009996253598953\\
0.002087979567023	0.002087799417573\\
};
\node[right, align=left]
at (axis cs:0.25,0.246) {1};
\end{axis}

\begin{axis}[%
width=5in,
height=5in,
at={(0in,0in)},
scale only axis,
xmin=0,
xmax=1,
ymin=0,
ymax=1,
axis line style={draw=none},
ticks=none,
axis x line*=bottom,
axis y line*=left
]
\addplot [color=green, dashed, forget plot]
  table[row sep=crcr]{%
0.7	0.71\\
0.7	0.6\\
};
\addplot [color=green, dashed, forget plot]
  table[row sep=crcr]{%
0.7	0.6\\
0.595	0.6\\
};
\end{axis}
\end{tikzpicture}%

%% file: Figures/SA/VOL_FD_STressNEW.tex
%
%
\begin{tikzpicture}

\begin{axis}[%
width=3.639in,
height=3.946in,
at={(0.857in,0.679in)},
scale only axis,
separate axis lines,
every outer x axis line/.append style={black},
every x tick label/.append style={font=\color{black}},
every x tick/.append style={black},
xmin=-101.429444112,
xmax=135.725842847023,
xtick={-100,  -50,    0,   50,  100},
xticklabel style={font=\large},
xlabel={\large{Finite Difference}},
every outer y axis line/.append style={black},
every y tick label/.append style={font=\color{black}},
every y tick/.append style={black},
ymin=-111.431604304865,
ymax=145.728003039888,
ytick={-100,  -50,    0,   50,  100},
ylabel={\large{Analytical sensitivity analysis}},
yticklabel style={font=\large},
axis background/.style={fill=white},
xmajorgrids,
ymajorgrids
]
\addplot[only marks, mark=o, mark options={}, mark size=1.7678pt, draw=blue, forget plot] table[row sep=crcr]{%
x	y\\
-0.04864705260843	-0.048582623192312\\
-2.28866338147782	-2.28886467581224\\
-8.42661029309966	-8.42671935536657\\
-4.13558518630453	-4.13555443064761\\
-8.14282611827366	-8.14279342909112\\
-1.95596658159047	-1.95587000228455\\
0.021133018890396	0.021287561330023\\
0.191037543118	0.191037895735326\\
0.834381353342906	0.834380882037771\\
-14.531677152263	-14.5315544975448\\
-56.6036105738021	-56.603722172317\\
-16.07649755897	-16.0761801954781\\
-53.3383827132639	-53.338367660923\\
-9.71826739259996	-9.71835075529085\\
3.05433422909118	3.05439852822235\\
0.500316673424095	0.500195114220064\\
1.63533695740625	1.63534364182984\\
10.8742387965322	10.8744646748103\\
-9.62840567808598	-9.62832154122322\\
-92.2085855563637	-92.2083347865198\\
15.6581081682816	15.6579157138981\\
-83.1616889627185	-83.1615144949947\\
5.86504756938666	5.86497685185383\\
19.9838905245997	19.9837294984099\\
3.13016789732501	3.13021148242264\\
3.67584652849473	3.67580613128957\\
33.1495248246938	33.149789216366\\
70.0522905390244	70.0525268620777\\
10.0779143394902	10.0778194783485\\
71.0939493728802	71.0939026332176\\
16.0735253302846	16.0733298114425\\
82.846872828668	82.846416160747\\
41.1725450248923	41.172347858175\\
5.01818067277782	5.01791921638855\\
1.74595334101468	1.74578834524805\\
33.1896262650844	33.1895664100627\\
103.434522316093	103.434779755329\\
22.6268239202909	22.6267934130733\\
31.5103934553917	31.5104326827342\\
34.6695451298729	34.6695092502054\\
123.38712986093	123.387352468373\\
39.6361538150813	39.6361502398633\\
2.26257543545216	2.26241660296187\\
3.69256667909212	3.69269308998555\\
33.2501113007311	33.2502063564837\\
70.201978815021	70.202071219869\\
10.1383848232217	10.1382941706898\\
71.1959073669277	71.195606506734\\
16.1811221914832	16.1811458200461\\
83.0726421554573	83.0723953186421\\
41.3056004617829	41.3054540457803\\
5.039204552304	5.03894792201589\\
1.65168603416532	1.65182806454206\\
10.9754982986487	10.9754376530041\\
-9.46680302149616	-9.46671287944028\\
-92.0971287996508	-92.0972762996071\\
15.9279406943824	15.9276073233395\\
-82.8284682938829	-82.8282921574144\\
6.39928111922927	6.399122035252\\
20.2229784918018	20.2226665463809\\
3.16110526910052	3.16100802675744\\
0.194209860637784	0.19430794454552\\
0.856522092362866	0.856508387018669\\
-14.4900805025827	-14.4898655174075\\
-56.5720911254175	-56.5723241806276\\
-15.9934133989736	-15.9936171053091\\
-53.1646255694795	-53.1646273121718\\
-9.46921136346646	-9.46939273015671\\
3.13591590384021	3.13552596546629\\
0.5070396582596	0.507005612334186\\
-0.048174115363508	-0.048074805298449\\
-2.28654607781209	-2.28657102558512\\
-8.42522058519535	-8.425162022866\\
-4.13080488215201	-4.13088686299964\\
-8.11819336377084	-8.11813214534914\\
-1.92318111658096	-1.92330270226108\\
0.028416252462193	0.028328089086815\\
};
\addplot [color=black, line width=2.0pt, forget plot]
  table[row sep=crcr]{%
-101.429444112	-101.429444112\\
-75.0788566721086	-75.0788566721086\\
-48.7282692322171	-48.7282692322171\\
-22.3776817923257	-22.3776817923257\\
3.97290564756581	3.97290564756581\\
30.3234930874573	30.3234930874573\\
56.6740805273487	56.6740805273487\\
83.0246679672402	83.0246679672402\\
109.375255407132	109.375255407132\\
135.725842847023	135.725842847023\\
};
\node[right, align=left]
at (axis cs:48,46) {1};
\end{axis}

\begin{axis}[%
width=5in,
height=5in,
at={(0in,0in)},
scale only axis,
xmin=0,
xmax=1,
ymin=0,
ymax=1,
axis line style={draw=none},
ticks=none,
axis x line*=bottom,
axis y line*=left
]
\addplot [color=green, dashed, forget plot]
  table[row sep=crcr]{%
0.665	0.6\\
0.665	0.665\\
};
\addplot [color=green, dashed, forget plot]
  table[row sep=crcr]{%
0.595	0.6\\
0.665	0.6\\
};
\end{axis}
\end{tikzpicture}%

%% file: Figures/SA/Compliance_Comp.tex
%
%
\begin{tikzpicture}

\begin{axis}[%
width=3.869in,
height=4.075in,
at={(0.656in,0.55in)},
scale only axis,
separate axis lines,
every outer x axis line/.append style={black},
every x tick label/.append style={font=\color{black}},
every x tick/.append style={black},
xmin=0,
xmax=80,
xtick={\empty},
xlabel style={font=\color{white!15!black}},
xlabel={\large{Design variable}},
every outer y axis line/.append style={black},
every y tick label/.append style={font=\color{black}},
every y tick/.append style={black},
ymode=log,
ymin=1e-08,
ymax=0.1,
yminorticks=true,
ylabel style={font=\color{white!15!black}},
ylabel={\large{Relative error}},
yticklabel style={font=\large},
axis background/.style={fill=white},
xmajorgrids,
ymajorgrids,
yminorgrids,
legend style={font=\fontsize{10}{15}\selectfont,legend cell align=left, align=left, draw=white!15!black},
legend pos=south east,
]
\addplot[only marks, mark=*, mark options={}, mark size=1.5000pt, color=blue, fill=blue] 
table[row sep=crcr]{%
x	y\\
1	0.00126924697760999\\
2	0.00105293479783602\\
3	0.000282286935396527\\
4	0.000273671418596598\\
5	0.000173417334937093\\
6	0.000742372722173464\\
7	0.00094211718017051\\
8	0.00154052256075184\\
9	0.000620524921009001\\
10	1.12075251886807e-05\\
11	0.0024493086529516\\
12	0.00852986465365983\\
13	0.00238687165521252\\
14	2.09755312237528e-05\\
15	0.000435885350025732\\
16	0.00135332908254649\\
17	0.00166826347867963\\
18	0.0013392380437845\\
19	0.000971630517836812\\
20	0.00457410339888427\\
21	0.0105716184348631\\
22	0.00439964391719073\\
23	0.000803301954781223\\
24	0.00115270808999985\\
25	0.00154536378890625\\
26	0.00191822137035624\\
27	0.00201952917823741\\
28	0.00211985723639671\\
29	0.00438888204137156\\
30	0.0123139753850694\\
31	0.00481532923680239\\
32	0.00229219080197792\\
33	0.00204245692359655\\
34	0.00180675807374246\\
35	0.00197910747022174\\
36	0.00241394371362531\\
37	0.00256179560600113\\
38	0.00281025488405878\\
39	0.0105027755483765\\
40	0.00402613855983902\\
41	0.00344894390662169\\
42	0.00299567181450509\\
43	0.00189353610290769\\
44	0.00193253325941972\\
45	0.00201605491351576\\
46	0.00211763977573303\\
47	0.00438942043798155\\
48	0.0123111638932811\\
49	0.00481364279337395\\
50	0.00229458437836475\\
51	0.00204234102802964\\
52	0.00179802287455386\\
53	0.00167131343186801\\
54	0.00133838135662217\\
55	0.000973004377018286\\
56	0.00456894015140011\\
57	0.0105717213942057\\
58	0.00439904967865013\\
59	0.000799469137659787\\
60	0.0011558458496542\\
61	0.00156501572281336\\
62	0.00162239321884532\\
63	0.000618428120792366\\
64	9.70668164880952e-06\\
65	0.00244432910563335\\
66	0.00851767725971749\\
67	0.00237961927981968\\
68	1.9575470429213e-05\\
69	0.000418955418106115\\
70	0.00149907882237675\\
71	0.00132963689417732\\
72	0.00106056751322068\\
73	0.00026243531214607\\
74	0.000285796297969692\\
75	0.000165402364395524\\
76	0.000740532985317609\\
77	0.00105790629653971\\
};
\addlegendentry{C\'ea's Method}

\addplot[only marks, mark=*, mark options={}, mark size=1.5000pt, color=red, fill=red]table[row sep=crcr]{%
x	y\\
1	3.11345321913092e-05\\
2	3.72833131385645e-05\\
3	4.3948477422016e-05\\
4	3.48054349522533e-05\\
5	1.82780669163408e-05\\
6	1.57313499633244e-05\\
7	2.86283242921099e-05\\
8	4.29200382114856e-05\\
9	4.42718502766408e-06\\
10	7.40171814406715e-07\\
11	2.9581822607046e-06\\
12	2.97224635650726e-06\\
13	3.68314153625014e-07\\
14	4.28548977816738e-06\\
15	1.16370044775477e-06\\
16	0.000103905343824901\\
17	8.64167487507362e-06\\
18	1.39390615098992e-07\\
19	7.42045503804479e-07\\
20	2.58290530302504e-07\\
21	6.26889345285849e-07\\
22	3.62293217520302e-08\\
23	3.7293967432272e-07\\
24	4.52873571096136e-06\\
25	6.9119335138932e-06\\
26	7.71387910839668e-06\\
27	1.74715753144058e-06\\
28	1.58666458621083e-06\\
29	7.49478451752564e-07\\
30	2.14793204546147e-06\\
31	5.49791563793164e-06\\
32	3.32931256937376e-06\\
33	3.4123368114943e-06\\
34	2.38493508610605e-05\\
35	2.0820808831787e-05\\
36	1.3286643202604e-06\\
37	1.19011630614959e-06\\
38	6.62819481890563e-06\\
39	5.68860638318104e-06\\
40	5.09056664593796e-06\\
41	1.69038998828789e-06\\
42	7.38486462934364e-07\\
43	2.77942396558058e-05\\
44	6.62400883610197e-06\\
45	1.65821659264816e-06\\
46	5.55640480211556e-07\\
47	1.74126030413231e-07\\
48	4.93417384578253e-06\\
49	7.06187294764811e-06\\
50	7.32431422929313e-07\\
51	3.32082792780378e-06\\
52	3.23676479321926e-05\\
53	5.35849221537296e-06\\
54	4.27589300318786e-07\\
55	9.57216212085935e-07\\
56	4.78042544231039e-06\\
57	5.35010503748976e-07\\
58	3.09904917380172e-07\\
59	2.66249116783396e-06\\
60	5.6059812087266e-07\\
61	1.35170873731452e-05\\
62	3.91289648530012e-05\\
63	2.75791173190304e-06\\
64	3.08940046827227e-07\\
65	7.80039454553984e-06\\
66	9.1218695514305e-06\\
67	7.35688829464285e-06\\
68	2.01005859644867e-06\\
69	1.70805616056798e-05\\
70	4.2646663003088e-05\\
71	9.09446579720165e-05\\
72	4.43819390986412e-05\\
73	2.39434019688231e-05\\
74	2.26919480918369e-05\\
75	1.04964355312242e-05\\
76	1.30613153562526e-05\\
77	8.62793165436817e-05\\
};
\addlegendentry{Discretized domain SA}

\end{axis}
\end{tikzpicture}%

%% file: Figures/SA/Comp_Stress_New.tex
%
%
\begin{tikzpicture}

\begin{axis}[%
width=3.561in,
height=3.776in,
at={(0.844in,0.51in)},
scale only axis,
separate axis lines,
every outer x axis line/.append style={black},
every x tick label/.append style={font=\color{black}},
every x tick/.append style={black},
xmin=0,
xmax=80,
xtick={\empty},
xlabel={\large{Design variable}},
every outer y axis line/.append style={black},
every y tick label/.append style={font=\color{black}},
every y tick/.append style={black},
ymode=log,
ymin=1e-08,
ymax=0.0124950954632237,
yminorticks=true,
ylabel={\large{Relative error}},
yticklabel style={font=\large},
axis background/.style={fill=white},
xmajorgrids,
ymajorgrids,
yminorgrids,
legend style={font=\fontsize{10}{15}\selectfont,legend cell align=left, align=left, draw=white!15!black},
legend pos=south east,
]
\addplot[only marks, mark=*, mark options={}, mark size=1.5000pt, color=blue, fill=blue] table[row sep=crcr]{%
x	y\\
1	0.00200964994306107\\
2	0.0016823085750994\\
3	0.00121688732285775\\
4	0.00131290000201218\\
5	0.00145702887685969\\
6	0.00149740341947668\\
7	0.001639532965449\\
8	0.00193985847969584\\
9	0.00187440543404829\\
10	0.00188602891504677\\
11	0.00390790224891498\\
12	0.00936311464780807\\
13	0.00339222881339333\\
14	0.00152967980347768\\
15	0.00145863743349063\\
16	0.00140634466456687\\
17	0.0018297279485979\\
18	0.0018709087866667\\
19	0.00205149887580893\\
20	0.00594943406929331\\
21	0.0114025791799559\\
22	0.00521861426572674\\
23	0.00170358465468359\\
24	0.00152286763482817\\
25	0.00145495439277881\\
26	0.00169390770899094\\
27	0.00201893257801267\\
28	0.00239461884633414\\
29	0.00486937760976738\\
30	0.0124950954632237\\
31	0.00519694942182324\\
32	0.00273663817763514\\
33	0.00209438139512886\\
34	0.0015448075454821\\
35	0.00169574386422654\\
36	0.00244755859757604\\
37	0.00270396802788227\\
38	0.00261952192573127\\
39	0.00930697594850347\\
40	0.00418051326215336\\
41	0.00401553116483827\\
42	0.00334969072198592\\
43	0.00153012634094981\\
44	0.0017188392092794\\
45	0.00201642007137362\\
46	0.00239158812857423\\
47	0.004865802717606\\
48	0.0124920603271029\\
49	0.00519477291923496\\
50	0.00273868925173071\\
51	0.00209693674595742\\
52	0.00150872936696617\\
53	0.00182471561296556\\
54	0.0018724213159442\\
55	0.00205283351029774\\
56	0.00594646796794697\\
57	0.0114030647938703\\
58	0.00522089201589204\\
59	0.00170228296285967\\
60	0.00152748200255267\\
61	0.00149444293797024\\
62	0.00203958328205666\\
63	0.00185844710928\\
64	0.00188339262027167\\
65	0.00390477906622881\\
66	0.00935174083082248\\
67	0.0033866460406594\\
68	0.00153074098901801\\
69	0.00144093152027912\\
70	0.00167480869957551\\
71	0.00203316892938059\\
72	0.00172019395410664\\
73	0.00127009319232447\\
74	0.00133105412779538\\
75	0.00143187004696066\\
76	0.00150119601899108\\
77	0.0015396960623037\\
};
\addlegendentry{C\'ea's method}

\addplot[only marks, mark=*, mark options={}, mark size=1.5000pt, color=red, fill=red] table[row sep=crcr]{%
x	y\\
1	-0.000182530480178254\\
2	4.39982137741345e-05\\
3	8.97527615841664e-06\\
4	1.10725789459321e-07\\
5	1.26680620525366e-06\\
6	-3.6592314933981e-05\\
7	0.000623354923792658\\
8	-9.4029608568015e-05\\
9	0.00381056926480168\\
10	-7.92674600843096e-06\\
11	2.56896226791024e-07\\
12	-3.12507994134718e-06\\
13	-1.28577410845824e-07\\
14	-4.14881946845828e-06\\
15	1.55729584320931e-05\\
16	-0.000109099864485052\\
17	3.72553069872239e-05\\
18	-0.000148978825847138\\
19	-5.77349542608116e-06\\
20	-2.07546107866288e-06\\
21	2.50340799926575e-06\\
22	-1.20899542593453e-06\\
23	-1.44842669187883e-06\\
24	-5.78919687837585e-06\\
25	-1.02288793989238e-06\\
26	-3.85336085235893e-05\\
27	1.68789581984011e-05\\
28	6.59990767699865e-06\\
29	5.94171825336516e-07\\
30	-3.24356988073218e-06\\
31	1.27869855753525e-05\\
32	-4.94159786887343e-06\\
33	-5.1651415703809e-06\\
34	-8.79693067994101e-05\\
35	-0.000834631348649416\\
36	-9.87403672815525e-07\\
37	3.0838469177119e-06\\
38	2.37896570560283e-06\\
39	-6.02459118593654e-07\\
40	-4.71927329062586e-08\\
41	6.48395242883707e-07\\
42	-9.75661743104807e-07\\
43	-0.000139446066066663\\
44	0.000206004541710496\\
45	7.21756868824744e-06\\
46	3.88190709868823e-06\\
47	5.2488881307784e-06\\
48	-6.59183026229509e-06\\
49	1.89440982305264e-06\\
50	-2.77421653896061e-06\\
51	-3.99516604850609e-06\\
52	-7.61196262590313e-05\\
53	-0.000169151235545352\\
54	-9.57190700828617e-06\\
55	-6.05765335498691e-06\\
56	-4.70085486820603e-07\\
57	3.43364053629677e-06\\
58	-1.01908544979195e-06\\
59	-1.44373419765024e-06\\
60	-1.09837027850115e-05\\
61	-5.78065855920033e-05\\
62	0.000156434980427371\\
63	-0.00272173587932789\\
64	-1.13377313371098e-05\\
65	8.3294894063357e-07\\
66	1.61151212368735e-06\\
67	-5.40857184967249e-10\\
68	6.00354127560243e-06\\
69	-8.85178046042742e-05\\
70	-0.000259155274708631\\
71	-3.57770067771744e-05\\
72	1.41751529371933e-05\\
73	-2.25043252440418e-06\\
74	1.0726085727014e-05\\
75	8.41729036874057e-07\\
76	5.06532918181133e-05\\
77	-0.000241600254776583\\
};
\addlegendentry{Discretized domain SA}

\end{axis}
\end{tikzpicture}%

%% file: Figures/SA/Compliance_diffE.tex
%
%
\definecolor{mycolor1}{rgb}{1.00000,0.00000,1.00000}%
\begin{tikzpicture}

\begin{axis}[%
width=3.869in,
height=4.075in,
at={(0.656in,0.55in)},
scale only axis,
separate axis lines,
every outer x axis line/.append style={black},
every x tick label/.append style={font=\color{black}},
every x tick/.append style={black},
xmin=0,
xmax=80,
xtick={\empty},
xlabel style={font=\color{white!15!black}},
xlabel={\large{Design variable}},
every outer y axis line/.append style={black},
every y tick label/.append style={font=\color{black}},
every y tick/.append style={black},
ymode=log,
ymin=1e-07,
ymax=0.1,
yminorticks=true,
ylabel style={font=\color{white!15!black}},
ylabel={\large{Relative error}},
yticklabel style={font=\large},
axis background/.style={fill=white},
xmajorgrids,
ymajorgrids,
yminorgrids,
legend style={font=\fontsize{10}{15}\selectfont,at={(0.5,0.03)}, anchor=south, legend cell align=left, align=left, draw=white!15!black}
]
\addplot[only marks, mark=o, mark options={}, mark size=1.5000pt, draw=blue] table[row sep=crcr]{%
x	y\\
1	0.00790432681170086\\
2	0.00427109478349889\\
3	0.00174486440460633\\
4	0.00454812046684453\\
5	0.000321713306272735\\
6	0.00128614732644345\\
7	0.00149426547208026\\
8	0.00406501089862886\\
9	0.00222087864929576\\
10	0.000190400178413474\\
11	0.000712533691386974\\
12	0.00429160744562025\\
13	0.00103995264270368\\
14	0.00050399764243421\\
15	0.000664115446389729\\
16	0.00544252345591433\\
17	0.00413411815459669\\
18	0.002585291776464\\
19	0.00155905313285653\\
20	0.00186326966935207\\
21	0.00436053312265724\\
22	0.00162946505182564\\
23	0.000896826630764742\\
24	0.0015564275882635\\
25	0.00344841810011543\\
26	0.004684550259505\\
27	0.00369627600815424\\
28	0.00313190149271887\\
29	0.00253573144325727\\
30	0.0040958729622115\\
31	0.00198061868700149\\
32	0.00230913557340478\\
33	0.00283798444771256\\
34	0.00279561741340457\\
35	0.00362083586633696\\
36	0.00370037761987801\\
37	0.00299878011496062\\
38	0.00246706556615758\\
39	0.00936739570073535\\
40	0.00241954324736547\\
41	0.00323304675830923\\
42	0.00375729675398533\\
43	0.00357850052640476\\
44	0.005798829901522\\
45	0.00354600742070934\\
46	0.00300728832505075\\
47	0.00239620714701949\\
48	0.000151772174772642\\
49	0.00194164105792493\\
50	0.00252234910961751\\
51	0.00301126244758722\\
52	0.00130528739932581\\
53	0.00435360814512712\\
54	0.00263152734382859\\
55	0.00155618419054222\\
56	0.00073109703138178\\
57	0.00356660771135738\\
58	0.00133824198639029\\
59	0.00084568584654642\\
60	0.00172356674300617\\
61	0.00318005136748353\\
62	0.00864854378279067\\
63	0.00136299802723961\\
64	1.09098878998568e-05\\
65	0.00034527079399498\\
66	0.00195497344094349\\
67	0.000695623532387073\\
68	0.000323865628096461\\
69	4.20877433529139e-05\\
70	0.00424459712026387\\
71	0.0118170968278923\\
72	0.00046234974304037\\
73	0.00299401473625637\\
74	0.00431527614897085\\
75	0.00172074648971953\\
76	0.00255956359837302\\
77	0.00559184943013624\\
};
\addlegendentry{1:1000}

\addplot[only marks, mark=+, mark options={}, mark size=1.5000pt, draw=red] table[row sep=crcr]{%
x	y\\
1	0.00111093896140119\\
2	0.00226751041298045\\
3	0.000596491398381014\\
4	0.00128921753543522\\
5	0.000316390339329985\\
6	0.00127659368381185\\
7	0.00116800903707412\\
8	0.00307346309542118\\
9	0.00127260499156511\\
10	0.000131603370504136\\
11	0.000226335334769705\\
12	0.00293237347845464\\
13	0.000806134355094992\\
14	0.000270008108070936\\
15	0.000656620318633151\\
16	0.00299087169127799\\
17	0.00319645638113421\\
18	0.00246619091747957\\
19	0.00150048853938425\\
20	0.00150853721934424\\
21	0.00419377501834815\\
22	0.00149185745633326\\
23	0.000957061827654224\\
24	0.00188025348251624\\
25	0.0029097642815073\\
26	0.00374646771608382\\
27	0.00342412406222464\\
28	0.00296089784664254\\
29	0.0028305351741653\\
30	0.00498095176054769\\
31	0.00255454697194517\\
32	0.00260574274194757\\
33	0.00321773014949809\\
34	0.00360471570065638\\
35	0.0035282197041221\\
36	0.00327296253380418\\
37	0.0029085019448706\\
38	0.00279906551608722\\
39	0.00473822747815532\\
40	0.00322426182205318\\
41	0.00307579174479789\\
42	0.0034489025811509\\
43	0.00378728763220206\\
44	0.00387522906752643\\
45	0.00337924487674545\\
46	0.00294841191915368\\
47	0.00280517881382007\\
48	0.00476551711098085\\
49	0.00255228321124436\\
50	0.00261918023087224\\
51	0.00322337037170947\\
52	0.00345091604475956\\
53	0.00331343364552782\\
54	0.00246932789885538\\
55	0.00149845465935317\\
56	0.00139429316536271\\
57	0.00412212770058022\\
58	0.00145807047156788\\
59	0.000949706935207977\\
60	0.00189648664570414\\
61	0.00296578142363467\\
62	0.00410254716461922\\
63	0.00118159470607942\\
64	0.000143745306235224\\
65	0.000159356428818103\\
66	0.00229782234719805\\
67	0.000791798818611749\\
68	0.000265913627175236\\
69	0.000555329517945979\\
70	0.00253124223575513\\
71	0.00125817713412919\\
72	0.00235343128231213\\
73	0.000810930488171956\\
74	0.00100960258583363\\
75	0.000135632355661642\\
76	0.00151030849421851\\
77	0.00203595301398451\\
};
\addlegendentry{10:1000}

\addplot[only marks, mark=asterisk, mark options={}, mark size=1.5000pt, draw=black] table[row sep=crcr]{%
x	y\\
1	0.00229414621715522\\
2	0.00213388095832595\\
3	0.00155908952375206\\
4	0.000879037357337227\\
5	0.000747340858421587\\
6	0.00150552192589242\\
7	0.00166168336644394\\
8	0.00293050651851832\\
9	0.00118258931956834\\
10	0.000278776808339045\\
11	0.000854263938980003\\
12	0.000125936155125708\\
13	0.000353115362303683\\
14	0.000197250514171717\\
15	0.00101157751696572\\
16	0.00286764101801068\\
17	0.00299342203538112\\
18	0.00230304824269271\\
19	0.00122397767363962\\
20	0.000157405815650489\\
21	8.58633518532858e-05\\
22	0.000117667761358976\\
23	0.0010636485467832\\
24	0.00213518149413394\\
25	0.00296809982605809\\
26	0.00307042816223344\\
27	0.00264982376554174\\
28	0.00208150164686156\\
29	0.00124656967603295\\
30	0.000671901373541697\\
31	0.00112485243106259\\
32	0.00180943184702842\\
33	0.00246306636493537\\
34	0.00296043114445775\\
35	0.00307989188543059\\
36	0.00210920657451546\\
37	0.00169103247234021\\
38	0.00174451415103949\\
39	0.0018665152080233\\
40	0.00160218405885951\\
41	0.0014053460839687\\
42	0.00186564331164815\\
43	0.00295950813073633\\
44	0.00308475371322164\\
45	0.00264634485113624\\
46	0.00207928284386533\\
47	0.00124710522472654\\
48	0.000669122830677547\\
49	0.00112317422981924\\
50	0.00181182030718761\\
51	0.00246294551075371\\
52	0.00295168034822813\\
53	0.00299647247579238\\
54	0.00230218835496744\\
55	0.00122535066551162\\
56	0.00016254904462287\\
57	8.59666067504152e-05\\
58	0.000117079896499229\\
59	0.00105981067653346\\
60	0.00213831568399575\\
61	0.00298777147813311\\
62	0.00301248704548764\\
63	0.00118049031543541\\
64	0.000280276910619355\\
65	0.000859229686237915\\
66	0.000138017475231671\\
67	0.000360344385750826\\
68	0.000195852364382694\\
69	0.000994632921112733\\
70	0.00301360206011115\\
71	0.00235447117466822\\
72	0.00214150288170057\\
73	0.00153926202343402\\
74	0.000866924273495647\\
75	0.000755346116332401\\
76	0.00150368261344048\\
77	0.00177738849920418\\
};
\addlegendentry{200:1000}

\addplot[only marks, mark=diamond, mark options={}, mark size=2.5883pt, draw=green] table[row sep=crcr]{%
x	y\\
1	0.00109794136893578\\
2	0.0011432481169375\\
3	0.00080225553869984\\
4	0.000481134519446534\\
5	0.000536832541769604\\
6	0.00102396606447609\\
7	0.000638862931512307\\
8	0.00245031821155887\\
9	0.000918664320323388\\
10	0.00015013046751716\\
11	0.000957736866914868\\
12	0.00176798384140363\\
13	0.000772657937493384\\
14	0.000176241814904697\\
15	0.000828886982106455\\
16	0.00248817302679198\\
17	0.00239606680365243\\
18	0.00177196301988477\\
19	0.00082152922733707\\
20	0.00100275586324214\\
21	0.00210666609136582\\
22	0.000879444253059887\\
23	0.000704940781658654\\
24	0.00164812195154347\\
25	0.00239689890814538\\
26	0.00238928969269323\\
27	0.00186180793860885\\
28	0.00121256926597813\\
29	9.44220418754835e-05\\
30	0.00217274756322735\\
31	0.00029008340643005\\
32	0.000901887140925803\\
33	0.00162293958104093\\
34	0.00227722903378448\\
35	0.00238963133950577\\
36	0.00104181302932229\\
37	0.000536397655623272\\
38	0.000406992493022396\\
39	0.00117621324486115\\
40	4.6031444670264e-06\\
41	7.82851178823189e-05\\
42	0.000660639831013738\\
43	0.00222409426338265\\
44	0.00239852553965991\\
45	0.00186249551261877\\
46	0.00121065545199706\\
47	9.35244476529717e-05\\
48	0.00217634888242272\\
49	0.000286655465697308\\
50	0.000900244533145486\\
51	0.0016208726770229\\
52	0.00216334026395306\\
53	0.00242462567912889\\
54	0.00177520850784732\\
55	0.000820217343654558\\
56	0.00100858801823539\\
57	0.00210975918806808\\
58	0.000880286545656306\\
59	0.000704206280683201\\
60	0.00164672928411167\\
61	0.00237494870899136\\
62	0.00253799048058487\\
63	0.000913933074951755\\
64	0.000149222764326727\\
65	0.000960556757089697\\
66	0.00178381327769579\\
67	0.000783029211284531\\
68	0.000188162645731826\\
69	0.000788100715143126\\
70	0.00255189348786053\\
71	0.00105313295926\\
72	0.001100726013289\\
73	0.000803610504575841\\
74	0.000512338112884233\\
75	0.000626848013262682\\
76	0.00107252976620415\\
77	0.00122137625084689\\
};
\addlegendentry{500:1000}

\addplot[only marks, mark=square, mark options={}, mark size=1.0607pt, draw=mycolor1] table[row sep=crcr]{%
x	y\\
1	0.000195085367515053\\
2	0.000582947724660167\\
3	0.000447047283956641\\
4	0.000319907108069803\\
5	0.000327010273887315\\
6	0.000597427798080792\\
7	4.46690311852435e-05\\
8	0.00234119637098279\\
9	0.000836194521581912\\
10	1.50235679631833e-06\\
11	0.000866603425777721\\
12	0.00218479698827098\\
13	0.00080531268934715\\
14	0.000188773807600074\\
15	0.00062444630946505\\
16	0.00249921845240919\\
17	0.00210981976498632\\
18	0.00148734471138018\\
19	0.000666045426912799\\
20	0.00118041950240093\\
21	0.00279675006388398\\
22	0.00113793620929839\\
23	0.000454373349545725\\
24	0.00124982961800892\\
25	0.00196848838404004\\
26	0.00206602754169237\\
27	0.00147580776052134\\
28	0.000820132323349631\\
29	0.000572812931392316\\
30	0.0031716482491986\\
31	0.000833746748379423\\
32	0.000370727110985468\\
33	0.00110059487102221\\
34	0.00185228565165879\\
35	0.00213500925081484\\
36	0.000576730878376733\\
37	5.90132249871996e-05\\
38	0.000115485890492049\\
39	0.00236937868922094\\
40	0.000682032310111381\\
41	0.000561776322257735\\
42	6.17564364247599e-05\\
43	0.00196367741580383\\
44	0.00208300358051868\\
45	0.00146669005863821\\
46	0.000814395586342212\\
47	0.000565008382365047\\
48	0.00319428027885206\\
49	0.000828554756150789\\
50	0.000376737674090095\\
51	0.00109289069509445\\
52	0.00184301035013445\\
53	0.00213813176436868\\
54	0.00148978758095762\\
55	0.000661408506141634\\
56	0.00117780473364237\\
57	0.00279941788251631\\
58	0.00113986299866298\\
59	0.000443321460950357\\
60	0.00123376119983518\\
61	0.00197958598392522\\
62	0.00225832790788785\\
63	0.000827013572561744\\
64	8.88123105633336e-07\\
65	0.000867399138836007\\
66	0.00220738137160781\\
67	0.000816932154453378\\
68	0.000191825489739193\\
69	0.000537515253816718\\
70	0.0023342225225859\\
71	0.000383047360087654\\
72	0.000583405689176081\\
73	0.000447304889305872\\
74	0.000286594738916363\\
75	0.00037995400625953\\
76	0.000768074638352064\\
77	0.000604962665175651\\
};
\addlegendentry{750:1000}

\end{axis}
\end{tikzpicture}%

%% file: Figures/SA/Stress_diffE.tex
%
%
\definecolor{mycolor1}{rgb}{1.00000,0.00000,1.00000}%
\begin{tikzpicture}

\begin{axis}[%
width=3.869in,
height=4.075in,
at={(0.656in,0.55in)},
scale only axis,
every outer x axis line/.append style={black},
every x tick label/.append style={font=\color{black}},
every x tick/.append style={black},
xmin=0,
xmax=80,
xtick={\empty},
xlabel style={font=\color{white!15!black}},
xlabel={\large{Design variable}},
every outer y axis line/.append style={black},
every y tick label/.append style={font=\color{black}},
every y tick/.append style={black},
ymode=log,
ymin=1e-07,
ymax=0.1,
yminorticks=true,
ylabel style={font=\color{white!15!black}},
ylabel={\large{Relative error}},
yticklabel style={font=\large},
axis background/.style={fill=white},
xmajorgrids,
ymajorgrids,
yminorgrids,
legend style={font=\fontsize{10}{15}\selectfont,at={(0.5,0.03)}, anchor=south, legend cell align=left, align=left, draw=white!15!black}
]
\addplot[only marks, mark=o, mark options={}, mark size=1.5000pt, draw=blue] table[row sep=crcr]{%
x	y\\
1	0.0146684732190374\\
2	0.00146154693393098\\
3	0.00408607916152852\\
4	0.00624422361360529\\
5	0.00205845775285144\\
6	0.00133057528848082\\
7	0.00409908016522342\\
8	0.00279108206070821\\
9	0.00354437094045355\\
10	0.00333259161739179\\
11	0.00655660384534714\\
12	0.0168881681419435\\
13	0.00276727207604919\\
14	0.000991821961480664\\
15	0.00117646618600734\\
16	0.00414288116074583\\
17	0.00279939718803289\\
18	0.00219092248803157\\
19	0.0025500451530991\\
20	0.00762554065513386\\
21	0.0313865770890582\\
22	0.00354470315358191\\
23	0.00121991500683088\\
24	0.000915890158070943\\
25	0.00184741855032557\\
26	0.00302600701269566\\
27	0.00204634614384965\\
28	0.00154011678128688\\
29	0.00317815994167136\\
30	0.0512924068271898\\
31	0.00151724833363867\\
32	0.000142334984077182\\
33	0.000799333850501734\\
34	0.000761327328859286\\
35	0.00170417814756154\\
36	0.000393763337267525\\
37	0.00256351433278707\\
38	0.00666975461145104\\
39	0.0101416518568683\\
40	0.0149734523410816\\
41	0.0071051495579098\\
42	0.00302777404161885\\
43	0.00142628712836509\\
44	0.00430387013605755\\
45	0.00187925930447074\\
46	0.00140634135306623\\
47	0.00303270116729913\\
48	0.0487768121767559\\
49	0.00147805909328209\\
50	0.000352383851622403\\
51	0.00097774077184343\\
52	0.000806133567931108\\
53	0.00306868350015341\\
54	0.00224184100239021\\
55	0.00254674677483898\\
56	0.00651816948229816\\
57	0.0309302172487726\\
58	0.00330173962205216\\
59	0.00116930926637544\\
60	0.00109032801170656\\
61	0.0015200497549016\\
62	0.00828306912225436\\
63	0.00255696969041857\\
64	0.0031366343528081\\
65	0.00553268263829834\\
66	0.0125894804159469\\
67	0.00249008185376047\\
68	0.0011610368633349\\
69	0.000563447388042062\\
70	0.00280534036021246\\
71	0.0187368324334235\\
72	0.0054708169156886\\
73	0.00283409834494606\\
74	0.000783953033541892\\
75	0.000422802966727227\\
76	0.00019646222991602\\
77	0.0024651033553976\\
};
\addlegendentry{1:1000}

\addplot[only marks, mark=+, mark options={}, mark size=1.5000pt, draw=red] table[row sep=crcr]{%
x	y\\
1	0.00549936630150735\\
2	0.00446329056970303\\
3	0.00675047615151733\\
4	0.00490889083936991\\
5	0.00282556029875023\\
6	0.00175666379352722\\
7	0.00186454250427328\\
8	0.00166750089579733\\
9	0.00269776527197067\\
10	0.00353172142135505\\
11	0.00671761774284591\\
12	0.0125796460149266\\
13	0.00313471212029329\\
14	0.00153576177884255\\
15	0.00129738665099654\\
16	0.00139796568480224\\
17	0.00171699574229613\\
18	0.00215202752743861\\
19	0.0027703367892141\\
20	0.0067665545549638\\
21	0.0211253973249953\\
22	0.00361378908711645\\
23	0.00151139193388671\\
24	0.00130399304031706\\
25	0.00121698135776613\\
26	0.00203116502589778\\
27	0.00191145123935409\\
28	0.00166711765026776\\
29	0.00300030596984015\\
30	0.0307241192107808\\
31	0.00207223942575913\\
32	0.00100473524618771\\
33	0.00147638934604646\\
34	0.00169281839575325\\
35	0.00167713860438997\\
36	0.000343219660053634\\
37	0.0019296842582727\\
38	0.0056420970466949\\
39	4.71266343623555e-05\\
40	0.010669101075694\\
41	0.0047283681692542\\
42	0.0018534775278329\\
43	0.00173171706378498\\
44	0.00217162602025735\\
45	0.00186498470554992\\
46	0.00165505865434874\\
47	0.00297997569903453\\
48	0.0304935237906362\\
49	0.00207120453025907\\
50	0.0010186323888982\\
51	0.00148543809313215\\
52	0.00151628102031716\\
53	0.00183458734424225\\
54	0.00215338601279159\\
55	0.00276776670726971\\
56	0.00665228072131593\\
57	0.021056964658136\\
58	0.00358402878388929\\
59	0.00150410793108539\\
60	0.00132204393789462\\
61	0.00128449449597043\\
62	0.0028464190561144\\
63	0.00259366207334716\\
64	0.00351802209223483\\
65	0.00664517254841912\\
66	0.0120389634905779\\
67	0.00312158632711451\\
68	0.00154050775901764\\
69	0.00119901797841802\\
70	0.00090993787668208\\
71	0.00526312539389823\\
72	0.00437798815620583\\
73	0.00654213852062685\\
74	0.00473390194192202\\
75	0.00271258047497974\\
76	0.00156924262073942\\
77	0.00108542231256202\\
};
\addlegendentry{10:1000}

\addplot[only marks, mark=asterisk, mark options={}, mark size=1.5000pt, draw=black] table[row sep=crcr]{%
x	y\\
1	0.00201369676865755\\
2	0.0016851435064706\\
3	0.00121836994179447\\
4	0.00131462597445669\\
5	0.00145915490769602\\
6	0.00149964899901582\\
7	0.00164222544820804\\
8	0.0019436288445914\\
9	0.00187792542767466\\
10	0.00188959274159509\\
11	0.0039232338633526\\
12	0.00945161116676903\\
13	0.00340377519769326\\
14	0.00153202330859129\\
15	0.00146076816461719\\
16	0.0014083252552756\\
17	0.00183308198994698\\
18	0.00187441564737124\\
19	0.0020557161752314\\
20	0.00598504168017147\\
21	0.0115340976415834\\
22	0.00524599106956022\\
23	0.00170649180794094\\
24	0.00152519029776959\\
25	0.00145707436953337\\
26	0.00169678190093341\\
27	0.00202301691276372\\
28	0.00240036680993533\\
29	0.00489320447005388\\
30	0.0126531983849589\\
31	0.00522409879905654\\
32	0.0027441479175915\\
33	0.00209877703470286\\
34	0.00154719766811415\\
35	0.00169862429595403\\
36	0.00245356383884451\\
37	0.00271129929448859\\
38	0.00262640184294464\\
39	0.00939440949169305\\
40	0.00419806332154443\\
41	0.00403172066481576\\
42	0.00336094886120337\\
43	0.00153247121552343\\
44	0.00172179870440301\\
45	0.0020204942365062\\
46	0.00239732153429603\\
47	0.00488959451990896\\
48	0.0126500859641096\\
49	0.00522189950135155\\
50	0.00274621026827543\\
51	0.0021013431295817\\
52	0.00151100907072492\\
53	0.00182805128668968\\
54	0.00187593385448063\\
55	0.002057056304412\\
56	0.0059820399770535\\
57	0.0115345945225823\\
58	0.00524829278579446\\
59	0.00170518567137657\\
60	0.00152981877319589\\
61	0.00149667964028907\\
62	0.00204375168382366\\
63	0.00186190736564143\\
64	0.00188694648134949\\
65	0.00392008613651248\\
66	0.00944002146500056\\
67	0.00339815438675886\\
68	0.00153308774927559\\
69	0.00144301080002498\\
70	0.00167761838944872\\
71	0.00203731112706362\\
72	0.00172315812028712\\
73	0.00127170838048114\\
74	0.001332828194265\\
75	0.00143392323869579\\
76	0.00150345299664439\\
77	0.00154207038199866\\
};
\addlegendentry{200:1000}

\addplot[only marks, mark=diamond, mark options={}, mark size=2.5883pt, draw=green] table[row sep=crcr]{%
x	y\\
1	2.30738352958112e-06\\
2	0.000191094572746023\\
3	0.000135496902518093\\
4	5.92418665272038e-05\\
5	7.44603631931638e-05\\
6	3.41941810848154e-05\\
7	3.17871934775613e-05\\
8	0.0010155031975738\\
9	0.000499311755747938\\
10	0.000186694919159204\\
11	0.000892019891466141\\
12	0.00311957715252329\\
13	0.000932340582048809\\
14	0.000204119005754715\\
15	0.000441826902392861\\
16	0.000936624636410168\\
17	0.000990050396741941\\
18	0.00080962180380359\\
19	0.000619939311886961\\
20	0.00182640320319064\\
21	0.00402318672715435\\
22	0.00177357923095064\\
23	0.000534284735627444\\
24	0.000671004625582314\\
25	0.000842080391599904\\
26	0.000966477098877143\\
27	0.00106951623512348\\
28	0.00116387304070018\\
29	0.00196511590993149\\
30	0.00473909082862668\\
31	0.00201695043168228\\
32	0.00117493935811581\\
33	0.000997476548893424\\
34	0.00085295768563231\\
35	0.000967279932217559\\
36	0.0013359339355477\\
37	0.00146034489434477\\
38	0.00150937628092077\\
39	0.00416540962961515\\
40	0.0018121443489563\\
41	0.00169187953863308\\
42	0.00146130739984915\\
43	0.000825827770020701\\
44	0.00102943584934125\\
45	0.00106639304598125\\
46	0.00116525155430463\\
47	0.00196025050924669\\
48	0.0047391801542774\\
49	0.00202000324361529\\
50	0.00117569203251066\\
51	0.00099708980067183\\
52	0.00081180326769031\\
53	0.000981416550540594\\
54	0.000813021062477124\\
55	0.00062052742510827\\
56	0.00182072669462926\\
57	0.00402274449455605\\
58	0.00177087486413663\\
59	0.000536180530991675\\
60	0.000674484375043224\\
61	0.000836985382981707\\
62	0.00110012642038546\\
63	0.00049046086464841\\
64	0.000190392554023268\\
65	0.000883980718785401\\
66	0.00310885658059793\\
67	0.000922646897078806\\
68	0.000193902968060865\\
69	0.000406861154757991\\
70	0.000940539840188406\\
71	0.000226000551641443\\
72	0.00017149112129205\\
73	0.000108169953875747\\
74	7.48762557071637e-06\\
75	4.0477472272126e-05\\
76	9.26651111001795e-05\\
77	6.33482292475374e-05\\
};
\addlegendentry{500:1000}

\addplot[only marks, mark=square, mark options={}, mark size=1.0607pt, draw=mycolor1] table[row sep=crcr]{%
x	y\\
1	9.28328962478349e-05\\
2	0.000111272573844473\\
3	3.33040316866376e-05\\
4	3.50803985361363e-05\\
5	1.85635162062831e-07\\
6	9.09404297150191e-05\\
7	7.52874445288264e-05\\
8	0.000615823186499639\\
9	0.000134580005411973\\
10	4.63587520237749e-07\\
11	0.000261687248627296\\
12	0.00105728912585225\\
13	0.000264725777511054\\
14	3.91352137498491e-05\\
15	8.68120250797217e-05\\
16	0.000480668854140509\\
17	0.000457363493329001\\
18	0.000317803736596827\\
19	0.000197892197001437\\
20	0.000596147361084435\\
21	0.00141006350065856\\
22	0.000578399651044149\\
23	0.0001337192858326\\
24	0.000229414321673359\\
25	0.000366534870239405\\
26	0.000508288435695097\\
27	0.000491743775075684\\
28	0.000504819739320569\\
29	0.000745176594034933\\
30	0.00172981562037549\\
31	0.000732118644851924\\
32	0.000458687367032949\\
33	0.000421091259896608\\
34	0.000426193546734328\\
35	0.000559016660355312\\
36	0.000597694764893296\\
37	0.000637498719489951\\
38	0.000652185436708842\\
39	0.00159536747424038\\
40	0.000700615852138634\\
41	0.00065592449509845\\
42	0.000590449265322667\\
43	0.000451382700062613\\
44	0.000501330205701416\\
45	0.000488400474342329\\
46	0.000503903567410322\\
47	0.000741596506907918\\
48	0.00172656064556097\\
49	0.000731344984098911\\
50	0.000459117369022881\\
51	0.000415678227952487\\
52	0.000409825098393923\\
53	0.000474767230679463\\
54	0.000316724013095765\\
55	0.000198015189141959\\
56	0.000595923059493375\\
57	0.00140822542453279\\
58	0.000578332675959183\\
59	0.00013214493763357\\
60	0.000227399945441837\\
61	0.000361526745395297\\
62	0.000607490412074243\\
63	0.000152199633072396\\
64	1.36986674983944e-06\\
65	0.00026342616743284\\
66	0.00105469628980653\\
67	0.00026562367578287\\
68	4.20118314574776e-05\\
69	6.56295693634934e-05\\
70	0.000557381851559122\\
71	3.59883807589111e-05\\
72	0.000124746204705576\\
73	3.49420264457191e-06\\
74	7.43859163093616e-05\\
75	3.10640803389743e-06\\
76	0.000168409130705537\\
77	4.31465452024081e-05\\
};
\addlegendentry{750:1000}

\end{axis}
\end{tikzpicture}%

%% file: Figures/SA/Convg_AddP.tex
%
%
\definecolor{mycolor1}{rgb}{1.00000,0.00000,1.00000}%
\begin{tikzpicture}

\begin{axis}[%
width=3.772in,
height=3.776in,
at={(0.633in,0.51in)},
scale only axis,
separate axis lines,
every outer x axis line/.append style={black},
every x tick label/.append style={font=\color{black}},
every x tick/.append style={black},
xmin=0,
xmax=80,
xtick={\empty},
xlabel={\large{Design variable}},
every outer y axis line/.append style={black},
every y tick label/.append style={font=\color{black}},
every y tick/.append style={black},
ymode=log,
ymin=1e-06,
ymax=10,
yminorticks=true,
ylabel={\large{Relative error}},
yticklabel style={font=\large},
axis background/.style={fill=white},
xmajorgrids,
ymajorgrids,
yminorgrids,
legend style={font=\fontsize{10}{15}\selectfont,legend cell align=left, align=left, draw=white!15!black}
]
\addplot[only marks, mark=*, mark options={}, mark size=1.5000pt, color=mycolor1, fill=mycolor1] table[row sep=crcr]{%
x	y\\
1	0.382251333567349\\
2	0.326683766238787\\
3	0.0394368367543019\\
4	0.0256960962683062\\
5	0.031016313456729\\
6	0.20709089115372\\
7	0.301336790680754\\
8	0.404144435716749\\
9	0.798820761366843\\
10	0.216111621515068\\
11	0.0381114070034355\\
12	0.0657414855780396\\
13	0.0320861702282796\\
14	0.188622291355266\\
15	0.224402292559982\\
16	0.140435460514009\\
17	2.87178622459875\\
18	0.482566780621645\\
19	0.25004368537959\\
20	0.0411316016578604\\
21	0.102932383988818\\
22	0.036490945272785\\
23	0.296105885615166\\
24	0.153552613148498\\
25	0.123328552243289\\
26	0.139845653107072\\
27	0.121530590130643\\
28	0.138329927256106\\
29	0.459471384530011\\
30	0.196292183895602\\
31	0.41959425622611\\
32	0.136797828416468\\
33	0.10620654734045\\
34	0.0958194548416271\\
35	0.111109315721569\\
36	0.120622825917416\\
37	0.125753867584053\\
38	0.15795902550805\\
39	0.0626101404779082\\
40	0.119745342372169\\
41	0.120593496539829\\
42	0.115504020418095\\
43	0.0905434928006934\\
44	0.139855295904718\\
45	0.121525487774335\\
46	0.138328015438321\\
47	0.459465506069259\\
48	0.19628659248959\\
49	0.419593206300429\\
50	0.136801789671719\\
51	0.106208287530565\\
52	0.0958057866926046\\
53	2.87285779668153\\
54	0.482537892882512\\
55	0.250048697685742\\
56	0.0411300225719538\\
57	0.102930850966101\\
58	0.036490469820008\\
59	0.296094207997115\\
60	0.153562318009728\\
61	0.123391690440091\\
62	0.404594277758176\\
63	0.798723051691101\\
64	0.216109361674408\\
65	0.0381106288420182\\
66	0.065738646076354\\
67	0.0320838226681253\\
68	0.188625160241017\\
69	0.22430390245949\\
70	0.141176072458015\\
71	0.382250627635499\\
72	0.32667307664126\\
73	0.0394394548487811\\
74	0.0256963606064317\\
75	0.0310159005435107\\
76	0.207085270902809\\
77	0.300937271712304\\
};
\addlegendentry{Zero}

\addplot[only marks, mark=+, mark options={}, mark size=1.5811pt, draw=green] table[row sep=crcr]{%
x	y\\
1	0.140894989985909\\
2	0.11815184201039\\
3	0.0133035799426596\\
4	0.00876806692522445\\
5	0.0111292309368067\\
6	0.0759140730870325\\
7	0.11106607439262\\
8	0.172662024242927\\
9	0.305796400922973\\
10	0.0791759548788767\\
11	0.0110785583762284\\
12	0.0134839432181139\\
13	0.00969542735512717\\
14	0.0695568272808091\\
15	0.085048336477084\\
16	0.0574686369888298\\
17	1.21562631099147\\
18	0.190247248906799\\
19	0.0928122492028169\\
20	0.0103918434956431\\
21	0.0190698090594673\\
22	0.0092769471191607\\
23	0.109315299586308\\
24	0.0593234038436426\\
25	0.0499964258020999\\
26	0.0586845216176053\\
27	0.0432172982754998\\
28	0.0450228499178604\\
29	0.116899740774911\\
30	0.0366029964032759\\
31	0.103517711688739\\
32	0.0424693525690684\\
33	0.0365445309704153\\
34	0.0392464914069705\\
35	0.04660148637959\\
36	0.0354362791874768\\
37	0.0350252978629307\\
38	0.0383324380727726\\
39	0.0113457883133081\\
40	0.0272021846980382\\
41	0.0309925402477069\\
42	0.0316677822078215\\
43	0.0372631395831332\\
44	0.0586763848945816\\
45	0.0432217496445577\\
46	0.0450248020112024\\
47	0.116901477614235\\
48	0.0366070714683138\\
49	0.103518559475493\\
50	0.0424662307281646\\
51	0.0365430558118326\\
52	0.0392582468856435\\
53	1.21527014476825\\
54	0.190244884032838\\
55	0.0928105562986258\\
56	0.0103934220537589\\
57	0.0190710134772893\\
58	0.00927737798034528\\
59	0.109327998592559\\
60	0.0593171439835755\\
61	0.0499443543942026\\
62	0.171860628303994\\
63	0.305793994403434\\
64	0.0791833008087759\\
65	0.0110793253304279\\
66	0.0134867420025513\\
67	0.0096978038110775\\
68	0.0695510618288685\\
69	0.0851252032557071\\
70	0.0568591512565203\\
71	0.14089003498073\\
72	0.11817707837392\\
73	0.0133007848191243\\
74	0.00876771742413743\\
75	0.0111295754515507\\
76	0.0759197036494474\\
77	0.111342519329352\\
};
\addlegendentry{One}

\addplot[only marks, mark=asterisk, mark options={}, mark size=1.5000pt, draw=red] table[row sep=crcr]{%
x	y\\
1	0.0439998382626749\\
2	0.0363341725887809\\
3	0.00371115537210585\\
4	0.00237199456970586\\
5	0.00313350984043992\\
6	0.022582538391152\\
7	0.0336368569195311\\
8	0.0479392111957425\\
9	0.0934500908028473\\
10	0.0242389927544838\\
11	0.00284482518797675\\
12	0.00221505741087904\\
13	0.00248613707380028\\
14	0.0204386918124713\\
15	0.0246031009230888\\
16	0.0142026236203544\\
17	0.345720379504385\\
18	0.056169903621211\\
19	0.0278133614728143\\
20	0.00238538057459932\\
21	0.00253534801282142\\
22	0.00208779755880189\\
23	0.0314001284315691\\
24	0.0165491947600348\\
25	0.0132879921519146\\
26	0.0169289216250831\\
27	0.0118733551681322\\
28	0.0119761715952071\\
29	0.0251719236235168\\
30	0.00428349158059412\\
31	0.0213732282348597\\
32	0.0109123777190399\\
33	0.0097519963124965\\
34	0.0109796087409474\\
35	0.0134171537084279\\
36	0.00866383315649069\\
37	0.0082504558287165\\
38	0.00767326593971471\\
39	0.00105087030338778\\
40	0.00503634391046433\\
41	0.00686066554047119\\
42	0.00737082024509326\\
43	0.0104850688798576\\
44	0.0169377584630822\\
45	0.0118686628547784\\
46	0.0119741539255064\\
47	0.0251692232485446\\
48	0.00427932846821619\\
49	0.0213724488538807\\
50	0.0109156636907487\\
51	0.00975352218327504\\
52	0.0109671949030988\\
53	0.346347778216203\\
54	0.0561604727492964\\
55	0.0278160435110471\\
56	0.00238380950432723\\
57	0.00253415716684336\\
58	0.00208736575052399\\
59	0.031387346310199\\
60	0.0165564350598207\\
61	0.0133439118381496\\
62	0.0486055930451952\\
63	0.0934157150544732\\
64	0.0242333177579659\\
65	0.00284406366266923\\
66	0.00221229545508005\\
67	0.00248377779559915\\
68	0.0204432117860043\\
69	0.0245182161192096\\
70	0.0148592474213319\\
71	0.0440025844526788\\
72	0.0363138689965042\\
73	0.00371390446744055\\
74	0.00237233347184098\\
75	0.00313315125064654\\
76	0.0225766999074706\\
77	0.0333167998093595\\
};
\addlegendentry{Two}

\addplot[only marks, mark=o, mark options={}, mark size=1.7678pt, draw=blue] table[row sep=crcr]{%
x	y\\
1	0.0133216739019405\\
2	0.0105147692962844\\
3	0.000941941262127596\\
4	0.000556602344444673\\
5	0.000780970984461668\\
6	0.00631019847299841\\
7	0.00958750300415788\\
8	0.0141220816547718\\
9	0.0269815170660798\\
10	0.00691382261479632\\
11	0.000709828489050697\\
12	0.000537733458064419\\
13	0.000616075183482077\\
14	0.00560471680594875\\
15	0.00669890359753276\\
16	0.00418454998783835\\
17	0.0945096067432409\\
18	0.0155425679842902\\
19	0.00765858162823926\\
20	0.000589791256141977\\
21	0.000638632120107827\\
22	0.000510129046403339\\
23	0.00820783886580424\\
24	0.00428441188868872\\
25	0.00332671610458616\\
26	0.00450601151266287\\
27	0.00287768512084644\\
28	0.00274228660984939\\
29	0.00588208342259355\\
30	0.00115954007980065\\
31	0.00498804978044343\\
32	0.00243483664796222\\
33	0.00230957471355757\\
34	0.0028725229820101\\
35	0.00362193623782799\\
36	0.00183414092502261\\
37	0.00166954642248843\\
38	0.00160312797480413\\
39	0.000315376642741461\\
40	0.00107477255717832\\
41	0.00137027606704022\\
42	0.00154622537532495\\
43	0.00284265078185277\\
44	0.00449734687501743\\
45	0.00288229554266093\\
46	0.0027442705956652\\
47	0.00588451588656201\\
48	0.00116367941517888\\
49	0.00498881518967442\\
50	0.0024315769406504\\
51	0.00230804112016008\\
52	0.00288472028241671\\
53	0.0939578818418045\\
54	0.01554846802935\\
55	0.00765615010860971\\
56	0.000591358137706323\\
57	0.000639816234658789\\
58	0.000510557050208057\\
59	0.00822051818505075\\
60	0.00427736851506871\\
61	0.00327171204006123\\
62	0.0134175889020772\\
63	0.0270046590402291\\
64	0.00691997142560569\\
65	0.000710586785676707\\
66	0.000540496330936289\\
67	0.000618437041093182\\
68	0.00559980663317268\\
69	0.00678142616602618\\
70	0.00353988452714025\\
71	0.013318198615482\\
72	0.0105365323848692\\
73	0.000939175869965165\\
74	0.000556259872573693\\
75	0.000781325641956841\\
76	0.00631596345539459\\
77	0.00989451006420479\\
};
\addlegendentry{Three}

\addplot[only marks, mark=*, mark options={}, mark size=1.5000pt, color=black, fill=black] table[row sep=crcr]{%
x	y\\
1	0.00343889364365015\\
2	0.0026931491205669\\
3	0.00012675853022944\\
4	3.90159727559788e-05\\
5	0.000105688843665576\\
6	0.00149151923209691\\
7	0.00257591310723267\\
8	0.0025951197515496\\
9	0.0074771995591398\\
10	0.00183100216055934\\
11	0.000109673660154154\\
12	4.24656056576429e-05\\
13	8.6031083167451e-05\\
14	0.00131799793065349\\
15	0.00163041202401822\\
16	7.43739562977131e-06\\
17	0.022491994584287\\
18	0.00397551218216937\\
19	0.00192902657276852\\
20	9.59347822234006e-05\\
21	4.48575843296548e-05\\
22	7.41994628778717e-05\\
23	0.00179384295420221\\
24	0.000861845961238195\\
25	0.000501058922888587\\
26	0.00100506727556753\\
27	0.000533056108130804\\
28	0.000418522696888436\\
29	0.00072494103220105\\
30	4.3682783148875e-05\\
31	0.000554606655314746\\
32	0.000332848246108612\\
33	0.000340185174627461\\
34	0.000410286088627679\\
35	0.000718898490664023\\
36	0.000189479017453693\\
37	0.000125499870801038\\
38	0.000119191961681847\\
39	1.12298696892907e-05\\
40	9.706148151776e-05\\
41	0.000123353222298905\\
42	0.000161621374043962\\
43	0.000336672210327329\\
44	0.00101376315191908\\
45	0.000528417607999573\\
46	0.000416522368093973\\
47	0.00072243238957454\\
48	3.95373316610409e-05\\
49	0.000553826894618869\\
50	0.000336102473359522\\
51	0.000341703089081647\\
52	0.000398015939829217\\
53	0.0230636617810502\\
54	0.00396859133176316\\
55	0.00193148768141352\\
56	9.43642257346088e-05\\
57	4.36713372714568e-05\\
58	7.37689047369668e-05\\
59	0.00178104510312393\\
60	0.000868864514786634\\
61	0.000556187443965141\\
62	0.00328932781347913\\
63	0.00745069858226208\\
64	0.00182494462540096\\
65	0.000108910989812074\\
66	3.97022728796313e-05\\
67	8.36686035317642e-05\\
68	0.00132275762883747\\
69	0.00154716974898984\\
70	0.000639815786911996\\
71	0.00344206766547907\\
72	0.00267172908871068\\
73	0.000129515150253514\\
74	3.93569784596502e-05\\
75	0.000105332504822604\\
76	0.00148569949848302\\
77	0.00226517091661847\\
};
\addlegendentry{Four}

\end{axis}
\end{tikzpicture}%

%% file: Figures/SA/Compare3_compliance.tex
%
%
\begin{tikzpicture}

\begin{axis}[%
width=3.561in,
height=3.776in,
at={(0.844in,0.51in)},
scale only axis,
separate axis lines,
every outer x axis line/.append style={black},
every x tick label/.append style={font=\color{black}},
every x tick/.append style={black},
xmin=0,
xmax=80,
xtick={\empty},
xlabel={\large{Design variable}},
every outer y axis line/.append style={black},
every y tick label/.append style={font=\color{black}},
every y tick/.append style={black},
ymode=log,
ymin=1e-08,
ymax=0.01,
yminorticks=true,
ylabel={\large{Relative error}},
yticklabel style={font=\large},
axis background/.style={fill=white},
xmajorgrids,
ymajorgrids,
yminorgrids,
legend style={font=\fontsize{10}{15}\selectfont,legend cell align=left, align=left, draw=white!15!black},
legend pos=south east
]
\addplot[only marks, mark=*, mark options={}, mark size=1.5811pt, color=blue, fill=blue] table[row sep=crcr]{%
x	y\\
1	0.00229414621715522\\
2	0.00213388095832595\\
3	0.00155908952375206\\
4	0.000879037357337227\\
5	0.000747340858421587\\
6	0.00150552192589242\\
7	0.00166168336644394\\
8	0.00293050651851832\\
9	0.00118258931956834\\
10	0.000278776808339045\\
11	0.000854263938980003\\
12	0.000125936155125708\\
13	0.000353115362303683\\
14	0.000197250514171717\\
15	0.00101157751696572\\
16	0.00286764101801068\\
17	0.00299342203538112\\
18	0.00230304824269271\\
19	0.00122397767363962\\
20	0.000157405815650489\\
21	8.58633518532858e-05\\
22	0.000117667761358976\\
23	0.0010636485467832\\
24	0.00213518149413394\\
25	0.00296809982605809\\
26	0.00307042816223344\\
27	0.00264982376554174\\
28	0.00208150164686156\\
29	0.00124656967603295\\
30	0.000671901373541697\\
31	0.00112485243106259\\
32	0.00180943184702842\\
33	0.00246306636493537\\
34	0.00296043114445775\\
35	0.00307989188543059\\
36	0.00210920657451546\\
37	0.00169103247234021\\
38	0.00174451415103949\\
39	0.0018665152080233\\
40	0.00160218405885951\\
41	0.0014053460839687\\
42	0.00186564331164815\\
43	0.00295950813073633\\
44	0.00308475371322164\\
45	0.00264634485113624\\
46	0.00207928284386533\\
47	0.00124710522472654\\
48	0.000669122830677547\\
49	0.00112317422981924\\
50	0.00181182030718761\\
51	0.00246294551075371\\
52	0.00295168034822813\\
53	0.00299647247579238\\
54	0.00230218835496744\\
55	0.00122535066551162\\
56	0.00016254904462287\\
57	8.59666067504152e-05\\
58	0.000117079896499229\\
59	0.00105981067653346\\
60	0.00213831568399575\\
61	0.00298777147813311\\
62	0.00301248704548764\\
63	0.00118049031543541\\
64	0.000280276910619355\\
65	0.000859229686237915\\
66	0.000138017475231671\\
67	0.000360344385750826\\
68	0.000195852364382694\\
69	0.000994632921112733\\
70	0.00301360206011115\\
71	0.00235447117466822\\
72	0.00214150288170057\\
73	0.00153926202343402\\
74	0.000866924273495647\\
75	0.000755346116332401\\
76	0.00150368261344048\\
77	0.00177738849920418\\
};
\addlegendentry{C\'ea's Method}

\addplot[only marks, mark=*, mark options={}, mark size=1.5000pt, color=red, fill=red] table[row sep=crcr]{%
x	y\\
1	0.00010814104795117\\
2	7.82807943622168e-05\\
3	4.08770090602076e-05\\
4	8.7753511143762e-05\\
5	8.15047173790478e-05\\
6	7.53583626053541e-05\\
7	6.50061416415517e-05\\
8	6.25663197611681e-05\\
9	3.48867838228564e-05\\
10	3.696405612585e-05\\
11	0.000327660434005118\\
12	0.00121323349331907\\
13	0.000290162310173659\\
14	5.9364362557623e-05\\
15	6.03194333113333e-05\\
16	0.000138069731168763\\
17	1.70761221733641e-05\\
18	1.72495740880399e-05\\
19	2.67852345695149e-05\\
20	0.000526634980385198\\
21	0.00150873679620245\\
22	0.000499926529608355\\
23	4.20967301263563e-05\\
24	3.75869178823021e-05\\
25	2.62085914068701e-05\\
26	1.37444617966803e-05\\
27	5.83249819809352e-05\\
28	0.000104555021749971\\
29	0.000586307389092454\\
30	0.00203463250476353\\
31	0.000679067187601462\\
32	0.000144573929869581\\
33	7.44502251386635e-05\\
34	4.73670045268646e-07\\
35	9.49303438634569e-06\\
36	0.000238236333522809\\
37	0.00034291352616358\\
38	0.000631363319915493\\
39	0.002397882024273\\
40	0.00097559007261191\\
41	0.000553782194772377\\
42	0.000379505357098923\\
43	1.06676027252649e-05\\
44	2.79592525044726e-05\\
45	5.48052366559096e-05\\
46	0.000102311109227597\\
47	0.000586822062015082\\
48	0.00203184247916433\\
49	0.00067733678189129\\
50	0.000146879688426745\\
51	7.42064342615495e-05\\
52	8.4118636103607e-06\\
53	1.41356726129164e-05\\
54	1.8213860626139e-05\\
55	2.55217415243832e-05\\
56	0.000521458262726837\\
57	0.00150888442316937\\
58	0.000499251713389026\\
59	4.61472859290818e-05\\
60	3.46689838318267e-05\\
61	6.78852354985738e-06\\
62	1.90560782437836e-05\\
63	3.71174717952394e-05\\
64	3.86068376994945e-05\\
65	0.000322608310759022\\
66	0.00120113622568436\\
67	0.00028274080386469\\
68	5.82463269279919e-05\\
69	7.75041973748941e-05\\
70	7.24998121706831e-06\\
71	0.000168769622125757\\
72	8.60841331424564e-05\\
73	6.05912850771958e-05\\
74	9.96566921742841e-05\\
75	7.31911575950121e-05\\
76	7.38425012308683e-05\\
77	0.000181229221768385\\
};
\addlegendentry{Parameterized shape SA}

\addplot[only marks, mark=*, mark options={}, mark size=1.5000pt, color=black, fill=black] table[row sep=crcr]{%
x	y\\
1	3.11345321913092e-05\\
2	3.72833131385645e-05\\
3	4.3948477422016e-05\\
4	3.48054349522533e-05\\
5	1.82780669163408e-05\\
6	1.57313499633244e-05\\
7	2.86283242921099e-05\\
8	4.29200382114856e-05\\
9	4.42718502766408e-06\\
10	7.40171814406715e-07\\
11	2.9581822607046e-06\\
12	2.97224635650726e-06\\
13	3.68314153625014e-07\\
14	4.28548977816738e-06\\
15	1.16370044775477e-06\\
16	0.000103905343824901\\
17	8.64167487507362e-06\\
18	1.39390615098992e-07\\
19	7.42045503804479e-07\\
20	2.58290530302504e-07\\
21	6.26889345285849e-07\\
22	3.62293217520302e-08\\
23	3.7293967432272e-07\\
24	4.52873571096136e-06\\
25	6.9119335138932e-06\\
26	7.71387910839668e-06\\
27	1.74715753144058e-06\\
28	1.58666458621083e-06\\
29	7.49478451752564e-07\\
30	2.14793204546147e-06\\
31	5.49791563793164e-06\\
32	3.32931256937376e-06\\
33	3.4123368114943e-06\\
34	2.38493508610605e-05\\
35	2.0820808831787e-05\\
36	1.3286643202604e-06\\
37	1.19011630614959e-06\\
38	6.62819481890563e-06\\
39	5.68860638318104e-06\\
40	5.09056664593796e-06\\
41	1.69038998828789e-06\\
42	7.38486462934364e-07\\
43	2.77942396558058e-05\\
44	6.62400883610197e-06\\
45	1.65821659264816e-06\\
46	5.55640480211556e-07\\
47	1.74126030413231e-07\\
48	4.93417384578253e-06\\
49	7.06187294764811e-06\\
50	7.32431422929313e-07\\
51	3.32082792780378e-06\\
52	3.23676479321926e-05\\
53	5.35849221537296e-06\\
54	4.27589300318786e-07\\
55	9.57216212085935e-07\\
56	4.78042544231039e-06\\
57	5.35010503748976e-07\\
58	3.09904917380172e-07\\
59	2.66249116783396e-06\\
60	5.6059812087266e-07\\
61	1.35170873731452e-05\\
62	3.91289648530012e-05\\
63	2.75791173190304e-06\\
64	3.08940046827227e-07\\
65	7.80039454553984e-06\\
66	9.1218695514305e-06\\
67	7.35688829464285e-06\\
68	2.01005859644867e-06\\
69	1.70805616056798e-05\\
70	4.2646663003088e-05\\
71	9.09446579720165e-05\\
72	4.43819390986412e-05\\
73	2.39434019688231e-05\\
74	2.26919480918369e-05\\
75	1.04964355312242e-05\\
76	1.30613153562526e-05\\
77	8.62793165436817e-05\\
};
\addlegendentry{Discretized domain SA}

\end{axis}
\end{tikzpicture}%

%% file: Figures/SA/Stress_ParaVsDom.tex
%
%
\begin{tikzpicture}

\begin{axis}[%
width=3.561in,
height=3.776in,
at={(0.844in,0.51in)},
scale only axis,
separate axis lines,
every outer x axis line/.append style={black},
every x tick label/.append style={font=\color{black}},
every x tick/.append style={black},
xmin=0,
xmax=80,
xtick={\empty},
xlabel={\large{Design variable}},
every outer y axis line/.append style={black},
every y tick label/.append style={font=\color{black}},
every y tick/.append style={black},
ymode=log,
ymin=1e-08,
ymax=1,
yminorticks=true,
ylabel={\large{Relative error}},
yticklabel style={font=\large},
axis background/.style={fill=white},
xmajorgrids,
ymajorgrids,
yminorgrids,
legend style={font=\fontsize{10}{15}\selectfont,legend cell align=left, align=left, draw=white!15!black},
legend pos=south east
]

\addplot[only marks, mark=*, mark options={}, mark size=1.5000pt, color=red, fill=red] table[row sep=crcr]{%
x	y\\
1	0.0133216739019405\\
2	0.0105147692962844\\
3	0.000941941262127596\\
4	0.000556602344444673\\
5	0.000780970984461668\\
6	0.00631019847299841\\
7	0.00958750300415788\\
8	0.0141220816547718\\
9	0.0269815170660798\\
10	0.00691382261479632\\
11	0.000709828489050697\\
12	0.000537733458064419\\
13	0.000616075183482077\\
14	0.00560471680594875\\
15	0.00669890359753276\\
16	0.00418454998783835\\
17	0.0945096067432409\\
18	0.0155425679842902\\
19	0.00765858162823926\\
20	0.000589791256141977\\
21	0.000638632120107827\\
22	0.000510129046403339\\
23	0.00820783886580424\\
24	0.00428441188868872\\
25	0.00332671610458616\\
26	0.00450601151266287\\
27	0.00287768512084644\\
28	0.00274228660984939\\
29	0.00588208342259355\\
30	0.00115954007980065\\
31	0.00498804978044343\\
32	0.00243483664796222\\
33	0.00230957471355757\\
34	0.0028725229820101\\
35	0.00362193623782799\\
36	0.00183414092502261\\
37	0.00166954642248843\\
38	0.00160312797480413\\
39	0.000315376642741461\\
40	0.00107477255717832\\
41	0.00137027606704022\\
42	0.00154622537532495\\
43	0.00284265078185277\\
44	0.00449734687501743\\
45	0.00288229554266093\\
46	0.0027442705956652\\
47	0.00588451588656201\\
48	0.00116367941517888\\
49	0.00498881518967442\\
50	0.0024315769406504\\
51	0.00230804112016008\\
52	0.00288472028241671\\
53	0.0939578818418045\\
54	0.01554846802935\\
55	0.00765615010860971\\
56	0.000591358137706323\\
57	0.000639816234658789\\
58	0.000510557050208057\\
59	0.00822051818505075\\
60	0.00427736851506871\\
61	0.00327171204006123\\
62	0.0134175889020772\\
63	0.0270046590402291\\
64	0.00691997142560569\\
65	0.000710586785676707\\
66	0.000540496330936289\\
67	0.000618437041093182\\
68	0.00559980663317268\\
69	0.00678142616602618\\
70	0.00353988452714025\\
71	0.013318198615482\\
72	0.0105365323848692\\
73	0.000939175869965165\\
74	0.000556259872573693\\
75	0.000781325641956841\\
76	0.00631596345539459\\
77	0.00989451006420479\\
};
\addlegendentry{Parameterized shape SA}

\addplot[only marks, mark=*, mark options={}, mark size=1.5000pt, color=black, fill=black] table[row sep=crcr]{%
x	y\\
1	1.90675225049618e-05\\
2	2.21886910259566e-05\\
3	2.57446573654322e-06\\
4	1.45851452981794e-06\\
5	4.07517436409293e-06\\
6	1.19817333617934e-06\\
7	0.000120091151176655\\
8	7.04410839058051e-05\\
9	1.5333399527203e-05\\
10	6.37423362980626e-07\\
11	5.22861968968198e-07\\
12	1.54110717312321e-06\\
13	1.02409441697596e-06\\
14	3.60909308334579e-06\\
15	4.65509467663607e-05\\
16	6.51517585853566e-05\\
17	0.00106808728311148\\
18	6.09194795740192e-06\\
19	5.04080321294701e-07\\
20	8.33735995000458e-07\\
21	3.31207837235444e-06\\
22	1.67760267699543e-07\\
23	6.437400786299e-07\\
24	3.23830450585338e-06\\
25	9.7006586596617e-06\\
26	2.18700803930449e-06\\
27	4.22550604826254e-06\\
28	3.70411063216868e-07\\
29	5.64430435475776e-06\\
30	3.65303629310436e-06\\
31	1.38775983853824e-05\\
32	1.86626318627071e-06\\
33	2.6842603208514e-06\\
34	3.07383602603739e-05\\
35	5.90369063315182e-05\\
36	5.99380172052299e-07\\
37	9.65908093440602e-07\\
38	3.59076842032741e-06\\
39	8.20502923471365e-07\\
40	2.59110989394488e-06\\
41	3.59556956298211e-07\\
42	2.39779064627975e-07\\
43	6.68447820828369e-05\\
44	2.18062787783108e-05\\
45	2.41311430219579e-06\\
46	6.26627247208334e-07\\
47	1.02309418841901e-05\\
48	8.72847473453365e-06\\
49	9.43238920012197e-06\\
50	8.57176816869212e-07\\
51	5.40004493765499e-07\\
52	4.05343171652566e-06\\
53	0.000346799297797243\\
54	1.21721158317747e-05\\
55	3.98580301854189e-06\\
56	5.84579752262756e-07\\
57	2.34111671069162e-06\\
58	8.27751523659724e-07\\
59	1.70888276239115e-06\\
60	1.92253948416084e-07\\
61	3.72434694173293e-05\\
62	0.00122744333429879\\
63	6.86196153542342e-05\\
64	2.46623091920054e-06\\
65	2.20137094458666e-06\\
66	6.78254389272013e-07\\
67	7.00597700286978e-07\\
68	8.1041365654483e-07\\
69	3.16143714128554e-05\\
70	0.000222930908385557\\
71	0.000113298433835554\\
72	9.46201798113968e-05\\
73	4.17213248282743e-06\\
74	6.29180317710826e-06\\
75	3.39124128657563e-06\\
76	8.88517220519894e-06\\
77	0.000268616738793664\\
};
\addlegendentry{Discretized domain SA}

\end{axis}
\end{tikzpicture}%

%% file: Figures/OptDes/ComplianceFirstRun_Conv.tex
%
%
\begin{tikzpicture}

\begin{axis}[%
width=3.678in,
height=3.558in,
at={(0.696in,0.691in)},
scale only axis,
separate axis lines,
every outer x axis line/.append style={black},
every x tick label/.append style={font=\color{black}},
every x tick/.append style={black},
xmin=0,
xmax=150,
xlabel={\large{Iteration number}},
xticklabel style={font=\large},
every outer y axis line/.append style={black},
every y tick label/.append style={font=\color{black}},
every y tick/.append style={black},
ymin=40,
ymax=50,
ylabel={\large{Compliance}},
yticklabel style={font=\large},
axis background/.style={fill=white},
xmajorgrids,
ymajorgrids,
legend style={font=\fontsize{10}{15}\selectfont,legend cell align=left, align=left, draw=white!15!black}
]
\addplot [color=black, line width=1.5pt]
  table[row sep=crcr]{%
1	49.4759562845563\\
2	49.3747310639346\\
3	49.2460616275212\\
4	49.0828044028419\\
5	48.8793242071745\\
6	46.905005153571\\
7	45.9643729566146\\
8	44.4895497334796\\
9	42.6679932392091\\
10	44.2241188741344\\
11	43.5970378065192\\
12	43.339037572761\\
13	43.0032100000711\\
14	43.1128333167905\\
15	43.0893210393611\\
16	43.0688860216944\\
17	43.0434804157607\\
18	43.0134127848044\\
19	42.9777281012819\\
20	42.9349703285952\\
21	42.5967236296546\\
22	42.8875873752688\\
23	42.6865543619013\\
24	42.4916967292551\\
25	41.6152284127246\\
26	42.1859777030532\\
27	41.7410176558337\\
28	41.6986146253536\\
29	41.5331890520136\\
30	41.4350990536834\\
31	41.4702837962625\\
32	41.4404665948274\\
33	41.4100502954166\\
34	41.3799858854641\\
35	41.3519741532157\\
36	41.3241253052391\\
37	41.2056531683999\\
38	41.2462231007503\\
39	41.2628111836053\\
40	41.200539405948\\
41	41.0960251892393\\
42	41.2123940476452\\
43	41.0470491132224\\
44	41.1340938702564\\
45	41.0683081477223\\
46	41.0615066255899\\
47	41.1053916624928\\
48	40.9957888595179\\
49	41.1027345848509\\
50	41.0362197567867\\
51	41.0333272071419\\
52	41.0323027633944\\
53	40.9860232684139\\
54	41.097761397852\\
55	41.0306348464952\\
56	41.0279558041294\\
57	41.0270758745532\\
58	41.0262028034805\\
59	41.0253244322684\\
60	40.9791053357978\\
61	41.089026243392\\
62	41.0242752548247\\
63	41.0224178494901\\
64	41.0219569774321\\
65	41.0214738254211\\
66	40.8917110810497\\
67	41.1550557825501\\
68	40.8989314850996\\
69	41.1592963578067\\
70	40.907650347941\\
71	41.1696396947704\\
72	41.0538355480495\\
73	40.797478086922\\
74	41.2455845546248\\
75	41.117959352565\\
76	40.9216382440506\\
77	41.2260096975382\\
78	41.1368258074076\\
79	40.9500224141606\\
80	41.1702496933068\\
81	40.9848529417215\\
82	41.168886296351\\
83	41.1318288020892\\
84	40.9799152844976\\
85	41.1404227346186\\
86	40.9925706842603\\
87	41.1343352873722\\
88	40.9979834904144\\
89	41.1309204871996\\
90	41.1240526120301\\
91	40.9948279430646\\
92	41.1254597974037\\
93	40.9967554924742\\
94	41.1249306282909\\
95	41.1211569306592\\
96	40.9938744756683\\
97	41.1222770138434\\
98	40.9949166564268\\
99	41.1224526003799\\
100	40.9951035676634\\
101	41.1226277356425\\
102	40.9952901238261\\
103	41.1228025982471\\
104	40.9954765093308\\
105	41.1229772669368\\
106	40.9956628466506\\
107	41.1231519854418\\
108	40.9958488981358\\
109	41.1233263258358\\
110	40.996035041473\\
111	41.1235007745166\\
112	41.1199770181058\\
113	40.992621582014\\
114	41.1210994336248\\
115	40.9936627695641\\
116	41.1212747199683\\
117	40.9938490745142\\
118	41.1214492978276\\
119	40.9940350779129\\
120	41.1216236308801\\
121	40.9942207399372\\
122	41.1217976161482\\
123	40.9944062410888\\
124	41.1219714614173\\
125	40.9945914507923\\
126	41.1221446625423\\
127	40.9947763020057\\
128	41.1211355401907\\
129	41.1194952222014\\
130	40.9940049136134\\
131	41.1200691715789\\
132	40.9945693569976\\
133	41.1202150639169\\
134	40.994722242278\\
135	41.1203574158544\\
136	40.9948730556857\\
137	41.1204985024449\\
138	40.9950231501341\\
139	41.1206391333398\\
140	40.9951729880563\\
141	41.1207795451993\\
142	40.9953227358543\\
143	41.1209198964547\\
144	41.1192744182499\\
145	40.9937745964887\\
146	41.1198513621676\\
147	40.9943367027276\\
148	41.1199951513907\\
149	40.994488323078\\
150	41.1201364126921\\
};
\addlegendentry{C\'eas Method}

\addplot [color=blue, dashed, line width=1.0pt, mark size=1.0pt, mark=o, mark options={solid, blue}]
  table[row sep=crcr]{%
1	49.4759564247351\\
2	49.3747316914615\\
3	49.2460662175887\\
4	49.0828199886734\\
5	48.8793547289894\\
6	46.921981169816\\
7	46.0263237726608\\
8	44.6596560195852\\
9	43.0956760250201\\
10	43.6767688099778\\
11	43.1853784806183\\
12	43.0420493026237\\
13	43.1100170033487\\
14	43.0900223976525\\
15	43.0720650859092\\
16	43.0511329870119\\
17	43.02613295374\\
18	42.9962506533147\\
19	42.9605099631644\\
20	42.6303920520367\\
21	42.7830696853778\\
22	42.8358309961411\\
23	42.6002244715126\\
24	42.489069955884\\
25	41.6595315905757\\
26	42.2477393061062\\
27	41.7878824404006\\
28	41.8487593031319\\
29	41.7377869543261\\
30	41.6952319585474\\
31	41.6569312671507\\
32	41.6191154572243\\
33	41.5803761666808\\
34	41.5385494926879\\
35	41.3886972598386\\
36	41.4391981596956\\
37	41.2283493779317\\
38	41.3127472101805\\
39	41.3404588239085\\
40	41.2337591146034\\
41	41.2061403829289\\
42	41.1831098419377\\
43	41.1599278212777\\
44	41.1374637917153\\
45	41.1156423884424\\
46	41.1272542737471\\
47	41.0174811279371\\
48	41.1309559330889\\
49	41.0150288886377\\
50	41.0085445088399\\
51	41.0055508991777\\
52	41.0526789611592\\
53	40.9396303751228\\
54	41.0483339692333\\
55	40.9800105270268\\
56	40.9284181436947\\
57	41.0405317055409\\
58	40.9748595639932\\
59	40.9718554447756\\
60	40.9708245171858\\
61	40.9219142763534\\
62	41.0350338431907\\
63	40.8393703826603\\
64	41.11375425719\\
65	40.8575105349029\\
66	41.1257028092271\\
67	40.8812110332037\\
68	41.1430916717574\\
69	40.923602244026\\
70	41.1321264940359\\
71	40.9472567841209\\
72	41.1263129165281\\
73	41.0908572826778\\
74	40.9387812126075\\
75	41.1000073113585\\
76	41.0770804949509\\
77	40.9334399653897\\
78	41.0811162729173\\
79	40.9405415012328\\
80	41.0768775963123\\
81	40.9438094064381\\
82	41.0752089311971\\
83	40.9452690706386\\
84	41.0747963872291\\
85	41.071283599473\\
86	40.9420863422711\\
87	41.072195938188\\
88	40.9430761655309\\
89	41.0723701312293\\
90	40.9432621923584\\
91	41.0725443041868\\
92	40.9434481998672\\
93	41.072718457075\\
94	40.943634188082\\
95	41.0728925899178\\
96	40.9438201570235\\
97	41.0730667027345\\
98	40.9440061067082\\
99	41.0732407955396\\
100	40.9441920371606\\
101	41.0734148683538\\
102	40.9443779484014\\
103	41.0735889211983\\
104	40.9445638404496\\
105	41.07376295409\\
106	41.0703673550995\\
107	40.9411774578016\\
108	41.0713060434646\\
109	40.9421276141112\\
110	41.0714801756194\\
111	40.9423135766116\\
112	41.0716542876588\\
113	40.9424995197603\\
114	41.0718283796041\\
115	40.942685443583\\
116	41.072002451477\\
117	40.9428713480982\\
118	41.072176503296\\
119	40.9430572333238\\
120	41.0723505350781\\
121	40.9432430992878\\
122	41.0725245468471\\
123	40.943428946009\\
124	41.07269853862\\
125	40.9436147735043\\
126	41.072872510414\\
127	40.9438005818005\\
128	41.0718072770209\\
129	41.0703061413074\\
130	40.9433866793427\\
131	41.0707046142484\\
132	40.9438120024398\\
133	41.0708330321375\\
134	40.9439491893656\\
135	41.0709614440469\\
136	40.94408637078\\
137	41.0710898499916\\
138	41.0698743244405\\
139	40.9429261324066\\
140	41.0702722128527\\
141	40.9433512889267\\
142	41.0704006080578\\
143	40.9434884512402\\
144	41.0705289973012\\
145	40.9436256080632\\
146	41.0706573805879\\
147	40.9437627594001\\
148	41.070785757924\\
149	40.9438999052566\\
150	41.0709141293151\\
};
\addlegendentry{Discretized domain SA}

\addplot [color=red, dashed, line width=2.0pt]
  table[row sep=crcr]{%
1	49.4759562845563\\
2	49.3747331630091\\
3	49.246067997575\\
4	49.0828217993113\\
5	48.8793564227128\\
6	46.9052177283271\\
7	45.9647290405605\\
8	44.4901607948256\\
9	42.66842731467\\
10	44.2244020167432\\
11	43.5973196004919\\
12	43.3393536445066\\
13	43.0033218274044\\
14	43.1131151256051\\
15	43.0895792659601\\
16	43.0691448266124\\
17	43.0437478419542\\
18	43.0136924676591\\
19	42.9780132670894\\
20	42.9352649358224\\
21	42.5971432014876\\
22	42.8878392747766\\
23	42.6869261610336\\
24	42.4921651359753\\
25	41.6158336028533\\
26	42.1865906392702\\
27	41.7414845539548\\
28	41.6991956547851\\
29	41.5336067059452\\
30	41.4353398380204\\
31	41.4706322996494\\
32	41.4407976899432\\
33	41.4103922647121\\
34	41.3803579254118\\
35	41.3523747207044\\
36	41.3248188201479\\
37	41.2042679324594\\
38	41.24647383173\\
39	41.2629061138651\\
40	41.1653404510799\\
41	41.1879658017582\\
42	41.1542425355983\\
43	41.0423551000351\\
44	41.1315004867673\\
45	41.067000540234\\
46	41.060171269769\\
47	41.1022062613526\\
48	40.9955988815059\\
49	41.0981946435217\\
50	40.9893572241694\\
51	41.0966173177429\\
52	41.0326090998299\\
53	40.9838708966392\\
54	41.0980827012782\\
55	41.0328029963632\\
56	41.0280250983752\\
57	41.0274963424368\\
58	41.0269381450791\\
59	41.0263014294875\\
60	41.0255810949386\\
61	41.0247859761557\\
62	40.9781628258725\\
63	41.089158032379\\
64	41.0250313900577\\
65	41.022683861995\\
66	41.0222855864714\\
67	41.0218488792221\\
68	40.9753141998846\\
69	41.0862743528544\\
70	40.8938986963923\\
71	41.1656912387741\\
72	40.9147794240473\\
73	41.1786440342086\\
74	40.9415422843287\\
75	41.1953816729083\\
76	41.1091750789125\\
77	40.9275997706951\\
78	41.1616863622322\\
79	40.9757414594946\\
80	41.1603670493434\\
81	40.9988588362397\\
82	41.153027059701\\
83	41.0084585143043\\
84	41.0248071612983\\
85	41.1574712273058\\
86	41.0221826433073\\
87	41.1538766639319\\
88	41.0248756681857\\
89	41.152670899912\\
90	41.0261270055518\\
91	41.1522631271905\\
92	41.0265358117666\\
93	41.1523659489959\\
94	41.0267465426549\\
95	41.1525391720103\\
96	41.0269313062348\\
97	41.1527123363349\\
98	41.0271159728727\\
99	41.1528854575445\\
100	41.0273005153051\\
101	41.1530584132325\\
102	41.0274850960048\\
103	41.1532314792654\\
104	41.0276695239676\\
105	41.1534043465586\\
106	41.0278540174065\\
107	41.1535772783064\\
108	41.0280384179104\\
109	41.1537501223209\\
110	41.0282228266281\\
111	41.1539229958612\\
112	41.0284070936324\\
113	41.1540957152961\\
114	41.0285914196659\\
115	41.1542685291236\\
116	41.0287756361799\\
117	41.1544412180522\\
118	41.1509461118079\\
119	41.025338720782\\
120	41.1520061083024\\
121	41.0263716189121\\
122	41.1521797481544\\
123	41.0265563632369\\
124	41.1523529280786\\
125	41.0267407989539\\
126	41.1525258552031\\
127	41.0269249710434\\
128	41.1515239166414\\
129	41.0278543071719\\
130	41.15167041221\\
131	41.0280064763972\\
132	41.1518119330422\\
133	41.0281558747348\\
134	41.1519514498784\\
135	41.028304001446\\
136	41.1520902619129\\
137	41.1504655311489\\
138	41.026770369076\\
139	41.1510293090481\\
140	41.0273253025759\\
141	41.1511711665166\\
142	41.0274747336207\\
143	41.1513104310841\\
144	41.0276225399643\\
145	41.1514489050526\\
146	41.0277697982806\\
147	41.1515869521969\\
148	41.0279166515466\\
149	41.1517242437361\\
150	41.0280628941706\\
};
\addlegendentry{Parameterized shape SA}

\end{axis}
\end{tikzpicture}%